\documentclass[12pt,a4paper]{article}

\pdfoutput=1

\usepackage[utf8]{inputenc}
\usepackage{graphicx}
\usepackage{url}
\usepackage{a4wide}
\usepackage{mathtools}
\usepackage{amssymb}
\usepackage{stmaryrd}

\newcommand{\noteperso}[1]{\begin{center}
 \fbox{\begin{minipage}{12cm}#1\end{minipage}}\end{center}}
\renewcommand{\noteperso}[1]{}

\graphicspath{{Figures/}}

\newcommand{\keywords}[1]{\bigskip\noindent\textbf{\textit{Keywords---}} #1}

\newcommand{\ie}{{\em i.e.}}

\newcommand{\graphstreamskip}{\medskip}
\newcommand{\notionskip}{\smallskip}

\newcommand{\relationskip}{\smallskip}
\newcommand{\exampleskip}{\smallskip}

\newcommand{\uniformity}{\Cup}

\newcommand{\substreameq}{\subseteq}

\newcommand{\distance}{\partial}
\newcommand{\latency}{\ell}
\newcommand{\timetoreach}{\mathcal{T}}
\newcommand{\cost}{c}
\newcommand{\closeness}{\mathcal{C}}
\newcommand{\betweenness}{\mathcal{B}}
\newcommand{\reaches}{\stackrel{\footnotesize\gamma}{\longleadsto}}
\renewcommand{\reaches}{\dashrightarrow}
\newcommand{\notreaches}{\longarrownot\dashrightarrow}
\newcommand{\weakreaches}{\mbox{\,-\,-\,-\,}}
\newcommand{\graphreaches}{\mbox{\,---\,}}

\newcommand{\sfp}{shortest fastest path}
\newcommand{\bfrac}[2]{\ensuremath{\frac{\sigma((i,#1),(j,#2),(t,v))}{\sigma((i,#1),(j,#2))}}}
\newcommand{\Xbfrac}[2]{\ensuremath{\frac{\sigma((i,#1),(j,#2),X)}{\sigma((i,#1),(j,#2))}}}

\newcommand{\presence}[2][]{\ensuremath{{T}^{#1}_{#2}}}
\newcommand{\T}[2][]{\ensuremath{{T}^{#1}_{#2}}}
\newcommand{\V}[2][]{\ensuremath{{V}^{#1}_{#2}}}
\newcommand{\E}[2][]{\ensuremath{{E}^{#1}_{#2}}}
\newcommand{\G}[2][]{\ensuremath{{G}^{#1}_{#2}}}

\newcommand{\linegraph}[1]{\widehat{#1}}
\newcommand{\myS}{\ensuremath{S = (T,V,W,E)}}
\newcommand{\myL}{\ensuremath{L = (T,V,E)}}

\newcommand{\cov}[1]{\mbox{cov}(#1)}
\newcommand*{\diff}{\mathop{}\!\mathrm{d}}

\newcommand{\demidelta}{\frac{\Delta}{2}}

\newcommand{\myscale}{0.45}

\begin{document}

\begin{center}
{\Large \bf
Stream Graphs and Link Streams\\
\medskip
for the Modeling of Interactions over Time
}\\
\medskip
Matthieu Latapy\,\footnote{Contact author. \url{Matthieu.Latapy@lip6.fr}},
Tiphaine Viard,
Cl\'emence Magnien
\\
\bigskip
\end{center}

\bigskip
\begin{abstract}
Graph theory provides a language for studying the structure of relations, and it is often used to study interactions over time too. However, it poorly captures the both temporal and structural nature of interactions, that calls for a dedicated formalism. In this paper, we generalize graph concepts in order to cope with both aspects in a consistent way. We start with elementary concepts like density, clusters, or paths, and derive from them more advanced concepts like cliques, degrees, clustering coefficients, or connected components. We obtain a language to directly deal with interactions over time, similar to the language provided by graphs to deal with relations. This formalism is self-consistent: usual relations between different concepts are preserved. It is also consistent with graph theory: graph concepts are special cases of the ones we introduce. This makes it easy to generalize higher-level objects such as quotient graphs, line graphs, $k$-cores, and centralities. This paper also considers discrete versus continuous time assumptions, instantaneous links, and extensions to more complex cases.
\end{abstract}


\newpage

\tableofcontents

\newpage



\noteperso{ai ajouté que les éléments de W sont des "temporal nodes"...}

\noteperso{DRAWING.PY : intégrer les couleurs, angles pour les chemins}

\noteperso{Parler des $(b,e,u,v)$ ? notation chevrons ? relier aux temps inter-contact ; définir les liens instantanés (b=e)}



\section{Introduction}


Friendship, dependencies, similarities, or connections are typical examples of {\bf relations} modeled by {\bf graphs} or networks, \ie\ sets of nodes and links: nodes represent individuals and two individuals are linked together if they are friends; nodes represent companies and they are linked together if they signed contracts with each other; nodes represent documents like web pages or articles, and they are linked together if they are similar; nodes represent computer devices and they are linked together if there is a wire between them; etc.

For decades, graph theory, social network analysis and network science have developped a wide set of tools for the study of such graphs. In particular, they developed a {\em language} for describing networks, with elementary yet powerful concepts such as node degree (their number of links), paths (sequences of links going from one node to another one), density (the fraction of pair of nodes actually linked together), or cliques (sets of nodes all pairwise linked together). This language forms the basis of network studies, and there is a global consensus on a wide set of concepts that are used in the field; with few variations, all courses and reference books on graphs and networks start with them, see for instance \cite{berge1962theory,Bondy:1976:GTA:1097029,wasserman1994social,west_introduction_2000,David:2010:NCM:1805895,Newman:2010:NI:1809753,DBLP:books/daglib/0030488,barabasi2016network,scott2017social}. Then, more advanced and specific concepts are defined on this common ground.

\graphstreamskip


Contacts, shopping, travels, or traffic are typical examples of {\bf interactions} that take place over time, \ie\ {\bf streams} of nodes and links active during specific periods of time: nodes are individuals linked together whenever they call each other; nodes are clients and products linked together when a client buys a product; nodes are places linked together when someone moves from one place to another; nodes are internet devices linked together when they exchange data; etc.

Such sequences of interactions play a key role in many areas, and they have been studied for a long time, see related work in Section~\ref{sec:related}. Although many variations exist, the most common approach is to model them by sequences of graphs (each graph then aggregates the interactions that occurred during a period of time), by labeled graphs (each link being labeled with its presence times), or other augmented graphs. This makes it possible to use graph theory to study these sequences of graphs, labeled graphs, and other variants. Other works deal directly with higher-level methods for studying graphs, like stochastic block models for instance, and extend them to cope with the dynamics. Finally, a few works define specific properties combining temporal and structural information, such as centrality measures for instance.

\notionskip


In this paper, we propose a different approach: {\bf we develop a formalism to directly cope with interactions over time, in a way similar to what graph theory does for relations}. This means that we do not transform interactions into graphs, but rather transform graph theory into a theory of interactions over time. We model them as {\em link streams} and {\em stream graphs} (depending on whether the dynamics is on links only, or on both nodes and links), so named in order to emphasize their streaming nature and the fact that they are {\em not} graphs or networks.
Then, we start with the most elementary graph concepts
and we define their equivalent for stream graphs and link streams. Finally, we elaborate on these basic concepts to extend more complex graph concepts.
With the aim to make our formalism as intuitive as possible, we put much effort in proposing simple definitions, explaining them with different points of view (especially combinatorial and probabilistic ones), and to provide illustrations and detailed examples of all key concepts we introduce.

In addition to these subjective features, we also put much emphasis on two more objective features to ensure the relevance of our definitions. First, we want our formalism to be a generalization of graph theory in a very precise sense: when the stream has no dynamics, it is equivalent to a classical graph and its properties should be the same as those of this graph (see the end of Section~\ref{sec:stream-graphs-and-link-streams}). Second, we want the relations that exist between various graph properties (between density and degree for instance) to still hold for stream properties. Similarly, if a graph concept is derived from another one (like clustering coefficient from density for instance) we want the corresponding stream concept to be derived from the corresponding other stream concept. {\bf These features ensure both the self-consistency of our formalism and its consistency with graph theory.}

\notionskip

After Section~\ref{sec:preliminaries} that introduces a few notations needed in the whole paper, we present our framework from Section~\ref{sec:stream-graphs-and-link-streams} to Section~\ref{sec:centralities}. Each of these sections is devoted to a key concept of graph theory that we redefine in the stream context. Therefore, they all have the same structure: first we recall the relevant graph concepts and their key properties, in italics; then we introduce equivalent concepts for stream graphs with detailed examples and discuss their properties; we introduce additional related concepts specific to stream graphs; we discuss the case of link streams, \ie\ when there is no dynamics on nodes; and we show that the newly introduced stream concepts are equivalent to the graph ones, whenever this makes sense. After these core sections, we show how our framework may be used under either discrete and continuous modeling of time in Section~\ref{sec:continuous-vs-discrete}; we show how it generalizes $\Delta$-analysis and may be used with instantaneous links in Section~\ref{sec:delta}; we show how it may be extended to bipartite streams and other particular cases in Section~\ref{sec:generalizations}; and we present related work in Section~\ref{sec:related-work}. We discuss our contributions and future work in Section~\ref{sec:conclusion}.

\section{Preliminary notations}
\label{sec:preliminaries}

In this paper, we rely on a few notations that we introduce below.

Given two finite sets $X$ and $Y$, one may consider the ordered pairs $(x,y)$ with $x\in X$ and $y \in Y$. Then, $(x,y) \not= (y,x)$ and $(x,x)$ exists if $x\in X$ and $x\in Y$. One may also consider unordered pairs $xy$ with $x\in X$ and $y \in Y$, with $x\ne y$. Then, $xy=yx$ and $xx$ does not exist. The {\bf set of ordered pairs} is denoted by $X\times Y$, and one often uses this notation for the set of unordered pairs too. In this paper, however, we use both notions intensively and need to make a clear distinction between them. We therefore denote the {\bf set of unordered pairs of distinct elements} by $X\otimes Y$.

Throughout this paper, we deal with {\bf set sizes}, denoted by $|X|$ for a given set $X$, but the meaning of this notation depends on the type of $X$. If $X$ is an interval $[\alpha,\omega]$ of $\mathbb{R}$, then $|X| = \omega-\alpha$. If it is an interval $[\alpha,\omega]$ of $\mathbb{N}$ then $|X| = \omega-\alpha +1$. If $X$ is the union of disjoint intervals of $\mathbb{R}$, then $|X|$ is the sum of these intervals' sizes. The same holds if it is the union of disjoint intervals of $\mathbb{N}$. If $X$ is the product of sets of these types, then its size is the product of their sizes.
Notice that, if $X$ contains just one element then depending on the context it may be seen as a (degenerate) interval of $\mathbb{R}$ or $\mathbb{N}$, thus having size $0$ or $1$, respectively.
For instance, the union of the intervals $[1,2]$ and $[3,3]$ of $\mathbb{R}$ has size $1$, while the union of the same intervals of $\mathbb{N}$ has size $3$.

Notice that $|X\times Y| = |X|\cdot |Y|$, and so $|X\times X| = n^2$ if $|X| = n$. This is different from $|X\otimes Y| = |(X\setminus Y) \times Y| + |(Y\setminus X) \times X| - |(X\setminus Y)\times(Y\setminus X)| + \frac{|X\cap Y|^2 - |X\cap Y|}{2}$, leading to $|X\otimes X| = \frac{n \cdot (n-1)}{2}$ if $|X| = n$, and $|X\otimes Y| = |X| \cdot |Y|$ if $X$ and $Y$ are disjoint.

\section{Stream graphs and link streams}
\label{sec:stream-graphs-and-link-streams}

{\em
A (simple undirected\,\footnote{Unless explicitly specified, we always consider simple and undirected graphs and stream graphs; we discuss more general cases in Section~\ref{sec:extensions}.}) graph $G=(V,E)$ is defined by a finite set of nodes $V$ and a set of links $E \subseteq V\otimes V$: $uv \in E$ means that $u$ and $v$ are linked together in $G$.

Graphs model relations between nodes. For instance, nodes may represent individuals and links may represent friendship relations. Nodes may represent computers and links may represent physical connections between them. Examples are countless, making graphs the key formalism for studying network structures.
}

\graphstreamskip

We define a (simple undirected\,\footnotemark[2]) {\bf stream graph} \myS\ by a finite set of nodes $V$, a set of time instants $T$, a set of temporal nodes $W \subseteq T \times V$, and a set of links $E \subseteq T\times V\otimes V$ such that $(t,uv) \in E$ implies $(t,u)\in W$ and $(t,v) \in W$. The set of time instants $T$ may be continuous or discrete, which has little influence on the following, as we explain in Section~\ref{sec:discrete-vs-continuous}. Until then, all the examples we give assume that $T$ is an interval of $\mathbb{R}^+$.

We define $v_t = 1$ if $(t,v)\in W$ and $v_t = 0$ otherwise, as well as $uv_t=1$ if $(t,uv)\in E$ and $uv_t=0$ otherwise. When $v_t=1$ we say that node $v$ is involved in $S$ at time $t$ or that $v$ is present at time $t$, and when $uv_t=1$ we say that nodes $u$ and $v$ are linked together at time $t$, or that link $uv$ is present at time $t$. We denote by $\presence{v}$ the set of time instants at which $v$ is present, by $\presence{uv}$ the set of time instants at which $uv$ is present, by $\V{t}$ the set of nodes present at time $t$, and by $\E{t}$ the set of links present at time $t$: $\presence{v} = \{t, v_t=1\}$, $\presence{uv} = \{t, uv_t=1\}$, $\V{t} = \{v, v_t=1\}$, and $\E{t} = \{uv, uv_t=1\}$. Notice that $\presence{uv} \subseteq \presence{u} \cap \presence{v}$.


\notionskip

If all nodes are present all the time, \ie\ $\presence{v} = T$ for all $v$ or, equivalently, $\V{t} = V$ for all $t$, then we say that $S$ is a {\bf link stream} and we denote it by $L=(T,V,E)$ (with $W=T\times V$ implicitly). Indeed, there is no dynamics on nodes in this case, and $S$ is fully defined by this triplet.
Link streams play an important role in many situations, and so we pay special attention to this case in all this paper.

\exampleskip

\begin{figure}[!h]
\centering
\ \hfill
\includegraphics[scale=\myscale]{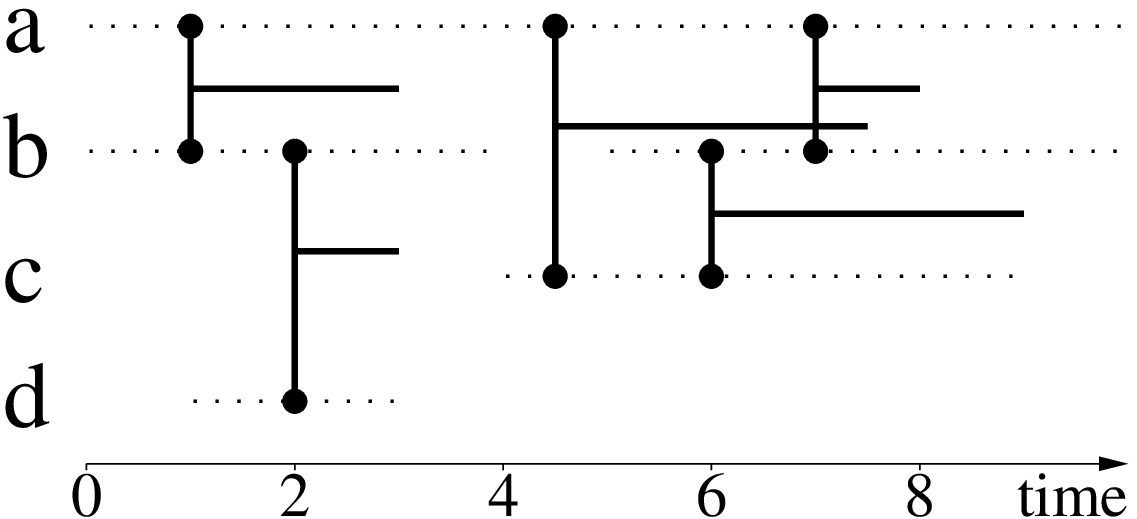}
\hfill
\includegraphics[scale=\myscale]{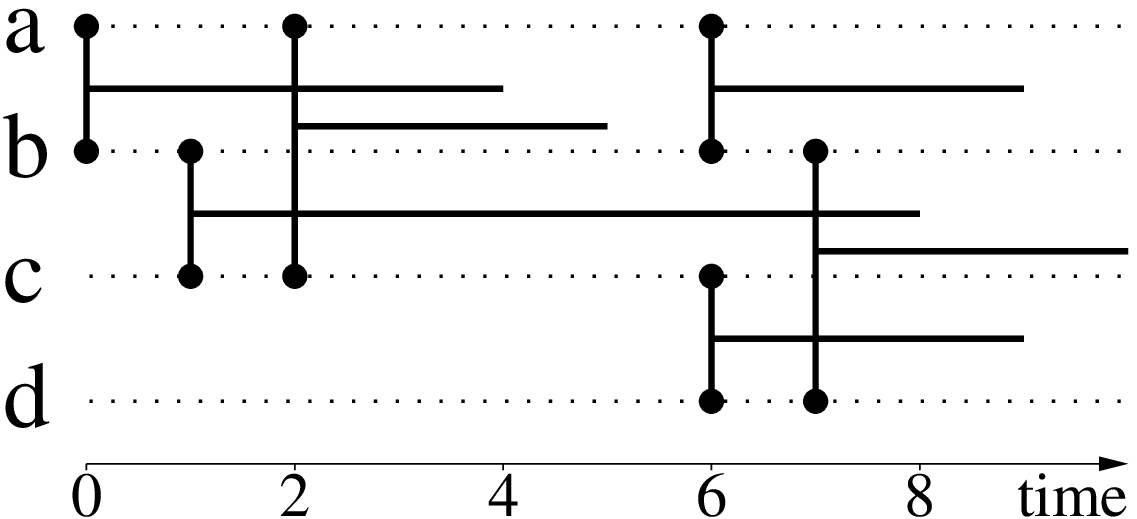}
\hfill \
\caption{
{\bf Simple examples of stream graphs and link streams.}
{\bf Left:} a stream graph \myS\ with $T = [0,10] \subseteq \mathbb{R}$, $V = \{a,b,c,d\}$, $W = [0,10]\times\{a\} \cup ([0,4]\cup[5,10])\times\{b\} \cup [4,9]\times\{c\} \cup [1,3]\times\{d\}$, and $E = ([1,3]\cup[7,8])\times\{ab\} \cup [4.5,7.5]\times\{ac\} \cup [6,9]\times\{bc\} \cup [2,3]\times\{bd\}$. In other words, $\presence{a} = [0,10]$, $\presence{b} = [0,4]\cup[5,10]$, $\presence{c} = [4,9]$, $\presence{d} = [1,3]$, $\presence{ab} = [1,3]\cup[7,8]$, $\presence{ac} = [4.5,7.5]$, $\presence{bc} = [6,9]$, $\presence{bd} = [2,3]$, and $\presence{ad} = \presence{cd} = \emptyset$.
{\bf Right:} a link stream $L=(T,V,E)$ with $T = [0,10] \subseteq \mathbb{R}$, $V = \{a,b,c,d\}$, and $E = ([0,4]\cup[6,9])\times\{ab\} \cup [2,5]\times\{ac\} \cup [1,8]\times\{bc\} \cup [7,10]\times\{bd\} \cup [6,9]\times\{cd\}$. In other words, $\presence{a} = \presence{b} = \presence{c} = \presence{d} = T$ and $\presence{ab} = [0,4]\cup[6,9]$, $\presence{ac} = [2,5]$, $\presence{bc} = [1,8]$, $\presence{bd} = [7,10]$ and $\presence{cd} = [6,9]$.
}
\label{fig:ex-ref}
\end{figure}

We illustrate these definitions in Figure~\ref{fig:ex-ref} with {\bf drawings} designed as follows. We display node names on a vertical axis on the left of the figure, and time on an horizontal axis at the bottom of the figure. Each node presence times are represented by an horizontal dotted line in front of its name, whenever the node is present. Each link presence times are represented by an horizontal solid line parallel to the two dotted lines of involved nodes, and a vertical solid line joining these two dotted lines (marked with bullets) when the two nodes start interacting. In Figure~\ref{fig:ex-ref}, for instance, in $S$ (leftmost example) the node $a$ arrives at time $0$ and stays until time $10$, and so $[0,10]\times\{a\} \subseteq W$, \ie\ $\presence{a} = [0,10]$. This is represented by a dotted line from time $0$ to $10$ in front of $a$ in the drawing. Likewise, $b$ arrives at time $0$, then leaves at time $4$, joins again at time $5$ and stays until time $10$, and so $([0,4]\cup[5,10])\times\{b\} \subseteq W$, \ie\ $\presence{b} = [0,4]\cup[5,10]$. This is represented by a dotted line from time $0$ to $4$ and another one from time $5$ to $10$ in front of $b$. These two nodes interact from time $1$ to time $3$ and from time $7$ to time $8$, and so $([1,3]\cup[7,8])\times\{ab\} \subseteq E$, \ie\ $\presence{ab} = [1,3]\cup[7,8]$. This is represented by a solid line at time $1$ between the dotted lines of $a$ and $b$, with an horizontal line starting from its middle until time $3$, and another such solid line at time $7$ with an horizontal line until time $8$.

\notionskip

Given a stream graph \myS, we define $\G{t} = (\V{t},\E{t})$, the {\bf graph induced by $S$ at time $t$}. In Figure~\ref{fig:ex-ref}, for instance, we obtain for $S$ at time $2$ the graph $G_2 = (\{a,b,d\},\{ab,bd\})$.

We also define $G(S) = (\{v, \presence{v}\ne \emptyset\}, \{uv, \presence{uv}\ne \emptyset\}) = (\bigcup_{t\in T} \V{t}, \bigcup_{t \in T}\E{t})$ the {\bf graph induced by $S$}: its nodes are those present in $S$ and they are linked together in $G(S)$ if there exists a time instant in $T$ such that they are linked together in $S$. In other words, it is the graph where there is a link between two nodes if they interacted at least once. In Figure~\ref{fig:ex-ref}, for instance, $G(S) = (\{a,b,c,d\},\{ab,ac,bc,bd\})$ and $G(L) = (\{a,b,c,d\},\{ab,ac,bc,bd,cd\})$.
One may in addition associate to each node $v$ or link $uv$ a weight capturing a quantity of interest, like for instance their presence duration $|\presence{v}|$ and $|\presence{uv}|$.

\notionskip

Stream graphs model interactions between nodes over time, as well as the dynamics of nodes themselves. For instance, nodes may represent individuals present in a given building and links may represent contacts between them. Nodes may represent on-line computers and links may represent data exchanges between them. Examples are countless, and we aim at making stream graphs the key formalism for studying jointly the dynamics and structure of interactions.

\notionskip

Since in a stream graph \myS\ nodes are not present all the time in general, $W$ may differ significantly from $T\times V$. To capture this, we define the {\bf coverage} of $S$ as follows:
$$
\cov{S} = \frac{|W|}{|T\times V|}.
$$
For instance, in Figure~\ref{fig:ex-ref} the stream graph $S$ has coverage $\cov{S}=\frac{26}{40}=0.65$.

Notice that $\cov{S} = 1$ if and only if all nodes are present all the time, and so it is equivalent to saying that $S$ is a link stream.

If in addition for all $u$ and $v$ in $V$, $\presence{uv} \in \{\emptyset,T\}$, \ie\ all existing links are present all the time, then there is no significant distinction between $S$ and $G(S)$, and we say that $S$ is a {\bf graph-equivalent stream}. This gives a formal ground to our wanted feature that stream graphs generalize graphs: we extend graph concepts to stream graphs in a way such that, if a stream graph has a given stream graph property and happens to be a graph-equivalent stream, then this graph has the corresponding graph property. In the following, we systematically check that this feature holds.

\section{Size, duration, uniformity and compactness}

{\em
The number of nodes of a graph $G=(V,E)$ is denoted by $n = |V|$ and its number of links by $m = |E|$.
}

\graphstreamskip

Given a stream graph $\myS$, we now define its number of nodes and links, as well as its duration. First notice that, unlike in graphs, some nodes may be present for much longer than others. In order to capture this, we define the {\bf contribution of node} $v$ as $n_v=\frac{|\presence{v}|}{|T|}$, which may be seen as the notion of coverage restricted to a node $v$. We then define the {\bf number of nodes} in $S$ as follows:
$$
n = \sum_{v\in V} n_v
 = \frac{|W|}{|T|}.
$$
Then, each node contributes to the total number of nodes proportionally to its involvement in $S$: $v$ in $V$ accounts for $1$ node only if it is present in $S$ all the time.

We define similarly the contribution of a pair of nodes $uv$ as $m_{uv} = \frac{|\presence{uv}|}{|T|}$ and the {\bf number of links} in $S$:
$$
m = \sum_{uv\in V\otimes V} m_{uv}
 = \frac{|E|}{|T|}.
$$
Like nodes, each link then contributes to $m$ proportionally to its presence in $S$: $uv$ in $V\otimes V$ accounts for $1$ link only if it is present in $S$ all the time.

Finally, we define the {\bf node and link contributions of a time instant} $t$ as $k_t = \frac{|\V{t}|}{|V|}$ and $l_t = \frac{|\E{t}|}{|V \otimes V|}$, leading to the following definition of the {\bf node duration} $k$ in $S$ and the {\bf link duration} $l$ in $S$:
$$
k = \int_{t\in T} k_t \diff t
 = \frac{|W|}{|V|}
\mbox{\ \ \ and\ \ \ }
l = \int_{t\in T} l_t \diff t
 = \frac{|E|}{|V\otimes V|}.
$$
Like the number of nodes $n$ and the number of links $m$, the node duration $k$ may be seen as a duration of $S$ where each time contributes proportionally to the number of nodes present at this time, and the link duration $l$ as a duration of $S$ where each time contributes proportionally to the number of links present at this time.


Notice that $n$ is the expected value of $|\V{t}|$ when one takes a random time $t$ in $T$. Likewise, $m$, $k$ and $l$ are the expected value of $|\E{t}|$, $|\T{v}|$ and $|\T{uv}|$ when one takes a random time $t$ in $T$, a random node $v$ in $V$ or a random pair of nodes in $V\otimes V$, respectively.

The following relation also hold: $\cov{S} = \frac{|W|}{|T\times V|} = \frac{n}{|V|} = \frac{k}{|T|}$, $n \cdot |T| = k \cdot |V| = |W|$, and $m \cdot |T| = l \cdot |V \otimes V| = |E|$.

\exampleskip

For the examples in Figure~\ref{fig:ex-ref}, we obtain for $S$ the following values: $n = \frac{|\presence{a}|}{10} + \frac{|\presence{b}|}{10} + \frac{|\presence{c}|}{10} + \frac{|\presence{d}|}{10} = 1 + 0.9 + 0.5 + 0.2 = 2.6$ nodes, $m = \frac{|\presence{ab}|}{10} + \frac{|\presence{ac}|}{10} + \frac{|\presence{bc}|}{10} + \frac{|\presence{bd}|}{10} = 0.3 + 0.3 + 0.3 + 0.1 = 1$ link, $k=\frac{26}{4}=6.5$ time units, and $l=\frac{10}{6}=1.66...$ time units. For $L$, we obtain $n = 4$ nodes, $m = 0.7 + 0.3 + 0.7 + 0.3 + 0.3 = 2.3$ links, $k=10$ time units and $l=\frac{23}{6}=3.833...$ time units.

\notionskip

In a link stream $\myL$, by definition $\presence{v} = T$ for all $v$ in $V$, and so $n_v=1$ and $n=|V|$. Likewise, for all $t$, $\V{t}=V$ and so $k_t=1$ and $k=|T|$. In a graph-equivalent stream, in addition $\presence{uv}\in\{\emptyset,T\}$ for all $uv$ in $V\otimes V$ and $\E{t}$ is the same for all $t$. Then, the number of nodes and links in the stream are equal to the number of nodes and links in the corresponding graph.

\notionskip

Notice now that, in a given stream graph, for two nodes $u$ and $v$ such that $|\T{u}|=|\T{v}|$ both $\T{u} = \T{v}$ or $\T{u} \cap \T{v} = \emptyset$ are possible, as well as all intermediary situations. This has a crucial influence on the possible existence of links between $u$ and $v$, and so on the structure of $S$. In order to capture this, we define the {\bf uniformity} of $S$ as follows:
$$
\uniformity(S) = \frac{\sum_{uv \in V\otimes V} |\presence{u} \cap \presence{v}|}{\sum_{uv \in V\otimes V} |\presence{u} \cup \presence{v}|}.
$$
If $S$ has uniformity $1$, then we say that it is uniform: for all $u$ and $v$ in $V$, $\presence{u} = \presence{v}$, \ie\ all nodes are present at the same times.

We also define for any pair of nodes $u$ and $v$ in $V$ the uniformity $\uniformity(u,v) = \frac{|\presence{u} \cap \presence{v}|}{|\presence{u} \cup \presence{v}|}$. It measures the overlap between the presence times of $u$ and $v$, thus their ability to be linked together.


Given a stream graph \myS, we define $S'=(T',V',W,E)$ such that $T'=[\min\{t, \exists(t,v)\in W\}, \max\{t, \exists(t,v)\in W\}]$ and $V' = \{v, \exists(t,v)\in W\}$. We then define the {\bf compactness} of $S$ as follows:
$$
c(S) = \frac{|W|}{|T' \times V'|} = \cov{S'}.
$$
If $S$ has a compactness of $1$, then we say that it is compact: for all $v$ in $V$, $\presence{v} = [b,e] \subseteq T$, \ie\ the presence times of all nodes is the same interval of $T$.

\exampleskip

For the examples in Figure~\ref{fig:ex-ref}, $S$ has uniformity $\uniformity(S) = \frac{\substack{|\presence{a}\cap\presence{b}| + |\presence{a}\cap\presence{c}| + |\presence{a}\cap\presence{d}|\\ + |\presence{b}\cap\presence{c}| + |\presence{b}\cap\presence{d}| + |\presence{c}\cap\presence{d}|}}{\substack{|\presence{a}\cup\presence{b}| + |\presence{a}\cup\presence{c}| + |\presence{a}\cup\presence{d}|\\ + |\presence{b}\cup\presence{c}| + |\presence{b}\cup\presence{d}| + |\presence{c}\cup\presence{d}|}} = \frac{\substack{(4+5)+5+2\\+4+2+0}}{\substack{10+10+10+10\\+(4+4)+(2+5)}} = \frac{22}{55} = 0.4$ and compactness $c(S) = \cov{S} = \frac{26}{40}$ since on this particular case $T'=T$ and $V'=V$, and so $S'=S$.

If $S$ is a link stream, then its uniformity and compactness are necessarily equal to $1$, like $L$ in Figure~\ref{fig:ex-ref}.

\section{Density}
\label{sec:density}

{\em
The density of graph $G=(V,E)$ is the probability when one takes a random element $uv$ in $V\otimes V$ that there is a link between $u$ and $v$ in $E$: $\delta(G) = \frac{2m}{n (n-1)}$. If $n \in \{0,1\}$ then $\delta(G)$ is defined to be $0$.
}

\graphstreamskip

We define the {\bf density} of stream graph \myS\ as the probability when one takes a random element $(t,uv)$ of $T \times V \otimes V$ such that $(t,u)$ and $(t,v)$ are in $W$, that $(t,uv)$ is in $E$:
$$
\delta(S) = \frac{\sum\limits_{uv\in V\otimes V}|\presence{uv}|}{\sum\limits_{uv\in V\otimes V}|\presence{u} \cap \presence{v}|} = \frac{\int\limits_{t\in T} |\E{t}| \diff t}{\int\limits_{t \in T} |\V{t}\otimes \V{t}| \diff t}
$$
If $\sum_{uv\in V\otimes V}|\presence{u} \cap \presence{v}| = \int_{t \in T} |\V{t}\otimes \V{t}| \diff t = 0$ then we define $\delta(S)$ to be $0$.

In other words, the density is the probability when one takes a random time and two random nodes such that a link may exist between them at this time, that the link indeed exists. It is the fraction of possible links that do exist.

\smallskip

Notice that $\sum_{uv\in V\otimes V}|\presence{uv}| = \int_{t\in T} |\E{t}| \diff t = |E|$. Also, $\sum_{uv\in V\otimes V}|\presence{u} \cap \presence{v}| = \int_{t \in T} |\V{t}\otimes \V{t}| \diff t$ is related to the uniformity $\uniformity(S)$ of $S$, but it cannot be directly derived from $|T|$, $|V|$, $|W|$ and $|E|$.

\exampleskip

For $S$ defined in Figure~\ref{fig:ex-ref} (left), $\sum_{uv\in V\otimes V}|\presence{uv}| = |\presence{ab}| + |\presence{ac}| + |\presence{bc}| + |\presence{bd}| = 3 + 3 + 3 + 1 = 10$, $\sum_{uv\in V\otimes V}|\presence{u} \cap \presence{v}| = |\presence{a} \cap \presence{b}| + |\presence{a} \cap \presence{c}| + |\presence{a} \cap \presence{d}| + |\presence{b} \cap \presence{c}| + |\presence{b} \cap \presence{d}| + |\presence{c} \cap \presence{d}| = 9 + 5 + 2 + 4 + 2 + 0 = 22$, and we obtain $\delta(S) = \frac{10}{22} \sim 0.45$. For $L$ defined in this figure (right), $\sum_{uv\in V\otimes V}|\presence{uv}| = 7 + 3 + 7 + 3 + 3 = 23$, $\sum_{uv\in V\otimes V}|\presence{u} \cap \presence{v}| = |V\otimes V|\cdot |T| = 60$ and we obtain $\delta(L) = \frac{23}{60} \sim 0.38$.

\relationskip

\begin{figure}[!h]
\centering
\ \hfill
\includegraphics[scale=\myscale]{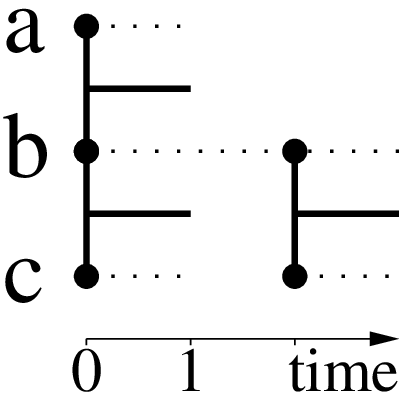}
\hfill
\includegraphics[scale=\myscale]{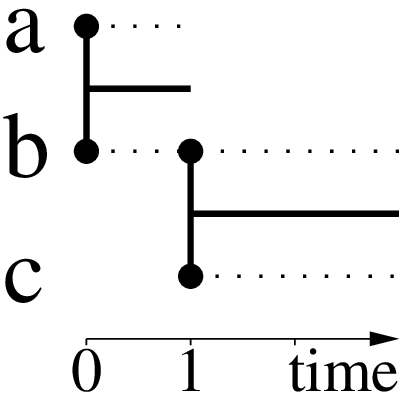}
\hfill\ 
\caption{
{\bf Two stream graphs with $n = 2$ nodes, $m=1$ link, but with different densities:}
Left: $\delta = 0.75$.
Right: $\delta = 1$.
}
\label{fig:density-relationS}
\end{figure}

Notice that there is in general no relation between the density $\delta$, the number of nodes $n$ and the number of links $m$ in a stream graph, see Figure~\ref{fig:density-relationS}.

However, the classical graph relation $\delta = \frac{2m}{n (n-1)}$ holds for a link stream $L=(T,V,E)$. Indeed, we then have $\presence{u} = \presence{v} = |T|$ for all $u$ and $v$, and $n = |V|$, which leads to:
$$
\delta(L)
= \frac{\sum_{uv\in V\otimes V}|\presence{uv}|}{\sum_{uv\in V\otimes V}|T|}
= \frac{2\cdot\sum_{uv\in V\otimes V}|\presence{uv}|}{n\cdot(n-1)\cdot|T|}
= \frac{2\cdot m}{n\cdot(n-1)}
$$
In addition, $\delta(L)$ is equal to the average density of $G_t$: $\frac{1}{|T|} \int_t \delta(G_t) \diff t = \frac{1}{|T|} \int_t \frac{|E_t|}{|V_t \otimes V_t|} \diff t = \frac{1}{|T|\cdot|V \otimes V|} \int_t |E_t|\diff t = \frac{\int_t |E_t|\diff t}{\int_t |V_t \otimes V_t|\diff t} = \delta(L)$, since, in $L$, $V_t=V$ for all $t$.

Finally, if we consider a graph-equivalent stream, then its density is equal to the density of the corresponding graph.

\notionskip

In addition to the global concept of density introduced above, we define the {\bf density of a pair of nodes} $uv$ in $V \otimes V$, the {\bf density of a node} $v$ in $V$, and the {\bf density at a time instant} $t$ in $T$ respectively as follows:
$$
\delta(uv) = \frac{|\presence{uv}|}{|\presence{u}\cap\presence{v}|}, \mbox{\ \ \ } \delta(v) = \frac{\sum_{u\in V, u\ne v}|\presence{uv}|}{\sum_{u\in V, u\ne v}|\presence{u}\cap\presence{v}|} \mbox{\ \ \ and\ \ \ } \delta(t) = \frac{|\E{t}|}{|\V{t} \otimes \V{t}|}.
$$
If $|\presence{u}\cap\presence{v}|=0$, $\sum_{u\in V, u\ne v}|\presence{u}\cap\presence{v}|=0$ or $|\V{t} \otimes \V{t}|=0$, respectively, then we define $\delta(uv)$, $\delta(v)$ and $\delta(t)$ to be $0$.

The density of $uv$ is the probability that there is a link between $u$ and $v$ whenever this is possible, \ie\ when they are both present. The density of $v$ is the probability that a link between $v$ and any other node exists whenever this is possible, and the density of $t$ is equal to $\delta(G_t)$, the density of the graph $G_t$, \ie\ the probability that a link exists between any two nodes present at time $t$.

\exampleskip

For $S$ defined in Figure~\ref{fig:ex-ref} (left), for instance, we obtain $\delta(ab) = \frac{|\presence{ab}|}{|\presence{a} \cap \presence{b}|} = \frac{3}{9} = \frac{1}{3}$ and $\delta(bd) = \frac{|\presence{bd}|}{|\presence{b} \cap \presence{d}|} = \frac{1}{2} = 0.5$. We also obtain $\delta(d) = \frac{|\presence{da}|+|\presence{db}|+|\presence{dc}|}{|\presence{d}\cap\presence{a}|+|\presence{d}\cap\presence{b}|+|\presence{d}\cap\presence{c}|} = \frac{0+1+0}{2+2+0} = 0.25$ and $\delta(2) = \frac{|\E{2}|}{|\V{2} \otimes \V{2}|} = \frac{2}{3\cdot 2 / 2} = \frac{2}{3}$.

\notionskip

Notice that $uv_t$ is strongly related to the concept of density: it is the probability that $u$ and $v$ are linked together at time $t$, which is equal to $1$ or $0$ depending on whether $(t,uv)$ is in $E$ or not. We then have
$\delta(uv) = \frac{\int_{t\in T} uv_t \diff t}{\int_{t\in T} u_t \cdot v_t \diff t}$,
$\delta(v) = \frac{\sum_{u\in V}\int_{t\in T}uv_t \diff t}{\sum_{u\in V}\int_{t\in T}u_t\cdot v_t \diff t}$,
and
$\delta(t) = \frac{\sum_{uv\in V\otimes V} uv_t}{\sum_{uv\in V\otimes V} u_t \cdot v_t}$.
Likewise, $\delta(S) = \frac{\sum_{uv\in V\otimes V}\int_{t\in T}uv_t \diff t}{\sum_{uv\in V\otimes V}\int_{t\in T}u_t\cdot v_t \diff t}$.

%
%
%
%

In a link stream $\myL$, $\presence{v} = T$ for all $v$ and $\V{t} = V$ for all $t$, and so
$\delta(uv) = \frac{|\presence{uv}|}{|T|} = m_{uv}$,
$\delta(t) = \frac{|\E{t}|}{|V \otimes V|} = l_t$,
and, as shown above, $\delta(L)$ is equal to the average of $\delta(t)$.
In a graph-equivalent stream, $\delta(uv) \in \{0,1\}$, and $\delta(t)$ is equal to the density of the induced graph.

The density $\delta(v)$ of node $v$ is strongly related to its degree, that we introduce in Section~\ref{sec:degree}.

\section{Substreams and clusters}
\label{sec:clusters}

{\em
A graph $G'=(V',E')$ is a subgraph of $G=(V,E)$ if $V' \subseteq V$ and $E' \subseteq E$. This is denoted by $G' \subseteq G$.

Given two graphs $G=(V,E)$ and $G'=(V',E')$, their intersection is the graph $G\cap G' = (V\cap V', E\cap E')$. It is their largest (with respect to inclusion) common subgraph.
Their union is $G\cup G' = (V\cup V', E\cup E')$; it is the smallest graph having both $G$ and $G'$ for subgraphs.

A cluster $C$ of $G=(V,E)$ is a subset of $V$. The set of links between nodes in $C$ is $E(C) = \{ uv \in E, u\in C \mbox{ and } v\in C\}$, and $G(C) = (C,E(C))$ denotes the subgraph of $G$ induced by $C$.

Given a cluster $C$, the properties of its induced subgraph are said to be the properties of $C$; for instance, $\delta(C)$ denotes $\delta(G(C))$.
}

\graphstreamskip

We say that a stream $S' = (T', V', W', E')$ is a {\bf substream} of \myS\ if $T' \subseteq T$, $V' \subseteq V$, $W' \subseteq W$, and $E' \subseteq E$. We denote this by $S' \substreameq S$.

Given two stream graphs \myS\ and $S' = (T', V', W', E')$, their {\bf intersection} is the stream graph $S\cap S' = (T\cap T', V\cap V', W\cap W', E\cap E')$. It is their largest (with respect to inclusion) common substream.
Their {\bf union} is $S\cup S' = (T\cup T',V\cup V',W\cup W', E\cup E')$; it is the smallest stream graph having both $S$ and $S'$ for substreams.

\notionskip

\begin{figure}[!h]
\centering
\ \hfill
\includegraphics[scale=\myscale]{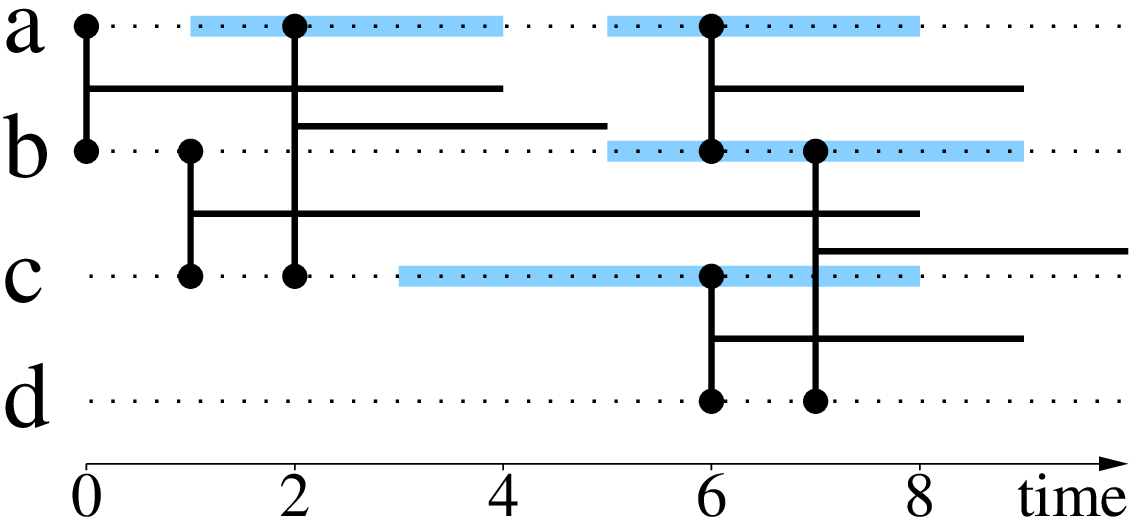}
\hfill
\includegraphics[scale=\myscale]{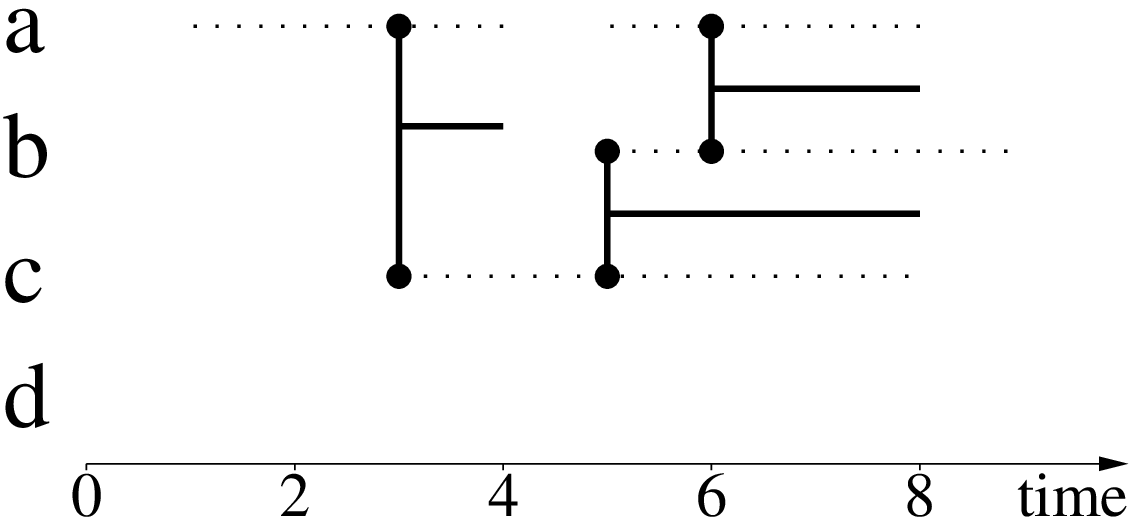}
\hfill\ 
\caption{
{\bf An example of cluster with its induced substream.} Left: the cluster, displayed in blue, is $C = ([1,4]\cup[5,8])\times\{a\} \cup [5,9]\times\{b\} \cup [3,8]\times\{c\}$.
Right: the substream induced by $C$ is $S(C) = ([0,10],\{a,b,c,d\},C,E(C))$ with $E(C) = [6,8]\times\{ab\} \cup [3,4]\times\{ac\} \cup [5,8]\times\{bc\}$.
}
\label{fig:node-substreams}
\end{figure}

We define a {\bf cluster} $C$ of \myS\ as a subset of $W$. We define the set of links between nodes involved in $C$ as $E(C) = \{ (t,uv) \in E, (t,u) \in C \mbox{ and } (t,v) \in C\}$, and we denote by $S(C) = (T,V,C,E(C))$ the {\bf substream of $S$ induced by $C$}. See Figure~\ref{fig:node-substreams}.

\notionskip

Given a cluster $C$, we say that the properties of its induced substream are the properties of $C$; for instance, we denote $\delta(S(C))$ by $\delta(C)$. For any $v$ in $V$, we also denote by $\presence[C]{v}$ the set of times at which $v$ is in $C$, and for any $u$ and $v$ in $V$ we denote by $\presence[C]{uv}$ the set of time instants at which $u$ and $v$ are in $C$ and are linked together. For any $t$ in $T$, we denote by $\V[C]{t}$ the set of nodes present at time $t$ in $C$ and by $\E[C]{t}$ the set of links between nodes in $C$ at time $t$.

In Figure~\ref{fig:node-substreams}, for instance, $\presence[C]{a}=[1,4]\cup[5,8]$, $\presence[C]{b} = [5,9]$, $\presence[C]{c} = [3,8]$ and $\presence[C]{d} = \emptyset$; $\presence[C]{ab} = [6,8]$, $\presence[C]{ac} = [3,4] \cup \{5\}$, and $\presence[C]{bc} = [5,8]$; $\V[C]{7} = \{a,b,c\}$ and $\E[C]{7} = \{ab,bc\}$.

\notionskip

Notice that the substreams of $S$ induced by its clusters are defined over the same set of nodes $V$ and the same time space $T$ as $S$. We therefore define the substream of $S$ induced by a subset $V'$ of $V$ as the substream induced by the node cluster $(T\times V') \cap W$, \ie\ $(T,V',(T\times V') \cap W, (T\times V'\otimes V') \cap E)$ of $S$. Likewise, we define the substream of $S$ induced by a subset $T'$ of $T$ as the substream induced by $(T'\times V) \cap W$, \ie\ $(T',V,(T'\times V) \cap W, (T'\times V \otimes V) \cap E)$ of $S$.

For the example in Figure~\ref{fig:node-substreams}, for instance, the substream induced by $\{a,b,c\}$ and $[6,9]$ is $([6,9],\{a,b,c\},[6,9]\times\{a,b,c\},E')$ with $E' = [6,9]\times\{ab\} \cup [6,8]\times\{bc\}$.

\section{Cliques}
\label{sec:cliques}

{\em
A clique of graph $G$ is a cluster $C$ of $G$ of density $1$. In other words, all pairs of nodes involved in $C$ are linked together in $G$. A clique $C$ is maximal if there is no other clique $C'$ such that $C \subset C'$.
}

\graphstreamskip

We define a {\bf clique} of stream graph $S$ as a cluster $C$ of $S$ of density $1$. In other words, all pairs of nodes involved in $C$ are linked in $S$ whenever both are involved in $C$. A clique $C$ is maximal if there is no other clique $C'$ such that $C \subset C'$.

We say that a clique is compact (resp. uniform) if its induced substream is compact (resp. uniform). It is then fully defined by a set of nodes and a time interval (resp. a time set) meaning that all pairs of nodes are linked together at all these times.

\exampleskip

\begin{figure}[!h]
\centering
\includegraphics[scale=\myscale]{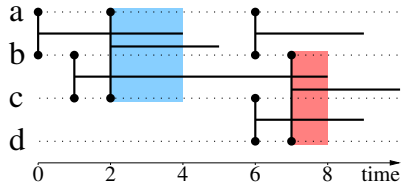}
\caption{{\bf Examples of maximal compact cliques.} We display the two maximal compact cliques involving three nodes of the link stream $L$ of Figure~\ref{fig:ex-ref} (right): $[2,4]\times\{a,b,c\}$ and $[7,8]\times\{b,c,d\}$. Its other maximal compact cliques are $[0,4]\times\{a,b\}$, $[6,9]\times\{a,b\}$, $[2,5]\times\{a,c\}$, $[1,8]\times\{b,c\}$, $[7,10]\times\{b,d\}$, $[6,9]\times\{c,d\}$ (involving two nodes each).}
\label{fig:cliques}
\end{figure}

For instance, in Figure~\ref{fig:cliques} the cluster $[0,1]\times\{a,b\}$ is a compact clique. However, it is not maximal, as it is included in $[0,4]\times\{a,b\}$, which is a maximal compact clique. This clique intersects another maximal compact clique, $[2,4]\times\{a,b,c\}$. There is a unique other maximal compact clique involving three nodes, $[8,9]\times\{b,c,d\}$. The maximal compact clique $[0,4]\times\{a,b\}$ is not a maximal clique because it is for instance included in the clique $[0,4]\times\{a,b\} \cup [6,9]\times\{c,d\}$ (which is not compact). This clique is not maximal either, as it is for instance included in the clique $[0,4]\times\{a,b\} \cup [6,9]\times\{c,d\} \cup [5,6]\times\{d\}$.

\relationskip

A clique in $S$ does not in general induce a clique in $G(S)$: for instance, $[0,1]\times\{a,b\}\cup[8,9]\times\{c,d\}$ is a clique for the example in Figure~\ref{fig:cliques} but $\{a,b,c,d\}$ is not a clique in its induced graph. Instead, for any $[b,e] \subseteq T$ and $X \subseteq V$, if $[b,e]\times X$ is a compact clique in $S$ then $X$ necessarily is a clique in $G(S)$. However, if $[b,e]\times X$ is maximal in $S$ then $X$ is not necessarily maximal in $G(S)$, see for instance $[0,4]\times\{a,b\}$ in Figure~\ref{fig:cliques} ($\{a,b\}$ is a clique in $G(S)$ but it is included in its other clique $\{a,b,c\}$). Conversely, if a cluster $X$ of $G(S)$ is a clique then in general there is no $[b,e]$ such that $[b,e] \times X$ is a compact clique in $S$. Finally, if one considers a graph-equivalent stream, then its maximal cliques are necessarily compact, and they correspond exactly to the maximal cliques of its induced graph.

\section{Neighborhood and degree}
\label{sec:neighborhood-and-degree}
\label{sec:degree}
\label{sec:neighborhood}

{\em
In the graph $G=(V,E)$, the neighborhood $N(v)$ of $v\in V$ is the cluster $N(v)=\{u, uv\in E\}$, and the degree $d(v)$ of $v$ is the number of nodes in this cluster, which is equal to the number of links involving $v$. We then have $\sum_{v\in V}d(v) = 2\cdot m$.

The average degree in $G$ is $d(G) = \frac{1}{n} \cdot \sum_{v\in V}d(v)$, and the following relation between density and average degree holds: $\delta(G) = \frac{d(G)}{n-1}$.
}

\graphstreamskip

\begin{figure}[!h]
\centering
\ \hfill
\includegraphics[scale=\myscale]{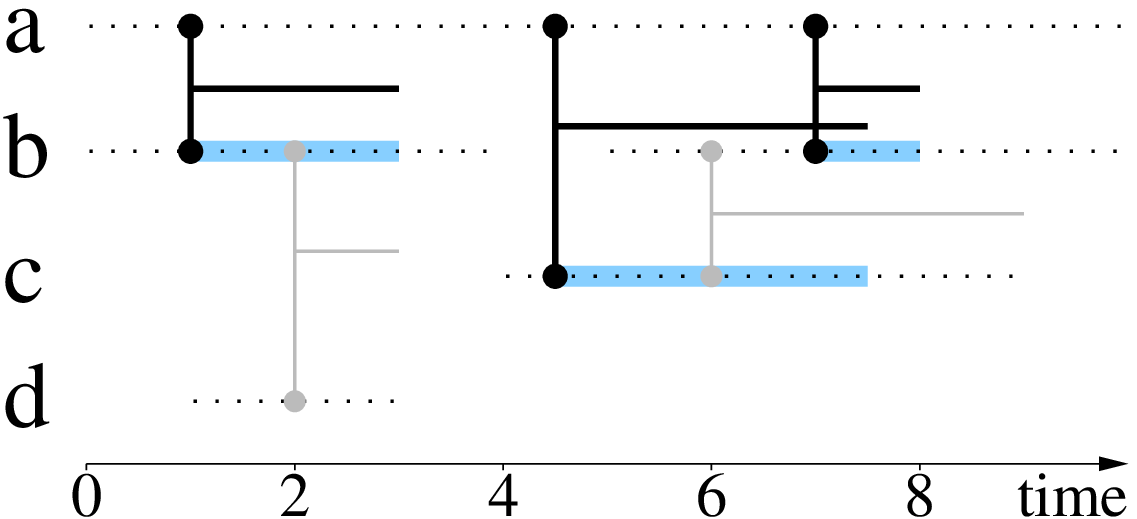}
\hfill
\includegraphics[scale=\myscale]{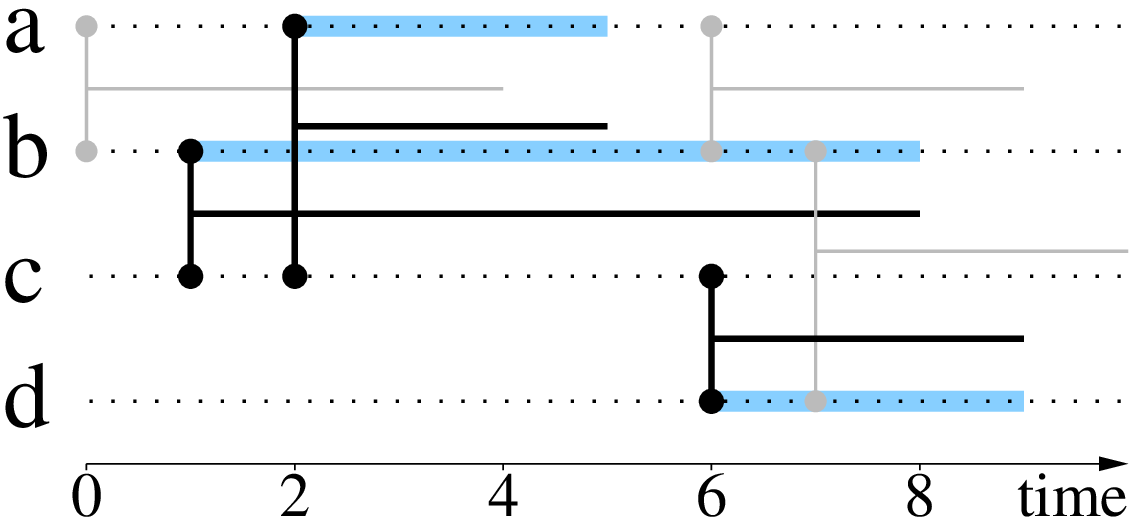}
\hfill\ 
\caption{
{\bf Two examples of neighborhoods and degrees of nodes.} We display in black the links involving the node under concern, and in grey the other links. Left: $N(a)=([1,3]\cup[7,8])\times\{b\} \cup [4.5,7.5]\times\{c\}$ is in blue, leading to $d(a)=\frac{3}{10} + \frac{3}{10} = 0.6$. Right: $N(c)=[2,5]\times\{a\} \cup [1,8]\times\{b\} \cup [6,9]\times\{d\}$ is in blue, leading to $d(c) = \frac{13}{10} = 1.3$.
}
\label{fig:neighborhood}
\end{figure}

In a stream graph \myS, we define the {\bf neighborhood of a node $v$} as the following cluster:
$$
N(v) = \{ (t,u), (t,uv)\in E\}
$$
and the {\bf degree $d(v)$ of $v$} as the number of nodes in this cluster. As with graphs, this is equal to the number of links involving $v$:
$$
d(v) = \frac{|N(v)|}{|T|} = \sum_{u\in V} \frac{|\presence{uv}|}{|T|} = \sum_{u\in V} m_{uv}.
$$
With this definition, each node $u$ contributes to the degree of $v$ proportionally to the duration of its links with $v$.
See Figure~\ref{fig:neighborhood} for an illustration.

As with graphs, the sum of the degree of all nodes in $S$ is equal to twice the number of links in $S$: $\sum_{v\in V} d(v) = \sum_{v\in V} \sum_{u\in V} \frac{|\presence{uv}|}{|T|} = 2\cdot m$.

\notionskip

We now define the {\bf average node degree of $S$} as follows:
$$
d(V)
= \frac{1}{n}\cdot \sum_{v\in V}n_v \cdot d(v)
= \sum_{v\in V}\frac{|\presence{v}|}{|W|}\cdot d(v)
$$
In this definition, the contribution of each node $v$ to the average node degree of $S$ is weighted by its presence duration $|\presence{v}|$.

\begin{figure}[!h]
\centering
\ \hfill
\includegraphics[scale=\myscale]{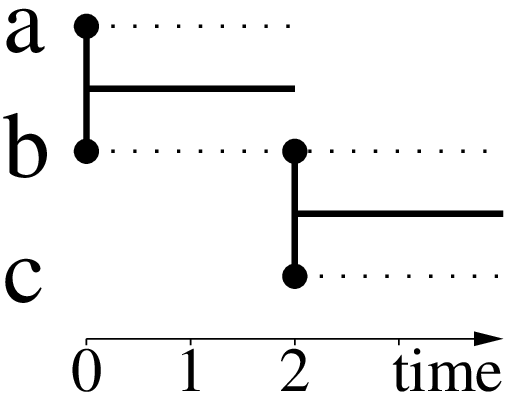}
\hfill
\includegraphics[scale=\myscale]{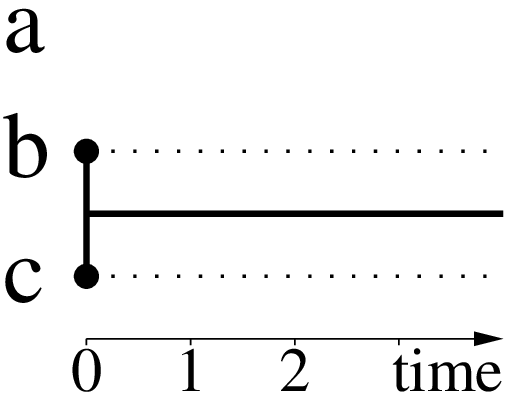}
\hfill\ 
\caption{
These two stream graphs have density $1$ (all possible links exist), $2$ nodes, and $1$ link. However, the leftmost one has average node degree $d(V) = \frac{|T_a|}{|W|}d(a) + \frac{|T_b|}{|W|}d(b) + \frac{|T_c|}{|W|}d(c) = \frac{2}{8} 0.5 + \frac{4}{8} 1 + \frac{2}{8} 0.5 = 0.75$ and the rightmost one has average node degree $1$.
}
\label{fig:density-degree-m-relation}
\end{figure}

As a consequence, there is no direct relation between the average node degree and the total number of links of $S$, as illustrated in Figure~\ref{fig:density-degree-m-relation}. Likewise, the usual relation between average node degree and density does not hold in general.

\notionskip

Instead, in a link stream $L=(T,V,E)$ we have $n_v = 1$ for all $v$, and so the following relation holds: $d(L) = \frac{1}{n} \cdot \sum_{v\in V} d(v) = \frac{2 \cdot m}{n}$. We have seen in Section~\ref{sec:density} that $\delta(L) = \frac{2 \cdot m}{n\cdot(n-1)}$ therefore the relation $\delta(L) = \frac{d(L)}{n-1}$ also holds. Going further, we have $\delta(v) = \frac{\sum_{u\in V, u\ne v}|\presence{uv}|}{\sum_{u\in V, u\ne v}|T|} = \frac{|N(v)|}{(|V|-1)\cdot|T|} = \frac{d(v)}{n-1}$.

\smallskip

Finally, if we consider a graph-equivalent stream, then the degree of any of its nodes is equal to the degree of this node in the corresponding graph, and the average node degree is preserved.

\notionskip

The definitions above generalize graph concepts to stream graphs. However, the temporal features of stream graphs make it natural to consider other generalizations, that we now introduce.

\notionskip


Given a stream graph \myS, we define the {\bf instantaneous neighborhood of} a node $v$ at time $t$ as $N_t(v) = \{u, (t,uv)\in E\}$, and the {\bf instantaneous degree of $v$} at time $t$ as the number of nodes in $N_t(v)$. If $v$ is not involved in $S$ at time $t$, then $N_t(v) = \emptyset$ and $d_t(v) = 0$. If $v$ is involved in $S$ at time $t$, then $N_t(v)$ and $d_t(v)$ are nothing but the neighborhood and the degree of $v$ in the graph $G_t$ induced by $S$ at time $t$.

The degree of $v$ is exactly the average instantaneous degree of $v$ at time $t$ for all $t$ in $T$: $d(v) = \int_t \frac{d_t(v)}{|T|} \diff t$. It is also natural to consider the average only for $t$ in $\presence{v}$, which is the expected instantaneous degree of $v$ when it is involved in $S$; we call it the {\bf expected degree of $v$} and denote it by $\widehat{d}(v) = \int_t \frac{d_t(v)}{|\presence{v}|} \diff t$.

We also consider these two ways to average instantaneous degrees over nodes; either over all nodes in $V$, leading to $\sum_v \frac{d_t(v)}{|V|}$ which we call the {\bf degree at $t$} and denote by $d(t)$, by analogy with $d(v)=\int_t \frac{d_t(v)}{|T|} \diff t$; or over nodes in $V_t$ only, leading to $\widehat{d}(t) = \sum_v \frac{1}{|\V{t}|}d_t(v)$, the {\bf expected degree at time $t$}, which is exactly the average degree of $G_t$.

\notionskip

Let us now consider ways to average $d(v)$ and $d(t)$ over $S$ as a whole.

\begin{figure}[!h]
\centering
\ \hfill
\includegraphics[scale=\myscale]{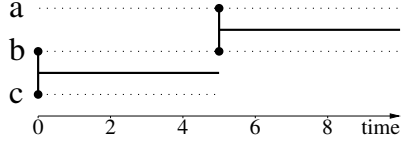}
\hfill \
\caption{
A simple stream graph \myS\ such that $d(V) \ne d(T)$. Indeed, we compute $d(V)$ with
$n=2.5$, $n_a= n_b = 1$, $n_c=0.5$,
$d(a) = 0.5$, $d(b) = 1$, and $d(c) = 0.5$,
leading to 
$d(V) = \frac{1}{n} \sum_{v\in V} n_v\cdot d(v)
= \frac{1}{2.5} (1\cdot 0.5 + 1 \cdot 1 + 0.5 \cdot 0.5)
= 0.7$;
and we compute $d(T)$ with
$k=\frac{25}{3}$, $k_t = 1$ for $t \in [0,5]$, $k_t = \frac{2}{3}$ for $t \in ]5,10]$,
and $d(t) = \frac{2}{3}$ for all $t$,
leading to
$d(T) = \frac{1}{k} \int_t k_t \cdot d(t) \diff t
= \frac{3}{25} (\int_0^5 1 \cdot \frac{2}{3} \diff t + \int_5^{10} \frac{2}{3} \cdot \frac{2}{3} \diff t)
= \frac{3}{25} (5 \cdot \frac{2}{3} + 5 \cdot \frac{4}{9} \diff t)
= \frac{3}{25} \cdot \frac{50}{9}
= \frac{2}{3}$.
}
\label{fig:dVdT}
\end{figure}

The weighted average of $d(v)$, $\sum_{v\in V}\frac{|\presence{v}|}{|W|} d(v) = \frac{1}{n} \sum_{v\in V} n_v\cdot d(v)$, is the average node degree of $S$, denoted by $d(V)$ and introduced above. Similarly, we introduce the weighted average of $d(t)$, $\int_t \frac{|\V{t}|}{|W|}d(t) \diff t = \frac{1}{k} \int_t k_t \cdot d(t) \diff t$, which we call the {\bf average time degree of $S$} and denote by $d(T)$. Notice that, in general, $d(V) \ne d(T)$, as illustrated in Figure~\ref{fig:dVdT}.

For averages over all $V$ and $T$, we obtain a unique quantity: $\sum_v \frac{1}{|V|} d(v) = \frac{2|E|}{|T\times V|} = \int_t \frac{1}{|T|} d(t) \diff t$, which is the average instantaneous degree of $v$ at time $t$ for a random $(t,v)$ in $T\times V$; we call it the {\bf degree of $S$} and denote it by $d(S)$.

Finally, it is also natural to consider the average instantaneous degree for $(t,v)$ in $W$ only: $\frac{\sum_v\int_t d_t(v) \diff t}{|W|} = \frac{\int_t\sum_v d_t(v) \diff t}{|W|} = \frac{2|E|}{|W|} = \frac{2m}{n}$. We call it the {\bf average expected degree of $S$} and denote it by $\widehat{d}(S)$.

In a link stream, we have $d(v) = \widehat{d}(v)$, $d(t) = \widehat{d}(t)$, and $d(V) = d(T) = d(S) = \widehat{d}(S)$.
In a graph-equivalent stream, we have in addition $d(t) = d(V)$, and, as already said, $d(V)$ is the average degree in the corresponding graph and $d(v)$ is the degree of $v$ in this graph.


\section{Clustering coefficient and transitivity ratio}
\label{sec:clustering-coefficient}

{\em
In the graph $G=(V,E)$, the clustering coefficient of a given node $v$ is the density of its neighborhood: $cc(v) = \delta(N(v))$. In other words, $cc(v)$ is the probability that two randomly chosen neighbors of $v$ are linked together in $G$. By definition of the density, if $d(v)<2$ then $cc(v)=0$. The clustering coefficient of $G$ as a whole is the average clustering coefficient of all its nodes: $cc(G) = \frac{1}{n}\cdot \sum_{v\in V} cc(v)$. It is the probability when one takes a random node $v$ that this node has more than one neighbor and that two of its neighbors chosen at random are linked together.

In $G$, the triplet $(u,v,w)$ in $V \times V \times V$ with $u\ne v\ne w$ is a connected triplet if there is both a link between $u$ and $v$ and between $v$ and $w$, \ie\ $uv\in E$ and $vw\in E$. The set of all connected triplets of $G$ is denoted by $\vee$. If in addition there is a link between $u$ and $w$, \ie\ $uw\in E$, then $(u,v,w)$ is a triangle and the set of all triangles of $G$ is denoted by $\triangledown$. The transitivity ratio of $G$ is the probability, when one takes a random connected triplet, that it is a triangle: $tr(G) = \frac{|\triangledown|}{|\vee|}$.
}

\graphstreamskip

\begin{figure}[!h]
\centering
\ \hfill
\includegraphics[scale=\myscale]{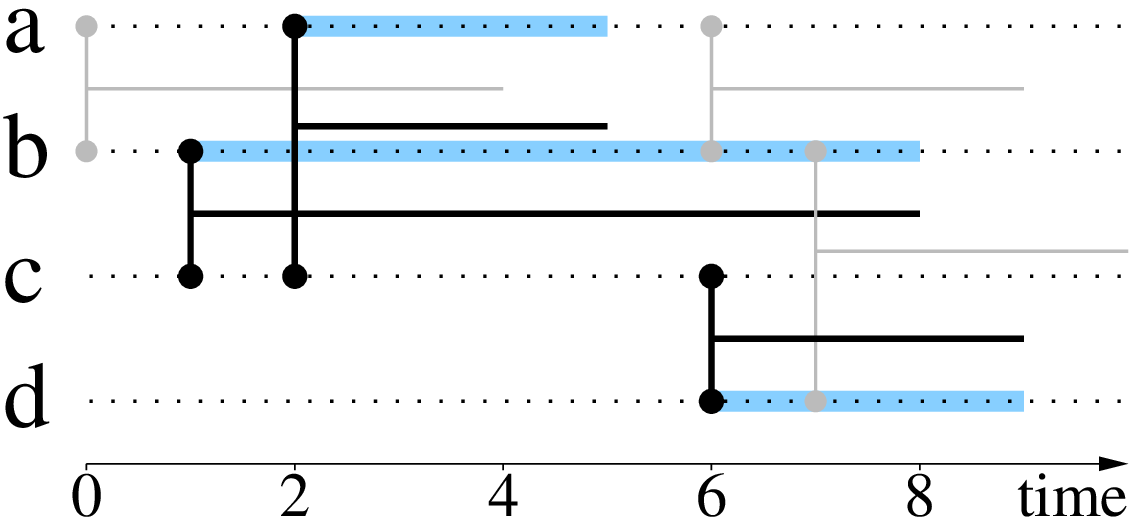}
\hfill
\includegraphics[scale=\myscale]{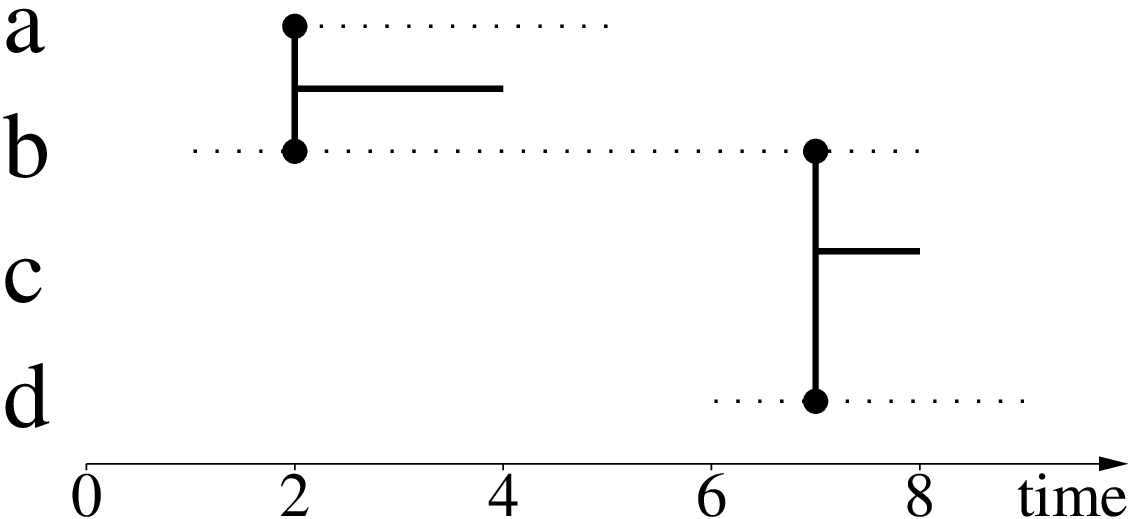}
\hfill \
\caption{
{\bf An example of clustering coefficient.}
{\bf Left:} we display in lack the links involving node $c$, in grey the other links, and in blue the neighborhood of $c$, like in Figure~\ref{fig:neighborhood}.
{\bf Right:} the substream induces by $N(c)$. The clustering coefficient of $c$, $cc(c)$, is the density of this substream. Here, we obtain $cc(c) = \frac{3}{5} = 0.6$.
}
\label{fig:cc}
\end{figure}

In a stream graph \myS, we define the {\bf clustering coefficient of a given node $v$} as the density of its neighborhood:
$$
cc(v) = \delta(N(v)) = \frac{\sum_{uw\in V\otimes V}|\presence{vu}\cap\presence{vw}\cap\presence{uw}|}{\sum_{uw\in V\otimes V}|\presence{vu}\cap\presence{vw}|}
$$
In other words, $cc(v)$ is the probability when one takes two random neighbors $u$ and $w$ of $v$ at time $t$, {\em i.e.} a random $(t,uw)$ in $T\times V\otimes V$ such that $(t,vu)$ and $(t,vw)$ are in $E$, that $u$ is linked to $w$ in $S$ at time $t$, {\em i.e.} that $(t,uw)$ is in $E$. By definition of density, if there is no such triplet then $cc(v)=0$. See Figure~\ref{fig:cc} for an illustration.

We define the {\bf node clustering coefficient of $S$} as the average clustering coefficient of all its nodes, weighted by their presence in $S$: 
$$
cc(V) = \frac{1}{n} \cdot \sum_{v\in V} n_v \cdot cc(v) =  \sum_{v\in V} \frac{|\presence{v}|}{|W|} \cdot cc(v)
$$

\notionskip

In $S$, we say that $(t,(u,v,w))$ in $T \times (V \times V \times V)$ with $u\ne v\ne w$ is a connected triplet if at time $t$ there is both a link between $u$ and $v$ and between $v$ and $w$, \ie\ $(t,uv)\in E$ and $(t,vw)\in E$. We denote by $\vee$ the set of all connected triplets of $S$. If in addition there is a link between $u$ and $w$ at time $t$, \ie\ $(t,uw)\in E$, then we say that $(t,(u,v,w))$ is a triangle and we denote the set of all triangles of $S$ by $\triangledown$. We define the {\bf transitivity ratio $tr(S)$ of $S$} as the probability, when one takes a random connected triplet, that it is a triangle:
$
tr(S) = \frac{|\triangledown|}{|\vee|}
$.

In Figure~\ref{fig:cc} for instance, the set $\vee$ of all connected triplets contains $[2,4] \times \{ (b,a,c), (c,a,b) \}$ because for all $t$ in $[2,4]$ the links $(t,ba)=(t,ab)$ and $(t,ac)=(t,ca)$ are in $E$. The set $\triangledown$ of all triangles also contains $[2,4] \times \{ (b,a,c), (c,a,b) \}$ since for all $t$ in $[2,4]$ the link $(t,bc)=(t,cb)$ also is in $E$. This leads to $\vee =
[2,4] \times \{ (b,a,c), (c,a,b) \} \cup\allowbreak
( [1,4] \cup [6,8] )\times \{ (a,b,c), (c,b,a) \} \cup\allowbreak
[7,8] \times \{ (c,b,d), (d,b,c) \} \cup\allowbreak
[7,9] \times \{ (a,b,d), (d,b,a) \} \cup\allowbreak
[2,5] \times \{ (a,c,b), (b,c,a) \} \cup\allowbreak
[6,8] \times \{ (b,c,d), (d,c,b) \} \cup\allowbreak
[7,9] \times \{ (b,d,c), (c,d,b) \}
$ and $\triangledown =
[2,4] \times \{ (b,a,c), (c,a,b), (a,b,c), (c,b,a), (a,c,b), (b,c,a) \} \cup\allowbreak
[7,8] \times \{ (c,b,d), (d,b,c), (b,c,d),\allowbreak (d,c,b),\allowbreak (b,d,c),\allowbreak (c,d,b) \}\allowbreak
$. We thus obtain $tr(S) = \frac{2\cdot6 + 1\cdot6}{\substack{2\cdot2 + (3+2)\cdot2 + 1\cdot2\\ + 2\cdot2 + 3\cdot2 + 2\cdot2 + 2\cdot2}} =\allowbreak
\frac{9}{17} \sim 0.52$.

\notionskip

In a link stream $L=(T,V,E)$, $n_v=1$ for all $v$, and so $cc(V) = \frac{1}{n}\sum_v cc(v)$. In a graph-equivalent stream, $cc(v)$ in the stream is equal to $cc(v)$ in the corresponding graph $G$, and $cc(V)$ is equal to $cc(G)$. Likewise, the transitivity ratio of a graph-equivalent stream is equal to the one of its corresponding graph.

\notionskip

Like with degrees in Section~\ref{sec:degree}, the temporal features of stream graphs make it natural to consider other generalizations of clustering coefficient, that we now introduce.

\notionskip

Given a stream graph \myS, we define the {\bf instantaneous clustering coefficient} of $v$ at time $t$ as
$cc_t(v) = \frac{\sum_{uw} vu_t \cdot vw_t \cdot uw_t}{\sum_{uw} vu_t \cdot vw_t}$.
If $v$ is not involved in $S$ at time $t$, then $cc_t(v) = 0$.
If $v$ is involved in $S$ at time $t$, then $cc_t(v)$ is exactly the clustering coefficient of $v$ in $G_t$.

Like for degrees, it is natural to consider the following ways to average the instantaneous clustering coefficient:
$\int_t \frac{cc_t(v)}{|\presence{v}|} \diff t$,
$\int_t \frac{cc_t(v)}{|T|} \diff t$,
$\sum_v \frac{cc_t(v)}{|\V{t}|} = cc(G_t)$,
and
$\sum_v \frac{cc_t(v)}{|V|}$.

Notice that $cc(v) \ne \int_t \frac{cc_t(v)}{|T|} \diff t$, but $cc(v)$ is related to $cc_t(v)$ by:
$
cc(v)
=\allowbreak \frac{\sum_{uw}|\presence{vu}\cap\presence{vw}\cap\presence{uw}|}{\sum_{uw}|\presence{vu}\cap\presence{vw}|}
=\allowbreak \frac{\sum_{uw}\int_t vu_t \cdot vw_t \cdot uw_t \diff t}{\sum_{uw}\int_t vu_t \cdot vw_t \diff t}
=\allowbreak \frac{\int_tcc_t(v)\sum_{uw} vu_t \cdot vw_t \diff t}{\int_t\sum_{uw} vu_t \cdot vw_t \diff t}
$.
It is then natural to define $cc(t)$ as such: $cc(t) = \frac{\sum_v cc_t(v) \sum_{uw} vu_t \cdot vw_t}{\sum_v\sum_{uw} vu_t \cdot vw_t}$, which is exactly $tr(G_t)$.

One may then consider the following ways to average $cc(v)$ and $cc(t)$:
$\sum_v \frac{1}{|V|}cc(v)$,
$\int_t \frac{1}{|T|}cc(t) \diff t$,
$cc(V) = \sum_v \frac{|\presence{v}|}{|W|}cc(v)$,
$cc(T) = \int_t \frac{|\V{t}|}{|W|}cc(t) \diff t$,
and
$
cc(S)
=
\int_t \frac{1}{|T|} \sum_v \frac{cc_t(v)}{|V|} \diff t
=
\sum_v \frac{1}{|V|} \int_t \frac{cc_t(v)}{|T|} \diff t
=
\frac{1}{|T\times V|} \sum_v \int_t cc_t(v) \diff t
$, thus introducing the {\bf time clustering coefficient of $S$}, $cc(T)$, and the {\bf clustering coefficient of $S$}, $cc(S)$, by extending the definition of $cc(V)$, like we did for $d(T)$ and $d(S)$ from $d(V)$ in Section~\ref{sec:degree}.

Finally, notice that $cc(t)$, $cc(v)$ and $tr(S)$ may be obtained from the definition of $cc_t(v) = \frac{\sum_{uw} vu_t \cdot vw_t \cdot uw_t}{\sum_{uw} vu_t \cdot vw_t}$ as follows:
$\frac{\sum_v\sum_{uw} vu_t \cdot vw_t \cdot uw_t}{\sum_v\sum_{uw} vu_t \cdot vw_t} = cc(t)$;
$\frac{\int_t\sum_{uw} vu_t \cdot vw_t \cdot uw_t \diff t}{\int_t\sum_{uw} vu_t \cdot vw_t \diff t} = cc(v)$; and
$\frac{\int_t\sum_v\sum_{uw} vu_t \cdot vw_t \cdot uw_t \diff t}{\int_t\sum_v\sum_{uw} vu_t \cdot vw_t \diff t} = tr(S)$.

\section{Neighborhoods and degrees in and of clusters}
\label{sec:cneighborhood-and-degree}
\label{sec:cdegree}
\label{sec:cneighborhood}

{\em
Given a graph $G=(V,E)$ and a cluster $C$ of $G$, the internal neighborhood of $v$ in $C$ is $N_C(v) = N(v) \cap C = \{ u\in C, uv\in E\}$ and its external neighborhood is $\overline{N_C}(v) = N(v) \setminus C = \{ u\not\in C, uv\in E\}$. The internal and external degree of $v$ in $C$, denoted respectively by $d_C(v)$ and $\overline{d_C}(v)$, are the number of nodes in $N_C(v)$ and $\overline{N_C}(v)$. The internal neighborhood and the internal degree of $v$ in $C$ are also its neighborhood and degree in $G(C)$.

The average degree in $C$, denoted by $d_C(C)$ or simply $d_C$, is the average degree of $G(C)$; it is equal to the average internal degree of nodes in $C$.

The neighborhood $N(C)$ of a cluster $C$ is $N(C) = \cup_{v\in C} N(v)$. Notice that $N(C)$ may intersect $C$ but it is not included in $C$ in general. The numbers of nodes in $N(C) \cap C$ and $N(C) \setminus C$ are often called the internal and external degrees of $C$, respectively, denoted by $d(C)$ and $\overline{d}(C)$.
}

\graphstreamskip

\begin{figure}[!h]
\centering
\includegraphics[scale=\myscale]{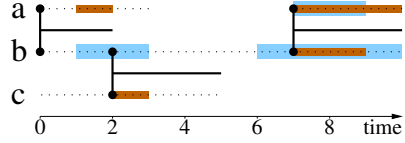}
\caption{
{\bf An example of cluster (in blue) with its neighborhood (in red).} $C=([1,3]\cup[6,10])\times\{b\} \cup [7,9]\times\{a\}$. We then have
$N_C(a)=[7,9]\times\{b\}$, $\overline{N_C}(a)=\emptyset$,
$N_C(b)=[7,9]\times\{a\}$, $\overline{N_C}(b)=([1,2]\cup[9,10])\times\{a\} \cup [2,3]\times\{c\}$,
$N_C(c)=\overline{N_C}(c)=\emptyset$, and
$N(C) = ([1,2]\cup [7,10])\times\{a\} \cup [7,9]\times\{b\} \cup [2,3]\times\{c\}$. The intersection of $N(C)$ with $C$ appears as overlaps between blue and red areas, leading to $d(C)=\frac{|[7,9]\times\{b\} \cup [7,9]\times\{a\}|}{10} = 0.4$ and $\overline{d}(C) = \frac{|([1,2]\cup[9,10])\times\{a\} \cup [2,3]\times\{c\}|}{10} = 0.3$.
}
\label{fig:cluster-neighborhood}
\end{figure}

Given a stream graph \myS\ and a cluster $C$ of $S$, we define the {\bf internal neighborhood of $v$} involved in $C$ as $N_C(v) = \cup_{(t,v)\in C} \{ (t,u)\in C, (t,uv)\in E\}$ and the {\bf external neighborhood of $v$} as $\overline{N_C}(v) = \cup_{(t,v)\in C} \{ (t,u)\not\in C, (t,uv)\in E\}$. Notice that, unlike for graphs, $N_C(v) \ne N(v) \cap C$ and $\overline{N_C}(v) \ne N(v) \setminus C$, and so $N_C(v) \cup \overline{N_C}(v) \ne N(v)$ in general. Indeed, we take into account the neighbors of $v$ {\em only when $v$ is involved in $C$}. See Figure~\ref{fig:cluster-neighborhood} for an illustration.

We define the {\bf internal and external degree of $v$} involved in $C$, denoted respectively by $d_C(v)$ and $\overline{d_C}(v)$, as the number of nodes in $N_C(v)$ and $\overline{N_C}(v)$. The internal neighborhood and the internal degree of $v$ are its neighborhood and degree in $S(C)$.

We define the {\bf average node degree in $C$}, denoted by $d_C$, as the average node degree of $S(C)$; it is the average internal degree of nodes involved in $C$, weighted by their presence in $C$: $d_C =
\sum_v \frac{|\presence[C]{v}|}{|C|} d_C(v)$.


We define the {\bf neighborhood $N(C)$ of cluster $C$} as $N(C) = \cup_{(t,v)\in C} \{ (t,u), (t,uv)\in E\}$, see Figure~\ref{fig:cluster-neighborhood}. Notice that $N(C)$ may intersect $C$ but it is not necessarily included in $C$. We call the numbers of nodes in $N(C) \cap C$ and $N(C) \setminus C$ the {\bf internal and external degrees of $C$}, respectively, denoted by $d(C)$ and $\overline{d}(C)$.

In a graph-equivalent stream, any compact cluster $C = T_C \times V_C$ induces the cluster $V_C$ in the corresponding graph, and the internal (resp. external) neighborhood of any node involved in $C$ is equal to $T_C$ times its internal (resp. external) neighborhood in $V_C$. Likewise, the neighborhood of $C$ in the stream is equal to $T_C$ times the neighborhood of $V_C$ in the graph.

\section{Relations between clusters and quotient stream}
\label{sec:quotient}

{\em
Let us consider a family $F = (C_1, C_2, \dots, C_k)$ of $k$ clusters of $G=(V,E)$. The quotient graph induced by $F$ is the graph $Q = (\{1,2,\dots,k\},E')$ where $ij$ is in $E'$ if $i\ne j$ and there is a $u$ in $C_i$ and a $v$ in $C_j$ such that $uv$ is in $E$.
Notice that, if $F = (\{v\})_{v\in V}$ then $Q$ is equivalent to $G$.

The intra-cluster density $\delta(F)$ of $F$ is the probability, when one takes a random pair of distinct nodes in a same cluster of $F$, that there is a link between them in $G$:
$\delta(F) = \frac{\sum_i|(C_i\otimes C_i)\cap E|}{\sum_{i}|C_i \otimes C_i|}$.
The inter-cluster density $\overline{\delta}(F)$ of $F$ is the probability, when one takes a random pair of distinct nodes in two different clusters of $F$, that there is a link between them in $G$:
$\overline{\delta}(F) = \frac{\sum_{i\ne j}|(C_i\otimes C_j)\cap E|}{\sum_{i\ne j}|C_i \otimes C_j|}$.

The density $\delta(C)$ of $C$ is equal to the intra-cluster density of the family composed of $C$ alone, or the inter-cluster density of the family $(C,C)$. The external density of $C$, denoted by $\overline{\delta}(C)$, is defined as the inter-cluster density of the family $(C,V\setminus C)$. It is the probability when one takes a random node $u$ in $C$ and a random node $v$ outside $C$ that there is a link between them in $G$.
}

\graphstreamskip

\begin{figure}[!h]
\centering
\ \hfill
\includegraphics[scale=\myscale]{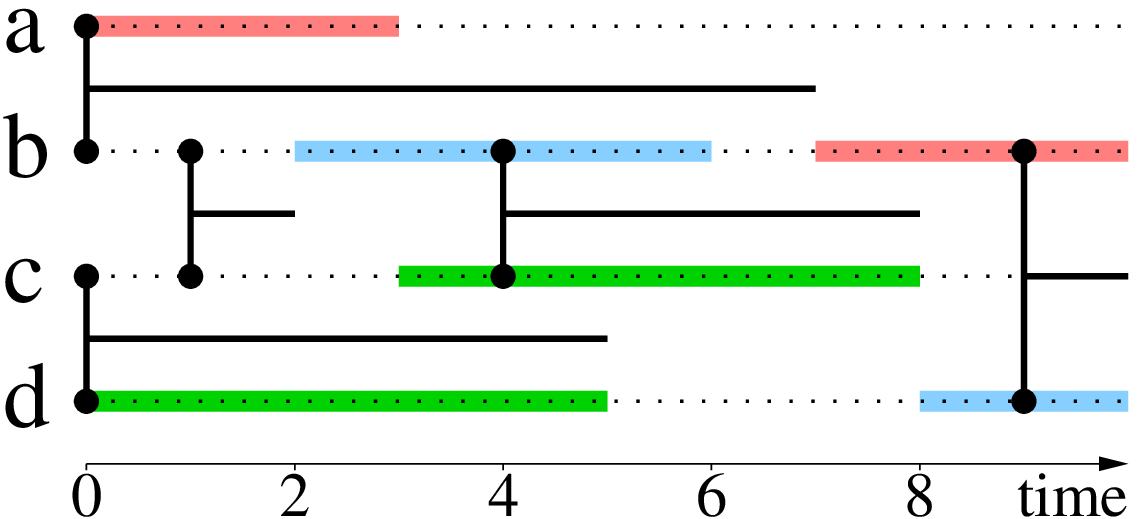}
\hfill
\includegraphics[scale=\myscale]{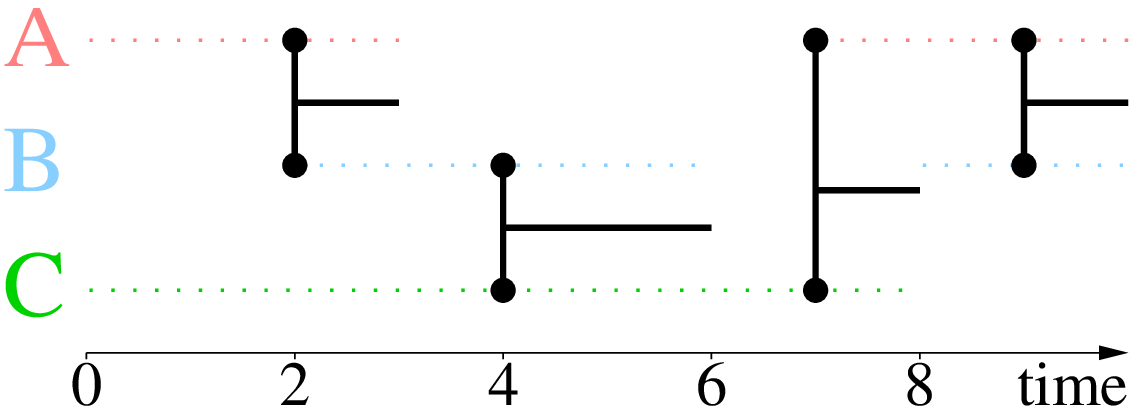}
\hfill \
\caption{
{\bf Example of quotient stream induced by a family of clusters.}
{\bf Left:} a stream graph and a family $F = (A,B,C)$ of clusters with $A=[0,3]\times\{a\} \cup [7,10]\times\{b\}$ (in red), $B=[2,6]\times\{b\} \cup [8,10]\times\{d\}$ (in blue), and $C=[3,8]\times\{c\} \cup [0,5]\times\{d\}$ (in green).
{\bf Right:} the induced quotient stream. For instance, there is a link between $A$ and $C$ from time $7$ to time $8$ because there is a link between $b$ and $c$ at these times, and $b$ is in $A$ and $c$ is on $C$ at these times.
}
\label{fig:quotient-nodes}
\end{figure}

Given a family $F = (C_1, C_2, \dots, C_k)$ of $k$ clusters of \myS, we define the {\bf quotient stream induced by $F$} as the stream graph $Q = (T,\{1,2,\dots,k\},W',E')$ where $(t,i)$ is in $W'$ when there is a $v$ such that $(t,v)$ is in $C_i$, and $(t,ij)$ is in $E'$ when $i\ne j$ and there is a $(t,u)$ in $C_i$ and $(t,v)$ in $C_j$ such that $(t,uv)$ is in $E$. See Figure~\ref{fig:quotient-nodes} for an illustration.
Notice that, if $F = (\presence{v}\times\{v\})_{v\in V}$ then $Q$ is equivalent to $S$.

The {\bf intra-cluster density $\delta(F)$ of $F$} is the probability, when one takes a random element $(t,uv)$ of $T \times V \otimes V$ such that $(t,u)$ and $(t,v)$ are in a same cluster of $F$, that there is a link $(t,uv)$ in $S$:
$$
\delta(F) = \frac{\sum_i\sum_{u\ne v}|\presence[C_i]{u} \cap \presence[C_i]{v} \cap \presence{uv}|}{\sum_i\sum_{u\ne v}|\presence[C_i]{u} \cap \presence[C_i]{v}|}
$$

The {\bf inter-cluster density} $\overline{\delta}(F)$ of $F$ is the probability, when one takes a random element $(t,uv)$ of $T \times V \otimes V$ such that $(t,u)$ and $(t,v)$ are in different clusters of $F$, that there is a link $(t,uv)$ in $S$:
$$
\overline{\delta}(F) = \frac{\sum_{i\ne j}\sum_{u\ne v}|\presence[C_i]{u} \cap \presence[C_j]{v} \cap \presence{uv}|}{\sum_{i\ne j}\sum_{u\ne v}|\presence[C_i]{u} \cap \presence[C_j]{v}|}
$$

As with graphs, the density $\delta(C)$ of $C$ is equal to the intra-cluster density of the family composed of $C$ alone, or the inter-cluster density of the family $(C,C)$. We define the {\bf external density of $C$}, denoted by $\overline{\delta}(C)$, as the inter-cluster density of the family $(C,W\setminus C)$. It is the probability when one takes a random $(t,u)$ in $C$ and a random $(t,v)$ in $W$ but outside $C$ that there is a link $(t,uv)$ between them in $S$.


%
%
%
%
%

\section{Line streams}
\label{sec:line-streams}

{\em
The line-graph $\linegraph{G}$ of $G=(V,E)$ is the graph $\linegraph{G} = (E,\linegraph{E})$ where each node is a link of $G$ and two nodes are linked together if they have an extremity in common: if $A=uv$ and $B=xy$ are two elements of $E$ then $AB$ is in $\linegraph{E}$ if $\{u,v\} \cap \{x,y\} \ne \emptyset$. In general, $\linegraph{\linegraph{\mbox{$G$}}} \ne G$.

The set of links in $G$ involving a given node $v$ corresponds to a cluster in $\linegraph{G}$ and this cluster has density $1$. If instead we consider a set $C$ of independent links (\ie\ if $uv$ and $xy$ are in $C$ then $\{u,v\}\cap\{x,y\}=\emptyset$) then the corresponding cluster in $\linegraph{G}$ has density $0$. Finally, if we consider a clique of $G$, then the cluster of $\linegraph{G}$ corresponding to the links of this clique has density lower than $1$, and it tends to $0$ when the size of the clique grows.
}

\graphstreamskip

\begin{figure}[!h]
\centering
\ \hfill
\includegraphics[scale=\myscale]{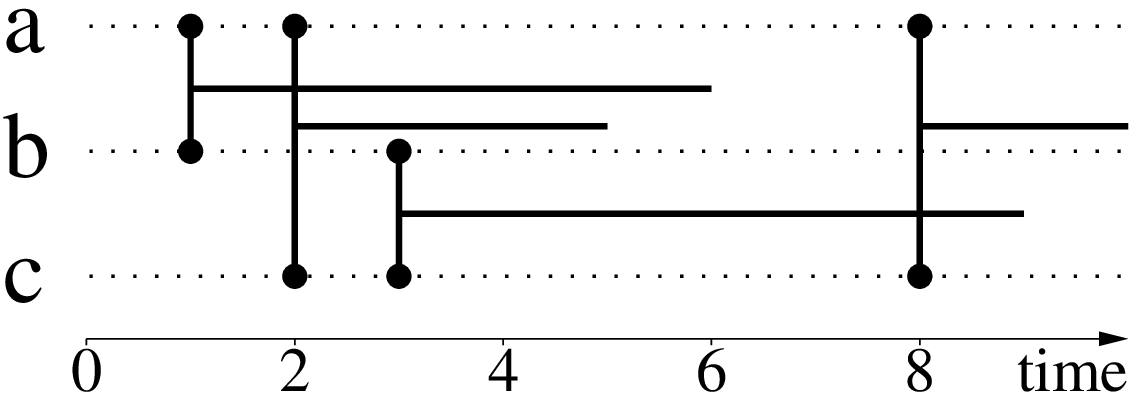}
\hfill
\includegraphics[scale=\myscale]{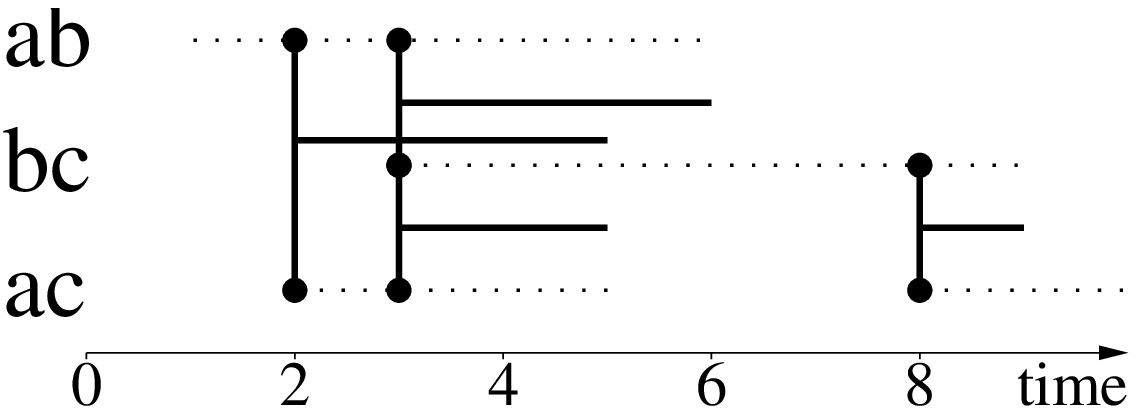}
\hfill \
\caption{
{\bf A stream graph and its line stream.} For instance, the node $ab$ is present in the line stream from time $1$ to time $6$ because $a$ and $b$ are linked together from time $1$ to time $6$ in the original stream. There is a link between nodes $ab$ and $bc$ in the line stream at time $4$ because $\{a,b\} \cap \{b,c\} = \{b\} \ne \emptyset$ and $(4,ab)$ and $(4,bc)$ are both present in the original stream.
}
\label{fig:line-stream}
\end{figure}

We define the {\bf line-stream} $\linegraph{S}$ of \myS\ as the stream graph $\linegraph{S} = (T, \linegraph{V}, \linegraph{W}, \linegraph{E})$. The set $\linegraph{V} = \{ uv, \exists (t,uv)\in E\}$ is the set of links in $G(S)$. The set $\linegraph{W}$ is such that each node $A=uv$ is present in $\linegraph{S}$ during the times at which the link $uv$ is present in $S$, leading to $\linegraph{W} = E$. Finally, for all $A=uv$ and $B=xy$ in $\linegraph{V}$ there is a link $(t,AB)$ in $\linegraph{E}$ if $\{u,v\} \cap \{x,y\} \ne \emptyset$ and $\{(t,uv),(t,xy)\} \subseteq E$. In other words, $A$ and $B$ are linked together at time $t$ if they have an extremity in common and are both present at time $t$. See Figure~\ref{fig:line-stream} for an illustration.
As with graphs, in general $\linegraph{\linegraph{\mbox{$S$}}}\ne S$.

The set of links in $S$ involving a given node $v$ corresponds to a cluster in $\linegraph{S}$, and this cluster has density $1$. If instead we consider a set $C$ of independent links (\ie\ if $(t,uv)$ and $(s,xy)$ are in $C$ then $\{u,v\}\cap\{x,y\}=\emptyset$ or $t\ne s$) then the corresponding cluster in $\linegraph{S}$ has density $0$. As with graphs, the density of a cluster of $\linegraph{G}$ corresponding to the links of a clique of $G$ tends to $0$.

For all $t$, the graph induced by $\linegraph{S}$ at time $t$ is the line graph of $G_t$. As a consequence, the line-stream of a graph-equivalent stream is a graph-equivalent stream too, and its corresponding graph is the line graph of the graph corresponding to the initial stream.

\section{$k$-cores}
\label{sec:k-cores}

{\em
The $k$-core of the graph $G=(V,E)$ is its largest cluster $C^k \subseteq V$ such that, for all $v$ in $C^k$, $d(v) \ge k$ in the sub-graph $G(C^k)$ of $G$ induced by $C^k$. This cluster is unique for a given $k$, and $C^{k+1} \subseteq C^k$ for all $k$. The $k$-core may be computed by iteratively removing from $G$ all elements of $V$ of degree lower than $k$. The $0$-core of $G$ is $V$, and the $k$-core contains all cliques of size $k+1$ of $G$. The core number of $v$ in $V$ is the largest $k$ such that $v \in C^k$. The $k$-shell of $G$ is $C^{k+1} \setminus C^k$.
}

\graphstreamskip

\begin{figure}[!h]
\centering
\ \hfill
\includegraphics[scale=\myscale]{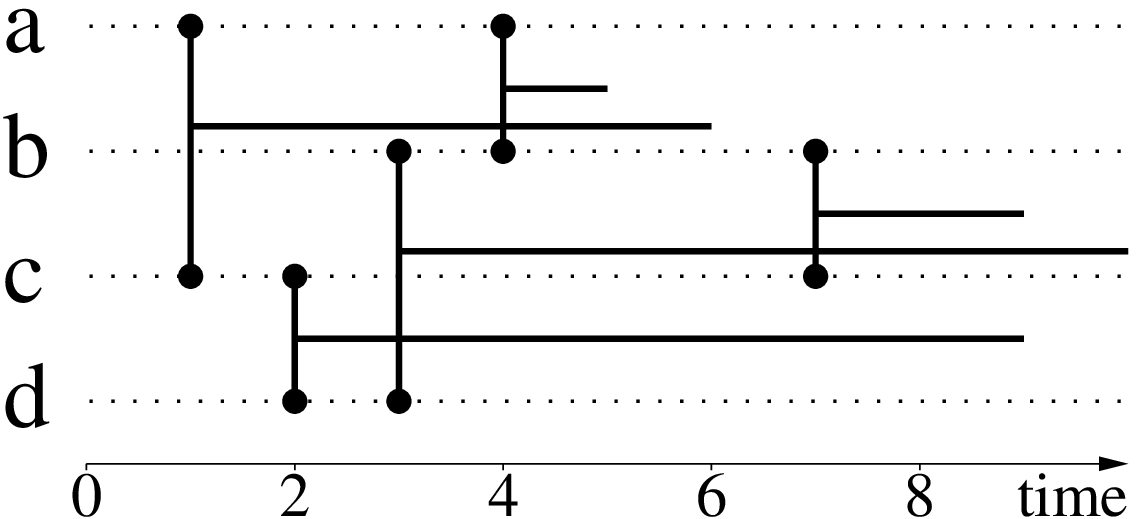}
\hfill
\includegraphics[scale=\myscale]{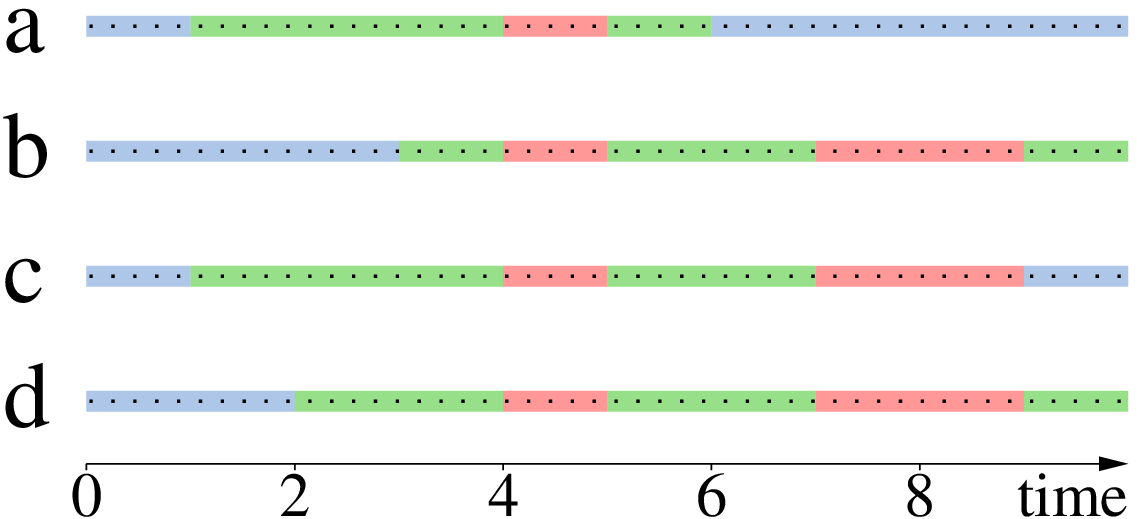}
\hfill \
\caption{
{\bf A link stream $L$, its $k$-shells and its $2$-core.} Each color corresponds to a $k$-shell of $L$: its $0$-shell in blue, its $1$-shell in green, and its $2$-shell in red. In this example, the $2$-shell also is the $2$-core of $L$. For instance, $(2,a)$ is not in the $2$-core since $d_2(a) = 1$ in $L$. As a consequence, although $d_2(c) = 2$ in $L$, since $(2,c)$ is linked to $(2,a)$, it cannot have instantaneous degree $2$ in the $2$-core, and so $(2,c)$ is not in the $2$-core either.
}
\label{fig:core}
\end{figure}

We define the {\bf $k$-core} of the stream graph \myS\ as its largest cluster $C^k \subseteq W$ such that, for all $(t,v)$ in $C^k$, $d_t(v) \ge k$ in the sub-stream $S(C^k)$ of $S$ induced by $C^k$. See Figure~\ref{fig:core} for an illustration.

This cluster is unique for a given $k$, and $C^{k+1} \subseteq C^k$ for all $k$. The $k$-core may be computed by iteratively removing from $S$ all elements of $W$ of instantaneous degree lower than $k$. The $0$-core of $S$ is $W$, and the $k$-core contains all compact cliques of $S$ involving $k+1$ nodes. We define the {\bf core number} of $(t,v)$ in $W$ as the largest $k$ such that $(t,v) \in C^k$, and the {\bf $k$-shell} of $S$ as $C^{k+1} \setminus C^k$.

Notice that, for all $t$, the set of nodes $v$ such that $(t,v) \in C^k$ is exactly $k$-core of $G_t$. As a consequence, the $k$-core of a graph-equivalent stream is $T$ times the $k$-core of the corresponding graph.

\section{Paths and distances}
\label{sec:paths}

{\em
In a graph $G=(V,E)$, a path $P$ from $u \in V$ to $v \in V$ is a sequence $(u_0,v_0)$, $(u_1,v_1)$, $\dots$, $(u_k,v_k)$ of elements of $ V\times V$ such that $u_0 = u$, $v_k = v$, and for all $i$, $u_i=v_{i-1}$ and $u_iv_i \in E$. The path $P$ involves $u$, $v$, and $v_i$ for all $i \in [1,k-1]$, and the integer $k+1$ is the length of $P$. If there exists a path from $u$ to $v$ in $G$ then $v$ is reachable from $u$, which is denoted by $u \graphreaches v$. Reachability is symmetric: $u \graphreaches v$ implies $v \graphreaches u$.

A subpath $Q$ of $P$ is a subsequence $(u_i,v_i)$, $(u_{i+1},v_{i+1})$, $\dots$, $(u_j,v_j)$ of the sequence defining $P$, with $j\ge i$. Then, $Q$ is a path from $u_i$ to $v_j$.

The path $P$ is a cycle if $k>0$ and $u=v$. In other words, it is a nonempty path from $v$ to itself. If $P$ has no subpath that is a cycle, then $P$ is a simple path. If $P$ is a cycle and has no subpath other than $P$ itself that is a cycle, then $P$ is a simple cycle. If there exists no simple cycle in $G$ then $G$ is acyclic.
If $Q$ is a subpath of $P$ and is a cycle from $u_i$ to $v_j$ (hence $v_{i-1}=u_i=v_j=u_{j+1}$) then $P' =  (u_0,v_0),\allowbreak \dots,\allowbreak (u_{i-1},v_{i-1}),\allowbreak (u_{j+1},v_{j+1}),\allowbreak \dots, (u_k,v_k)$ also is a path from $u$ to $v$. If one iteratively removes the cycles of $P$ in this way, one eventually obtains a simple path from $u$ to $v$.

The path $P$ is a shortest path from $u$ to $v$ if there is no path in $G$ of length lower than $k$. Then, $k$ is called the distance between $u$ and $v$ and it is denoted by $\distance(u,v)$. If there is no path between $u$ and $v$ then their distance is infinite. The diameter of $G$ is the largest finite distance between two nodes in $V$.
}


\graphstreamskip

In a stream graph \myS, a {\bf path} $P$ from $(\alpha,u) \in W$ to $(\omega,v) \in W$ is a sequence $(t_0,u_0,v_0)$, $(t_1,u_1,v_1)$, $\dots$, $(t_k,u_k,v_k)$ of elements of $T\times V\times V$ such that $u_0 = u$, $v_k = v$, $t_0 \ge\alpha$, $t_k \le\omega$, for all $i$, $t_i \le t_{i+1}$, $v_i = u_{i+1}$, and $(t_i,u_iv_i) \in E$, $[\alpha,t_0]\times\{u\} \subseteq W$, $[t_k,\omega]\times\{v\} \subseteq W$, and for all $i$, $[t_i,t_{i+1}]\times\{v_i\} \subseteq W$.

\begin{figure}[!h]
\centering
\includegraphics[scale=\myscale]{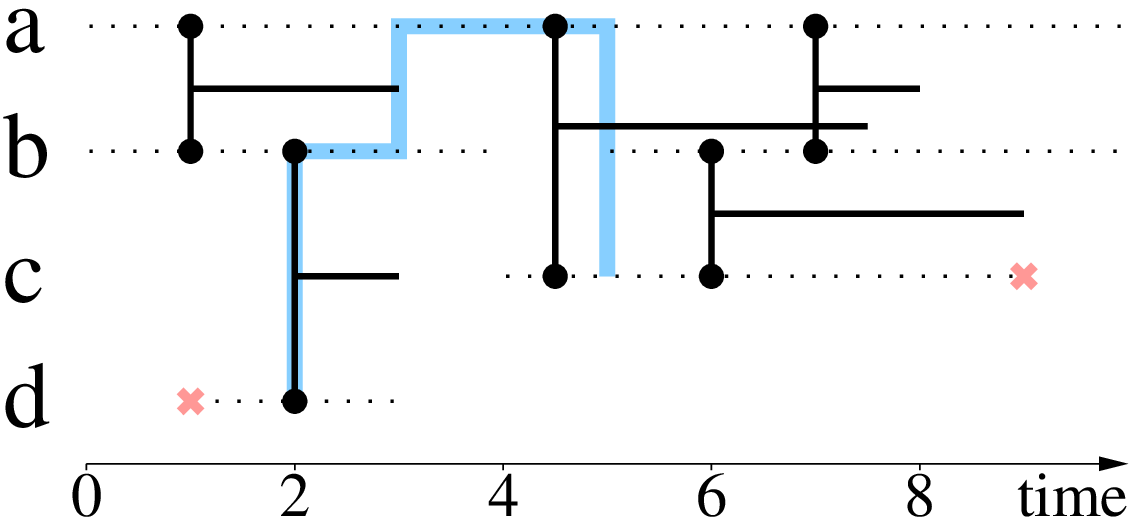}
\hfill
\includegraphics[scale=\myscale]{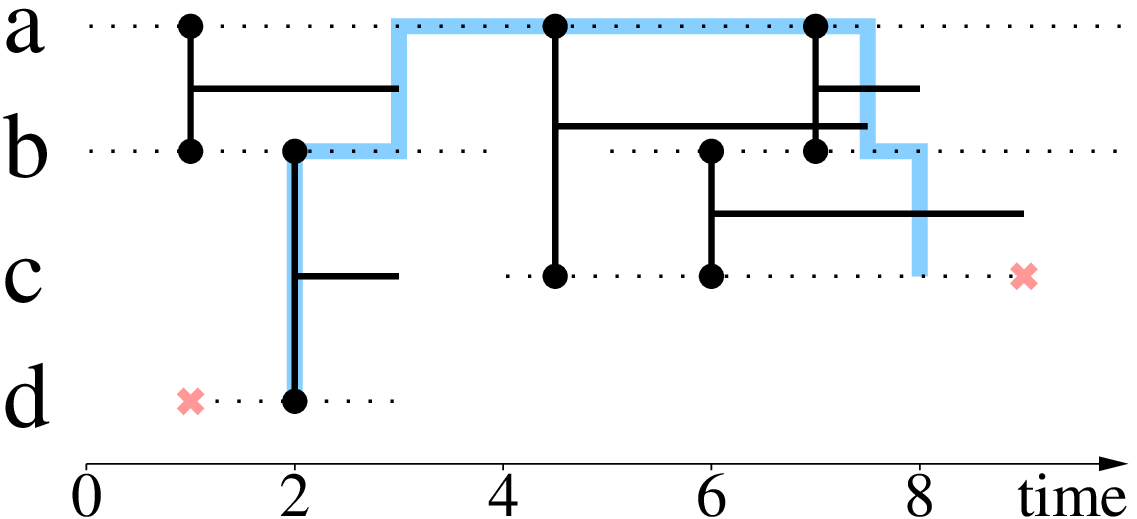}
\hfill
\includegraphics[scale=\myscale]{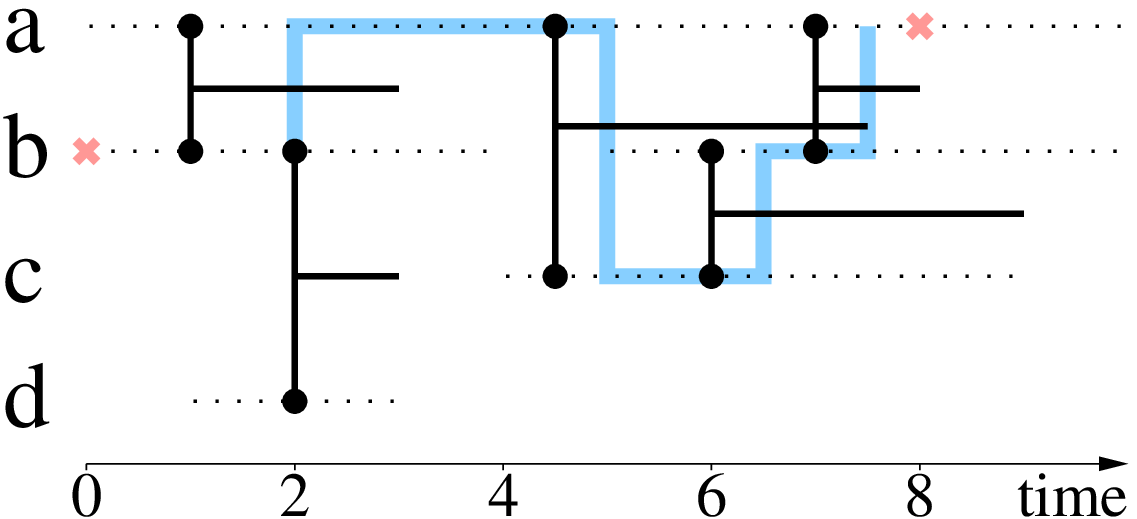}
\caption{
{\bf Paths in a stream graph.}
{\bf Left:}
a path $P_1$ from $(1,d)$ to $(9,c)$: $P_1 = (2,d,b), (3,b,a), (5,a,c)$. This path has length $3$ and duration $3$.
{\bf Center:}
another path $P_2$ from $(1,d)$ to $(9,c)$: $P_2 = (2,d,b), (3,b,a), (7.5,a,b), (8,b,c)$. This path has length $4$ and duration $6$.
{\bf Right:}
a path $P_3$ from $(0,b)$ to $(8,a)$: $P_3 = (2,b,a), (5,a,c), (6.5,c,b), (7.5,b,a)$. This path has length $4$ and duration $5.5$.
}
\label{fig:paths}
\end{figure}

We say that $P$ {\bf involves} $(t_0,u)$, $(t_k,v)$, and $(t,v_i)$ for all $i \in [1,k-1]$ and $t \in [t_i,t_{i+1}]$. We say that path $P$ {\bf starts at} $t_0$, {\bf arrives at} $t_k$, has {\bf length} $k+1$ and {\bf duration} $t_k-t_0$. See Figure~\ref{fig:paths} for an illustration.


If there exists a path from $(\alpha,u)$ to $(\omega,v)$ in $S$, we say that $(\omega,v)$ is {\bf reachable} from $(\alpha,u)$, which we denote by $(\alpha,u) \reaches (\omega,v)$. Notice that reachability is asymmetric: if $(\alpha,u) \reaches (\omega,v)$ then in general $(\omega,v) \notreaches (\alpha,u)$ (in particular this is always true if $\alpha\ne\omega$). We say that $v$ is reachable from $u$ if there exists $\alpha$ and $\omega$ such that $(\alpha,u) \reaches (\omega,v)$, which we also denote by $u \reaches v$. Reachability is asymmetric in this case too: in Figure~\ref{fig:paths}, for instance, $d \reaches c$ (through $P_1$) but $c \notreaches d$. We discuss reachability in more details and we give more complex examples in Section~\ref{sec:connectedness}.

\notionskip

A {\bf subpath} $Q$ of path $P$ is a subsequence $(t_i,u_i,v_i)$, $(t_{i+1},u_{i+1},v_{i+1})$, $\dots$, $(t_j,u_j,v_j)$ of the sequence defining $P$, with $j\ge i$. Then, $Q$ is a path from $(t_i,u_i)$ to $(t_j,v_j)$.
In Figure~\ref{fig:paths}, for instance, $Q_1 = (5, a, c)$, $Q_2 = (3,b,a), (7.5,a,b)$ and $Q_3 = (5,a,c), (6.5,c,b), (7.5,b,a)$ are subpaths of $P_1$, $P_2$ and $P_3$, respectively.

The path $P$ is a {\bf cycle} if $u=v$ and $[\alpha,\omega]\times\{v\} \subseteq W$. In other words, it is a path from $v$ at time $\alpha$ to itself at time $\omega$ such that $v$ is present at all times from $\alpha$ to $\omega$. This means that there is a path of length and duration $0$ (\ie\ the empty sequence) from $(\alpha,v)$ to $(\omega,v)$ in $S$. For instance, $Q_3$ defined above is a cycle, but $Q_2$ is not since $b$ is not present from time $3$ to time $7.5$.

If $P$ has no subpath that is a cycle, then we say that $P$ is a {\bf simple path}. If $P$ is a cycle and has no subpath other than $P$ itself that is a cycle, then $P$ is a simple cycle. If there exists no simple cycle in $S$ then $S$ is {\bf acyclic}.

If $Q$ is a subpath of $P$ and is a cycle from $(t_i,u_i)$ to $(t_j,v_j)$ (hence $t_j \ge t_i$, $v_{i-1}=u_i=v_j=u_{j+1}$, and $[t_{i-1},t_{j+1}]\times\{u_i\} \subseteq W$) then $P' =  (t_0,u_0,v_0),\allowbreak \dots,\allowbreak (t_{i-1},u_{i-1},v_{i-1}),\allowbreak (t_{j+1},u_{j+1},v_{j+1}),\allowbreak \dots, (t_k,u_k,v_k)$ also is a path from $(\alpha,u)$ to $(\omega,v)$. If one iteratively removes the cycles of $P$ in this way, one eventually obtains a simple path from $(\alpha,u)$ to $(\omega,v)$. In Figure~\ref{fig:paths}, for instance, $P_1$ and $P_2$ are simple paths but $P_3$ is not. Instead, the path $(2,b,a)$ obtained by removing $Q_3$ from $P_3$ is simple path.

\notionskip

Paths in stream graphs are quite different from paths in graphs. First, as already said, their temporal nature makes them asymmetric: the existence of a path from $u$ to $v$ does not imply the existence of a path from $v$ to $u$. In addition, paths in stream graphs have a length like in graphs but also a duration. This leads to the following set of definitions, that capture different notions for the cost of reaching a node from another one.

We say that $P$ is a {\bf shortest path} from $(\alpha,u)$ to $(\omega,v)$ if it has minimal length, and we call this length the {\bf distance} from $(\alpha,u)$ to $(\omega,v)$, denoted by $\distance((\alpha,u),(\omega,v))$. The distance $\distance(u,v)$ from $u$ to $v$ is the minimal such distance for all $\alpha$ and $\omega$ in $T$, and a shortest path from $u$ to $v$ is a path from $u$ to $v$ with length $\distance(u,v)$. For instance, in Figure~\ref{fig:paths}, the path $P_1$ is a shortest path from $(1,d)$ to $(9,c)$ but $P_2$ is not. It is impossible to reach $c$ from $d$ with a shorter path, therefore $P_1$ also is a shortest path from $d$ to $c$ and $\distance(d,c)=3$.

We say that $P$ is a {\bf fastest path} from $(\alpha,u)$ to $(\omega,v)$ if it has minimal duration, and we call this duration the {\bf latency} from $(\alpha,u)$ to $(\omega,v)$, denoted by $\latency((\alpha,u),(\omega,v))$. The latency $\latency(u,v)$ from $u$ to $v$ is the minimal such latency for all $\alpha$ and $\omega$ in $T$, and a fastest path from $u$ to $v$ is a path from $u$ to $v$ with duration $\latency(u,v)$. For instance, in Figure~\ref{fig:paths}, the path $P_1$ is not a fastest path from $(1,d)$ to $(9,c)$ since it has duration $3$ and there is another path from $(1,d)$ to $(9,c)$ having duration $1.5$, namely $(3,d,b), (3,b,a), (4.5,a,c)$. This is a fastest path from $(1,d)$ to $(9,c)$ as no faster path exists. Since there is no other path from $d$ to $c$ with lower duration, it also is a fastest path from $d$ to $c$ and $\latency(d,c)=1.5$.

We denote by $\timetoreach_\alpha(u,(t,v))$ the {\bf time to reach} $(t,v)$ from $u$ at time $\alpha$ as follows: $\timetoreach_\alpha(u,(t,v)) = \omega-\alpha$ where $\omega\le t$ is the minimal value such that there is a path from $(\alpha,u)$ to $(\omega,v)$ in $S$ and $[\omega,t] \subseteq T_v$. We call such a path a {\bf foremost path} from $(\alpha,u)$ to $(t,v)$. For instance, in Figure~\ref{fig:paths}, the times to reach $(5,a)$, $(3,b)$, $(10,b)$, and $(5,c)$ from $(1,d)$ are $1$, $1$, $5$ and $3.5$, respectively. Corresponding foremost paths are $F_{5,a}= (2,d,b), (2,b,a)$, $F_{3,b} = (2,d,b)$, $F_{10,b}$ in $\{(x,d,b), (y,b,a), (z,a,c), (6,c,b),\allowbreak x \in [2,3],\allowbreak y \in [x,3],\allowbreak z \in [4.5,6]\}$, and $F_{5,c}$ in $\{(x,d,b), (y,b,a), (4.5,a,c),\allowbreak x \in [2,3],\allowbreak y \in [x,3]\}$.

If there is no path from $(\alpha,u)$ to $(\omega,v)$ then we assert that $\distance((\alpha,u),(\omega,v))$, $\latency((\alpha,u),(\omega,v))$, and $\timetoreach_\alpha(u,(\omega,v))$ are infinite.
We respectively define the {\bf diameter}, the {\bf lapse}, and the {\bf flood time} of $S$ as the largest finite distance, the largest finite latency, and the largest finite time needed to reach an element of $W$ from an element of $W$.

One may combine the notions above by considering for instance {\bf fastest shortest paths} (the ones of minimum duration among those of minimal length) or {\bf shortest fastest paths} (the ones of minimal length among those of minimal duration). For instance, in Figure~\ref{fig:paths}, the unique fastest shortest path from $(1,d)$ to $(9,c)$ is $(3,d,b), (3,b,a), (4.5,a,c)$. The fastest shortest paths from $(0,a)$ to $(9,c)$ are $(x,a,c)$ for $x$ in $[4.5,7.5]$. The fastest shortest paths from $(7.6,a)$ to $(9,c)$ are $(x,a,b),(x,b,c)$ for $x$ in $[7.6,8]$. The fastest shortest paths from $(0,b)$ to $(6,b)$ are $(3,b,a), (x,a,c), (6,c,b)$ for $x$ in $[4.5,6]$. We discuss shortest fastest paths in more details and consider more complex examples in Section~\ref{sec:centralities} for betweenness definitions.

\notionskip

Many extensions of the concept of path in streams make sense and have been considered in the literature (see Section~\ref{sec:related} for references). We present two of the most common ones below.

First, one may capture the fact that transmission through a link has a cost, leading to the following notion: for a given $\gamma$, a {\bf $\gamma$-path} $P$ from $(\alpha,u) \in W$ to $(\omega,v) \in W$ is a sequence $(t_0,u_0,v_0)$, $(t_1,u_1,v_1)$, $\dots$, $(t_k,u_k,v_k)$ of elements of $T\times V\times V$ such that $u_0 = u$, $v_k = v$, $t_0 \ge\alpha$, $t_k \le\omega-\gamma$, for all $i$, $t_i \ge t_{i-1}+\gamma$, $u_i = v_{i-1}$, $[t_i,t_i+\gamma]\times\{u_iv_i\} \subseteq E$, and $[t_i,t_{i+1}]\times\{v_i\} \subseteq W$. The paths discussed since the beginning of this section are equivalent to $\gamma$-paths with $\gamma=0$, and concepts like reachability, cycles, distances, latencies, and others may easily be extended to this more general case. Notice also that $\gamma$ may be a function of the links, involved nodes, time, and other complex features, thus capturing the fact that different links may induce different delays, that delay may vary over time, etc.

Another natural generalization consists in capturing the fact that nodes cannot forward information without delay. One then needs to add the constraint $t_{i+1} \ge t_i + \gamma'$ to the previous definition, where $\gamma'$ captures the delay induced by node forwarding. Similarly, one may want to impose non-null delays on links and/or nodes but without bounds on these delays. The condition above then becomes $t_{i+1}>t_i$.

\relationskip

If $P = (t_0,u_0,v_0), \dots, (t_k,u_k,v_k)$ is a path of length $k$ in $S$, then $(u_0,v_0), \dots, (u_k,v_k)$ is a path of length $k$ in the induced graph $G(S)$. If it is a cycle in $S$, it is also a cycle in $G(S)$. However, the converse claims are false: paths in $G(S)$ do not correspond to paths in $S$, and in particular a node may be reachable from another node in $G(S)$ but not in $S$. Notice also that the distance between two nodes in $G(S)$ is bounded by the size of $V$, whereas it is unbounded in $S$. For instance, if $T=[0,x]$ for a given integer $x$, $V=\{a,b\}$, $\T{a} = \bigcup_{i=0,1,\dots} [2i,2i+1]$, $\T{b} = \bigcup_{i=0,1,\dots} [2i+1,2i+2]$, and $\T{ab} = \{i, i=1,\dots\}$, then the path $(0,a,b),(1,b,a),(2,a,b),\dots$ of length $x$ is a shortest path from $(0,a)$ to $(x,a)$ or $(x,b)$.

In a link stream \myL, since nodes are always present, the definition of path is much simpler:
a path $P$ from $(\alpha,u)\in T\times V$ to $(\omega,v)\in T\times V$ is a sequence $(t_0,u_0,v_0)$, $(t_1,u_1,v_1)$, $\dots$, $(t_k,u_k,v_k)$ of elements of $T\times V\times V$ such that $u_0 = u$, $v_k = v$, $t_0 \ge\alpha$, $t_k \le\omega$, $t_i \ge t_{i-1}$, $u_i = v_{i-1}$, and $(t_i,u_iv_i) \in E$. In this case, as with graphs, the distance is bounded by the size of $V$. In addition, if $(\alpha,u) \reaches (\omega,v)$ then for all $\alpha'\le\alpha$ and $\omega'\ge\omega$, $(\alpha',u) \reaches (\omega',v)$. However, the existence of a path between two given nodes in $G(L)$ still does not imply in general the existence of a path between them in $L$. For instance, if $T=[0,1]$, $V=\{a,b,c,d\}$, and $E = \{(0,ab), (0,cd), (1,bc)\}$ then there is a path between $a$ and $d$ in $G(L)$ but not in $L$.

In a graph-equivalent stream, there is a path from a node to another one in the stream if and only if there is a path between them in the corresponding graph, and the shortest paths have the same length. As a consequence, the distance between two nodes is the same in the stream and its corresponding graph, and a path is a cycle in the stream if and only if the corresponding path is a cycle in the graph.

\section{Connectedness and connected components}
\label{sec:connectedness}

{\em
A graph $G=(V,E)$ is connected if for all $u$ and $v$ in $V$ there is a path between $u$ and $v$ in $G$. A cluster $C$ is connected if $G(C)$ is connected, and it is a maximal connected cluster if it is included in no other connected cluster. These clusters are called the connected components of $G$, and they form a partition\,\footnote{A partition of a set $X$ into $k$ parts is a family $(P_1, P_2, \cdots, P_k)$ of $k$ subsets of $X$ such that $\cup_i P_i = X$ and $P_i \cap P_j = \emptyset$ for all $i\ne j$.} of $V$.
The reachability graph of $G$ is the graph $R = (V,E')$ where $uv \in E'$ if $u \graphreaches v$ in $G$. The connected components of $G$ are exactly but the cliques of $R$.
}

\graphstreamskip

\begin{figure}[!h]
\centering
\ \hfill
\includegraphics[scale=\myscale]{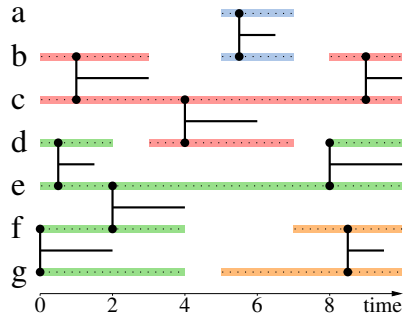}
\hfill \
\caption{
{\bf Weakly connected components of a stream graph.}
This stream graph has four weakly connected components, each displayed with a different color: $[5,7]\times\{a,b\}$ in blue, $([0,3]\cup[8,10])\times\{b\}\cup[0,10]\times\{c\}\cup[3,7]\times\{d\}$ in pink, $([0,2]\cup[8,10])\times\{d\}\cup[0,10]\times\{e\}\cup[0,4]\times\{f,g\}$ in green, and $[7,10]\times\{f\}\cup[5,10]\times\{g\}$ in orange.
}
\label{fig:weakly-connected}
\end{figure}

Given a stream graph \myS, we say that $(\omega,v)$ is {\bf weakly reachable} from $(\alpha,u)$, which we denote by $(\alpha,u) \weakreaches (\omega,v)$, if there is a sequence $(t_0,u_0,v_0)$, $(t_1,u_1,v_1)$, $\dots$, $(t_k,u_k,v_k)$ of elements of $T\times V\times V$ such that $u_0 = u$, $v_k = v$, for all $i$, $v_i = u_{i+1}$, and $(t_i,u_iv_i) \in E$, $[\alpha,t_0]\times\{u\} \subseteq W$, $[t_k,\omega]\times\{v\} \subseteq W$, and for all $i$, $[t_i,t_{i+1}]\times\{v_i\} \subseteq W$.
This sequence is similar to a path from $(\alpha,u)$ to $(\omega,v)$, except for time constraints: we do not necessarily have $t_0\ge\alpha$, $t_{i+1}\ge t_i$, nor $\omega\ge t_k$. As a consequence, weak reachability is symmetric: if $(\alpha,u) \weakreaches (\omega,v)$ then $(\omega,v) \weakreaches (\alpha,u)$. In Figure~\ref{fig:weakly-connected} for instance, we have $(9,d) \weakreaches (3,g)$ through the sequence $(8,d,e), (3,e,f), (1,f,g)$.

We say that $S$ is {\bf weakly connected} if for all $(\alpha,u)$ and $(\omega,v)$ in $W$, $(\alpha,u) \weakreaches (\omega,v)$. We say that a cluster $C\subseteq W$ is weakly connected if its induced substream $S(C)$ is weakly connected. It is a weakly connected component of $S$ if it is a maximal weakly connected cluster of $S$. Intuitively, this corresponds to the disconnected parts of a drawing of $S$, see Figure~\ref{fig:weakly-connected} for an illustration.

\notionskip

\begin{figure}[!h]
\centering
\ \hfill
\includegraphics[scale=\myscale]{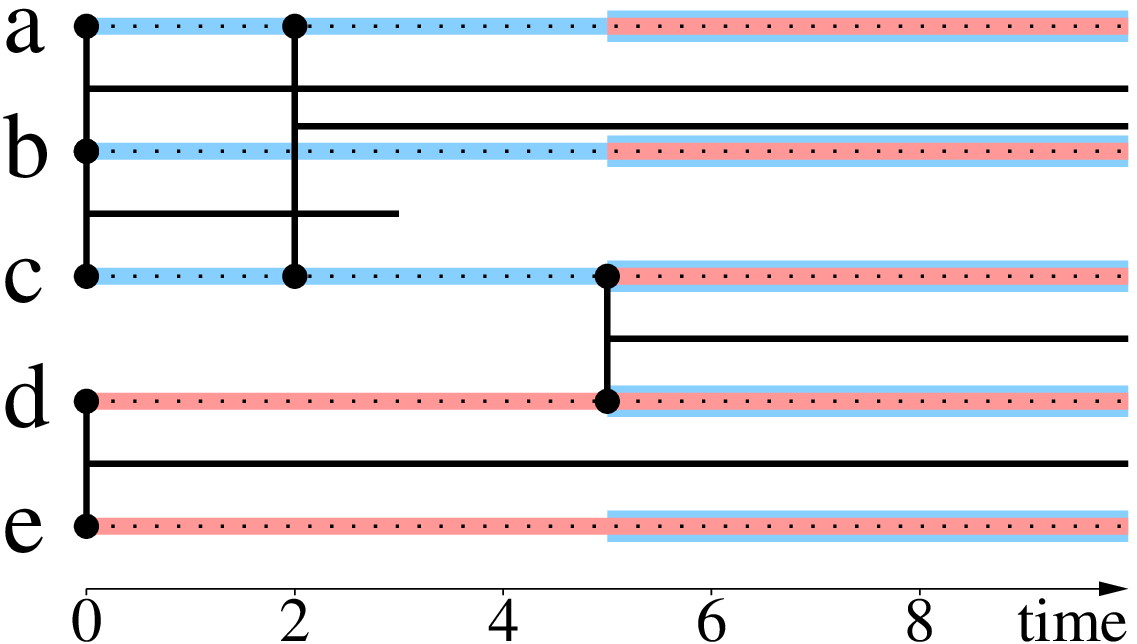}
\hfill
\includegraphics[scale=\myscale]{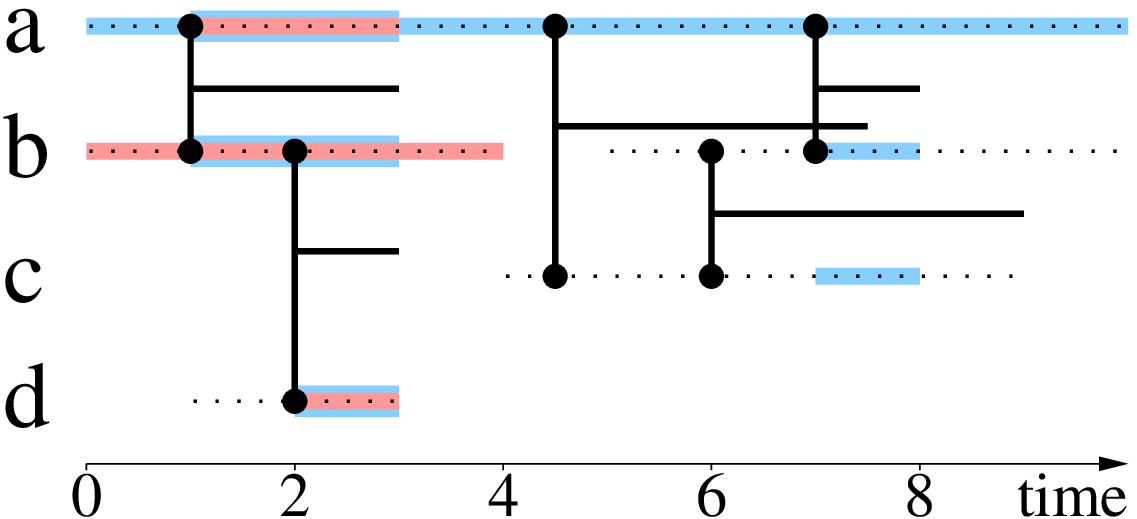}
\hfill \
\caption{
{\bf Strongly connected clusters in a link stream (left) and a stream graph (right).}
{\bf Left:} this link stream is not strongly connected, since $(0,a) \notreaches (0,d)$, for instance. It has only two maximal strongly connected clusters, namely $[0,10]\times\{a,b,c\} \cup [5,10]\times\{d,e\}$ (in blue) and $[0,10]\times\{d,e\} \cup [5,10]\times\{a,b,c\}$ (in pink), which overlap. It also contains an infinity of strongly connected clusters which are not maximal and may have an intricate structure, like for instance $[0,4]\times\{a,b,c\} \cup [4,5]\times\{c\} \cup [5,9]\times\{a,b\} \cup [9,9.5]\times\{c\} \cup [9,10]\times\{d\}$.
{\bf Right:} this stream graph is not strongly connected, since $(0,a) \notreaches (1,d)$, for instance. The cluster $[2,3]\times\{a,b,d\}$ is strongly connected but not maximal as it is included in $[2,3]\times\{a,b,d\} \cup [1,2]\times\{a,b\}$, which is strongly connected too. This cluster is not a maximal strongly connected cluster either, as it is included in $[2,3]\times\{a,b,d\} \cup [1,2]\times\{a,b\} \cup [0,1]\times\{a\}$ and $[2,3]\times\{a,b,d\} \cup [1,2]\times\{a,b\} \cup [0,1]\times\{b\}$ which are both strongly connected. Notice however that the union of these two clusters is not strongly connected, as $(0,a) \notreaches (0,b)$ for instance. They are not maximal either, but they are included (among others) respectively in $[2,3]\times\{a,b,d\} \cup [1,2]\times\{a,b\} \cup [0,10]\times\{a\} \cup [7,8]\times\{a,b,c\}$ (in blue) and $[2,3]\times\{a,b,d\} \cup [1,2]\times\{a,b\} \cup [0,4]\times\{b\}$ (in pink) which both are maximal strongly connected clusters of this stream graph.
}
\label{fig:connected-clusters}
\end{figure}

We say that \myS\ is {\bf strongly connected} if for all $(\alpha,u)$ and $(\omega,v)$ in $W$ with $\alpha \le \omega$ there is a path from $(\alpha,u)$ to $(\omega,v)$ in $S$. We say that a cluster $C$ is strongly connected if $S(C)$ is strongly connected. We say that $C$ is a {\em maximal} strongly connected cluster if it is included in no other strongly connected cluster. See Figure~\ref{fig:connected-clusters} for an illustration. The examples in this figure show that the maximal connected clusters of $S$ do not in general lead to a partition of $W$.

\notionskip

If $S$ is strongly connected then there is a path between $u$ and $v$ in $G_t$ for all $(t,u)$ and $(t,v)$ in $W$, \ie\ $G_t$ is a connected graph for all $t$. However, $G_t$ may be connected for all $t$ even though $S$ is not strongly connected. This happens for instance if $T=[0,3]$, $V=\{a,b\}$, $W = [0,1]\times\{a\} \cup [2,3]\times\{b\}$ and $E = \emptyset$.

If $S$ is compact, though, it is strongly connected if and only if $G_t$ is connected for all $t$ in $T$. Indeed, as already said, if $S$ is strongly connected then $G_t$ necessarily is connected. Conversely, if $G_t$ is connected for all $t$ in $T$ then $S$ necessarily is strongly connected: assume there exist $(\alpha,v)$ and $(\omega,u)$ in $W$ with $\omega \ge \alpha$ such that $(\alpha,v) \notreaches (\omega,u)$; since $S$ is compact, $(\alpha,v) \reaches (\omega,v)$, and so this implies that $(\omega,v) \notreaches (\omega,u)$, which contradicts the fact that $G_\omega$ is connected.


A cluster $C$ is a {\bf maximal strongly connected {\em compact} cluster} if it is compact, strongly connected, and included in no other strongly connected compact cluster.
For instance, the link stream of Figure~\ref{fig:connected-clusters} (left) has three maximal strongly connected compact clusters, namely $[0,10]\times\{a,b,c\}$, $[0,10]\times\{d,e\}$, and $[5,10]\times\{a,b,c,d,e\}$. These clusters overlap, and so maximal strongly connected compact clusters do not result in partition of $W$.

\notionskip

If $C = T_C \times V_C$ is a maximal strongly connected compact cluster, then, even though $V_C$ necessarily is a connected cluster of $G_t$, it is not in general a connected component of $G_t$. In Figure~\ref{fig:connected-clusters} (left) for instance, $\{a,b,c\}$ is not a connected component of $G_6$ (it is included in the connected component $\{a,b,c,d,e\}$ of $G_6$), although $[0,10]\times\{a,b,c\}$ is a maximal strongly connected compact cluster.

\begin{figure}[!h]
\centering
\ \hfill
\includegraphics[scale=\myscale]{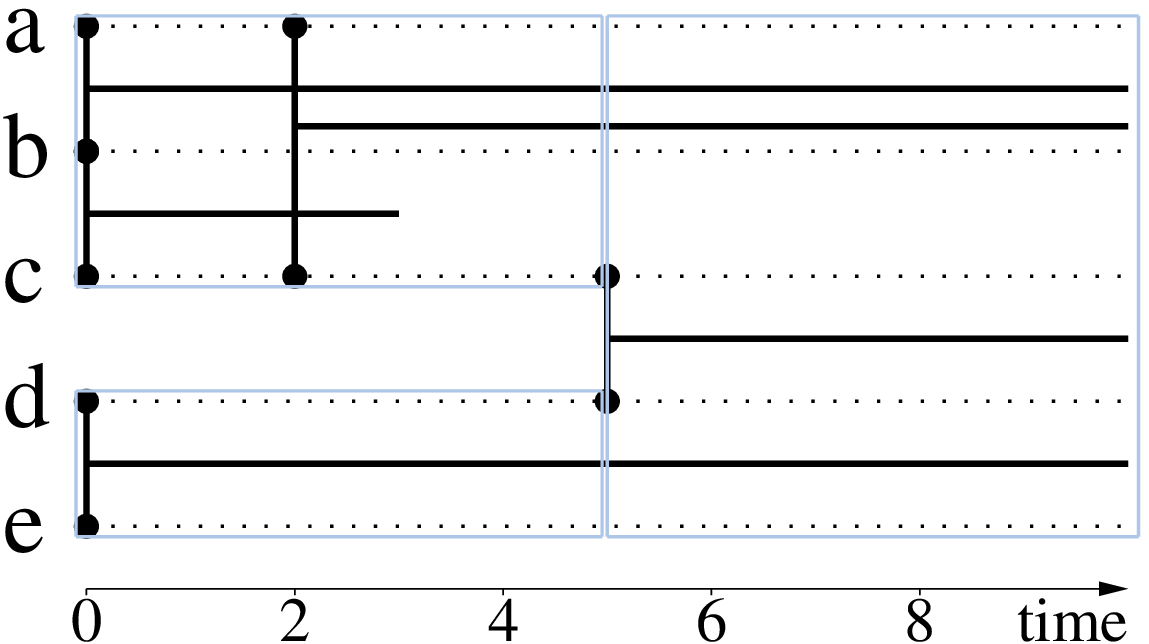}
\hfill
\includegraphics[scale=\myscale]{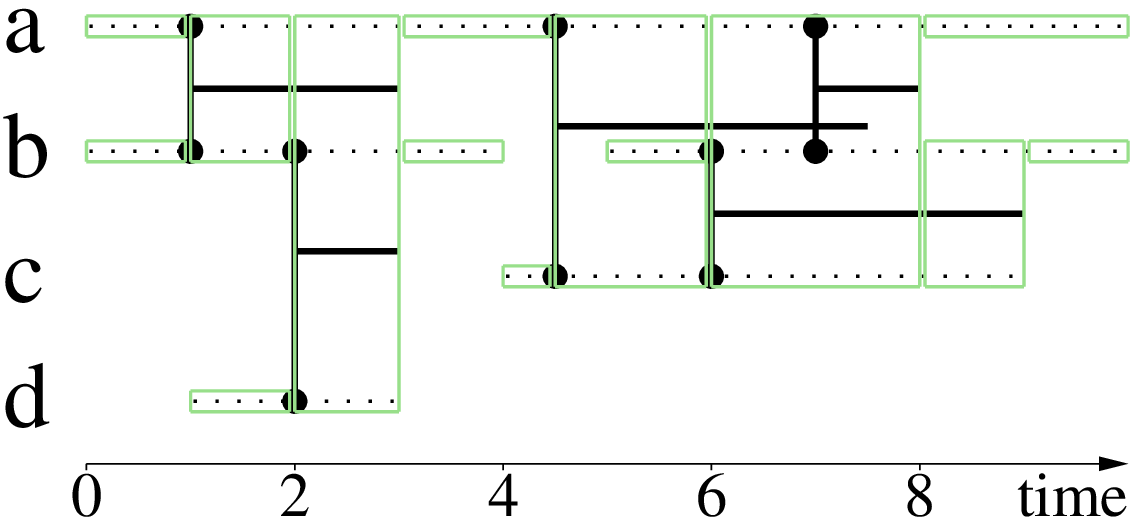}
\hfill \
\caption{
{\bf Connected components in a link stream (left) and a stream graph (right).}
We indicate each component $C = T_C \times V_C$ with a rectangle.
{\bf Left:} in this link stream, the connected components are $[0,5[\times\{a,b,c\}$, $[0,5[\times\{d,e\}$, and $[5,10]\times\{a,b,c,d,e\}$.
{\bf Right:} in this stream graph, the connected components are $[0,1[\times\{a\}$, $[0,1[\times\{b\}$, $[1,2[\times\{a,b\}$, $[1,2[\times\{d\}$, $[2,3]\times\{a,b,d\}$, $]3,4]\times\{b\}$, $]3,4.5[\times\{a\}$, $[4,4.5[\times\{c\}$, $[4.5,6[\times\{a,c\}$, $[5,6[\times\{b\}$, $[6,8]\times\{a,b,c\}$, $]8,10]\times\{a\}$, $]8,9]\times\{b,c\}$, and $]9,10]\times\{b\}$.
}
\label{fig:connected-components}
\end{figure}

This leads to the following definition of {\bf strongly connected components} of $S$: a strongly connected component $C$ of $S$ is a maximal compact cluster $C = T_C \times V_C$ such that $V_C$ is a connected component of $G_t$ for all $t$ in $T_C$. This implies that $C$ is a (not necessarily maximal) strongly connected compact cluster.
For instance, the maximal strongly connected compact cluster $[0,10]\times\{a,b,c\}$ of the link stream of Figure~\ref{fig:connected-clusters} (left) is not a connected component because $\{a,b,c\}$ is not a connected component of $G_6$. We display in Figure~\ref{fig:connected-components} the connected components of our two examples.

The set of all strongly connected components of $S$ is a partition of $W$. Indeed, each $(t,v)$ in $W$ clearly is in a connected component of $S$. Conversely, if $(t,v)$ is in two distinct connected components $C=T_C\times V_C$ and $D = T_D\times V_D$ of $S$, then it means that $V_C$ and $V_D$ are two connected components of $G_t$ to which $v$ belongs, which implies that $V_C = V_D$. But then, $(T_C\cup T_D) \times V_C$ also is a strongly connected component, which contradicts the hypothesis.

Notice that the maximal clusters of $S$ such that for all $t$ the set of nodes involved in them at time $t$ is a connected component of $G_t$ (but are not necessarily compact) do not lead to a partition of $W$. For instance, the two maximal strongly connected clusters of the link stream of Figure~\ref{fig:connected-clusters} (left) both have these properties but they overlap.

\notionskip

Given a stream graph \myS, we define its {\bf reachability stream graph} $R = (T,V,W,E')$ where $E'$ is the set of all $(t,uv)$ in $T\times V\otimes V$ such that $v \graphreaches u$ in $G_t$. In other words, there is a link between $u$ and $v$ at time $t$ in $R$ if there is a path in $S$ from $u$ to $v$ at time $t$. The strongly connected compact clusters of $S$ are exactly the compact cliques of $R$.

\relationskip

In a link stream $\myL$, the weakly connected components of $L$ are exactly the compact clusters $C=T\times V_C$ such that $V_C$ is a connected component of $G(L)$. However, strong connectivity in link stream has only a few additional properties compared to strong connectivity in stream graphs in general, as illustrated in the figures of this section (the leftmost example is a link stream). Just notice that for all $v$ in $V$, for all $\alpha$ and $\omega$ in $T$ with $\omega\ge\alpha$, $(\alpha,v) \reaches (\omega,v)$ (thanks to an empty path). As a consequence, for all $(\alpha,u)$ and $(\omega,v)$ in $T\times V$, if $(\alpha,u) \reaches (\omega,v)$ then for all $\alpha'\le\alpha$, $(\alpha',u) \reaches (\omega,v)$ and for all $\omega'\ge\omega$, $(\alpha,u)\reaches(\omega',v)$.

In a graph-equivalent stream, the strongly connected components are equivalent to the connected component of the corresponding graph.

\section{Trees and cascades}
\label{sec:trees}

{\em
A graph $G=(V,E)$ is a tree of root $r$, with $r\in V$, if for all $v$ in $V$ there is a unique simple path from $r$ to $v$ in $G$. Then, $G$ is connected and acyclic, and any connected acyclic graph (with a distinguished node $r$) is a tree (of root $r$).
This also implies that for all $v \ne r$ in $V$ there is a unique $u\ne v$ such that $u$ is the last node before $v$ on the simple path from $r$ to $v$, called the predecessor of $v$ and denoted by $p(v)$. In addition, the predecessor of the root is the root itself: $p(r)=r$.

Given a graph $G = (V,E)$, a subgraph $G' = (V',E')$ of $G$ is a shortest-path tree of root $r$ if it is a tree of root $r$ and for all $v$ in $V'$, the simple path from $r$ to $v$ in $G'$ is a shortest path from $r$ to $v$ in $G$. A cascade is a maximal shortest-path tree.
}

\graphstreamskip

\begin{figure}[!h]
\centering
\ \hfill
\includegraphics[scale=\myscale]{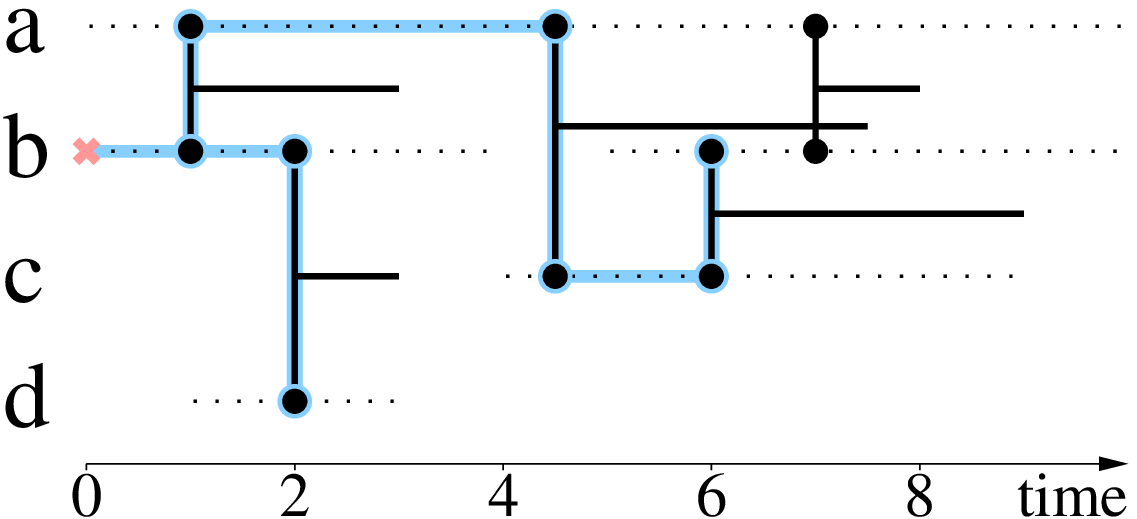}
\hfill
\includegraphics[scale=\myscale]{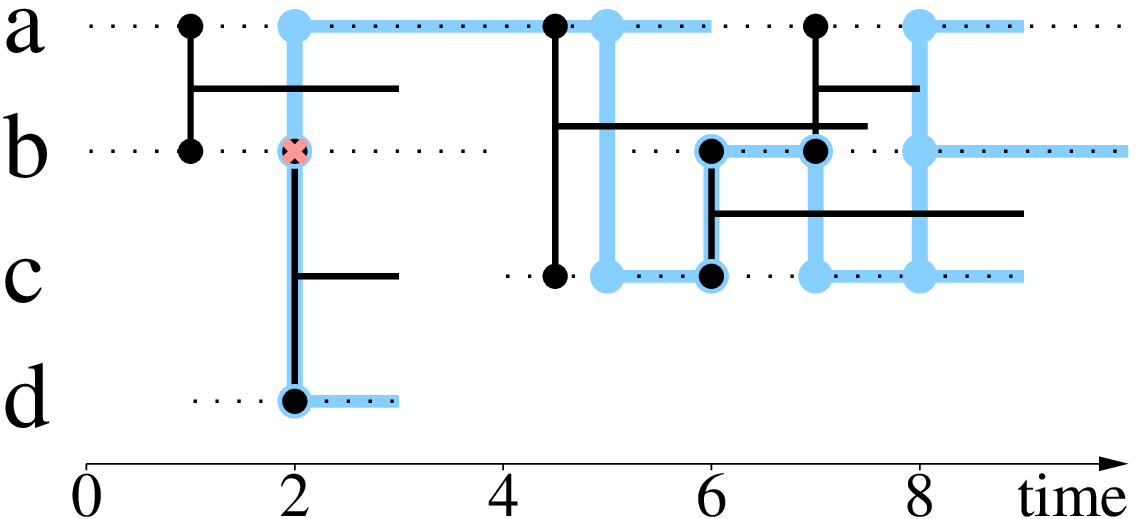}
\hfill \
\caption{
{\bf Examples of trees.} We consider a stream graph \myS\ and display two of its substreams that are trees.
{\bf Left:} the tree $S_1 = (T,V,W_1,E_1)$ (in blue) of root $(0,b)$ (in pink), with $W_1=[1,4.5]\times\{a\}\cup([0,2]\cup\{6\})\times\{b\}\cup[4.5,6]\times\{c\}\cup\{(2,d)\}$ and $E_1=\{(1,ab),(2,bd),(4.5,ac),(6,bc)\}$.
{\bf Right:} the tree $S_2 = (T,V,W_2,E_2)$ (in blue) of root $(2,b)$ (in pink), with $W_2=([2,6]\cup[8,9])\times\{a\}\cup([6,7]\cup[8,10])\times\{b\}\cup([5,6]\cup[7,9])\times\{c\}\cup[2,3]\times\{d\}$ and $E_2=\{(2,ab),(2,bd),(5,ac),(6,bc),(7,bc),(8,bc),(8,ab)\}$.
}
\label{fig:trees}
\end{figure}

We say that a stream graph $\myS$ is a {\bf tree} of root $r$, with $r\in W$, if for all $(t,v)$ in $W$ there is a unique simple path from $r$ to $(t,v)$ in $S$. Then, $S$ necessarily is weakly connected and acyclic, but the converse is not true. Notice also that a tree is not strongly connected in general. See Figure~\ref{fig:trees} for an illustration.

If \myS\ is a {\bf tree of root $r$}, then for all $(t,v) \ne r$ in $W$ either there is a unique last $(t',u)$ with $u\ne v$ before $(t,v)$ involved in the simple path from $r$ to $(t,v)$, and we call it the {\bf predecessor of $v$} at time $t$, or the simple path from $r$ to $(t,v)$ is the empty sequence, and we say that the predecessor of $v$  at time $t$ is $r$. We denote the predecessor by $p(t,v)$. In Figure~\ref{fig:trees}, for instance, $p(5,c)=(4.5,a)$, $p(6,b)=(6,c)$, and $p(1,b)=(0,b)=r$ in $S_1$ and $p(9,c)=(7,b)$ in $S_2$.

If $S$ is a tree of root $r$ and $\alpha$ is the first time at which any node is involved in $S$, \ie\ $\alpha = \min\{t, (t,v)\in W\}$, then necessarily $r$ is in $(\{\alpha\}\times V) \cap W$. In other words, the root of $S$ necessarily is one of the very first node occurrences in $S$. Moreover, it also is a tree of root $r'$ for all $r'$ in $(\{\alpha\}\times V) \cap W$. In Figure~\ref{fig:trees} (right), for instance, $S_2$ also is a tree of root $(2,a)$ and a tree of root $(2,d)$.

Given a stream graph \myS, we say that a substream $S' = (T,V,W',E')$ of $S$ is a {\bf shortest-path tree} of root $r$ if it is a tree of root $r$ and for all $(t,v)$ in $W'$, the simple path from $r$ to $(t,v)$ in $S'$ is a shortest path from $r$ to $(t,v)$ in $S$. We define similarly {\bf fastest-path trees} and {\bf foremost-path trees}. In Figure~\ref{fig:trees}, for instance, $S_1$ is a foremost-path tree of $S$.

For a given $r$ in $W$, we denote by $R(r)$ the cluster of all elements of $W$ reachable from $r$, and we call it the {\bf reachable cluster} of $r$.
We say that the substream $S'$ is a {\bf cascade} of root $r$ if it is a maximal foremost-path tree, in the sense that it is included in no other foremost-path tree with the same root. 

\relationskip

If $S$ is a graph-equivalent stream and $S' \substreameq S$ is a tree of root $r = (t,v)$, then its induced graph $G(S')$ is a tree of root $v$. If $S'$ is a shortest path tree of $S$ then $G(S')$ is a shortest path tree of $G(S)$. The same holds for cascades.

\section{Closeness and betweenness centralities}
\label{sec:centralities}

{\em
In a graph $G=(V,E)$, the closeness of a node $v$ measures its proximity to other nodes:
$
\closeness(v) = \sum_{u\ne v} \frac{1}{\distance(v,u)}
$.
The betweenness of $v$ measures how frequently $v$ is involved in shortest paths in $G$:
$
\betweenness(v) = \sum_{u\in V, w\in V} \frac{\sigma(u,w,v)}{\sigma(u,w)}
$
where $\frac{\sigma(u,w,v)}{\sigma(u,w)}$ is the fraction of all shortest paths from $u$ to $w$ that involve $v$ if there is a path from $u$ to $w$, $0$ otherwise.
In other words, the betweenness of $v$ in $V$ is the number of pairs of elements $u$ and $w$ of $V$, each counted with a weight equal to the fraction of shortest path between them that involve $v$.
The betweenness of any cluster $X\subseteq V$ is
$
\betweenness(X) = \sum_{u\in V, w\in V} \frac{\sigma(u,w,X)}{\sigma(u,w)}
$
where $\frac{\sigma(u,w,X)}{\sigma(u,w)}$ is the fraction of all shortest paths from $u$ to $w$ that involve an element of $X$, if $u \graphreaches w$, $0$ otherwise. Notice that $\betweenness(v) = \betweenness(\{v\})$.
}

\graphstreamskip

In the stream graph \myS, we define a general concept of {\bf closeness} of a node $v$ at a time instant $t$, with $(t,v)\in W$, as follows:
$$
\closeness_t(v) = \sum_{u \in V}\int_{\substack{s\in T\\(s,u)\ne(t,v)}} \frac{1}{\cost_t(v,(s,u))} \diff s
$$
where $\cost_t(v,(s,u))$ represents the cost to reach $(s,u)$ from $v$ at time $t$.

The cost $\cost_t(v,(s,u))$ may be captured in various ways, the most basic being the time to reach $(s,u)$ from $v$ at time $t$: $\cost_t(v,(s,u)) = \timetoreach_t(v,(s,u))$. Notice however that we must have $\cost_t(v,(s,u)) \ne 0$ for all $(s,u) \ne (t,v)$. To ensure this, one may for instance define $\cost_t(v,(s,u))$ as the length of a non-empty shortest foremost path from $(t,v)$ to $(s,u)$, \ie\ $\distance((t,v),(t+\timetoreach_t(v,(s,u)),u))$ if it is different from $0$. One may also combine both approaches by assuming that traversing a link has a cost $\gamma$, leading to the following cost function: $\cost_t(v,(s,u)) = \timetoreach_t(v,(s,u)) + \gamma \cdot \distance((t,v),(t+\timetoreach_t(v,(s,u)),u))$.

\notionskip


We now define the {\bf betweenness} of a node $v\in V$ at a time instant $t\in T$, with $(t,v)\in W$, as follows:
$$
\betweenness(t,v)
= \sum_{u\in V, w\in V} \int_{i \in T_u, j\in T_w} \bfrac{u}{w} \diff i \diff j
$$
where $\bfrac{u}{w}$ is the fraction of all \sfp s from $u$ at time $i$ to $w$ at time $j$ that involve $v$ at time $t$ if there is a path from $(i,u)$ to $(j,w)$, $0$ otherwise. In other words, the betweenness of $(t,v)$ in $W$ is the number of pairs of elements $(i,u)$ and $(j,w)$ of $W$, each counted with a weight equal to the fraction of \sfp s between them that involve $(t,v)$.

We extend the definition to any cluster $X \subseteq W$ as follows:
$$
\betweenness(X)
= \sum_{u\in V, w\in V} \int_{i \in T_u, j\in T_w} \Xbfrac{u}{w} \diff i \diff j
$$
where $\Xbfrac{u}{w}$ is the fraction of all \sfp s from $(i,u)$ to $(j,v)$ that involve at least an element of $X$ if $(i,u) \reaches (j,w)$, $0$ otherwise.
Then, $\betweenness(t,v) = \betweenness(\{(t,v)\})$. We also use this approach to define the betweenness of node $v$ as $\betweenness(v) = \betweenness(T_v \times \{v\})$, and the one of time $t$ as $\betweenness(t) = \betweenness(\{t\}\times V_t)$.

\notionskip

Instead of \sfp s, one may consider the fraction of fastest shortest paths, of shortest paths, of fastest simple paths, or other classes of paths. However, considering \sfp s has the advantage of putting more emphasis on time than distance, and to avoid considering as equivalent fastest paths with very different lengths (in particular the non-simple ones).

\notionskip

In a graph-equivalent stream, the betweenness of any node $v$ is equal to $\frac{|T|^2}{2}$ times its betweenness in the corresponding graph. Indeed, for any $(i,u)$ and $(j,w)$ with $j \ge i$, the fraction of paths involving $T \times \{v\}$ in a graph-equivalent stream is the fraction of paths between $u$ and $w$ in the corresponding graph that involve $v$.

\exampleskip

The rest of this section is devoted to detailed examples of betweenness centralities in various link streams, representative of what happens in stream graphs in general, in order to illustrate this concept in concrete situations.

\exampleskip

\begin{figure}[!h]
\centering
\ \hfill
\includegraphics[scale=\myscale]{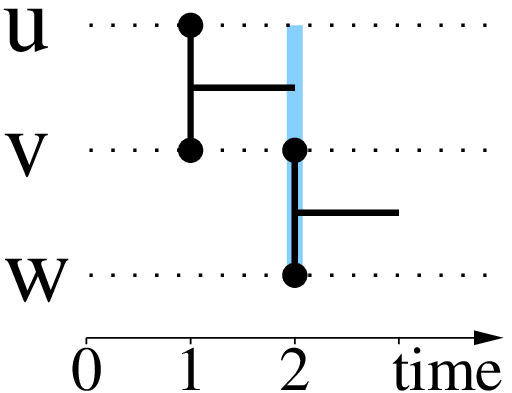}
\hfill
\includegraphics[scale=\myscale]{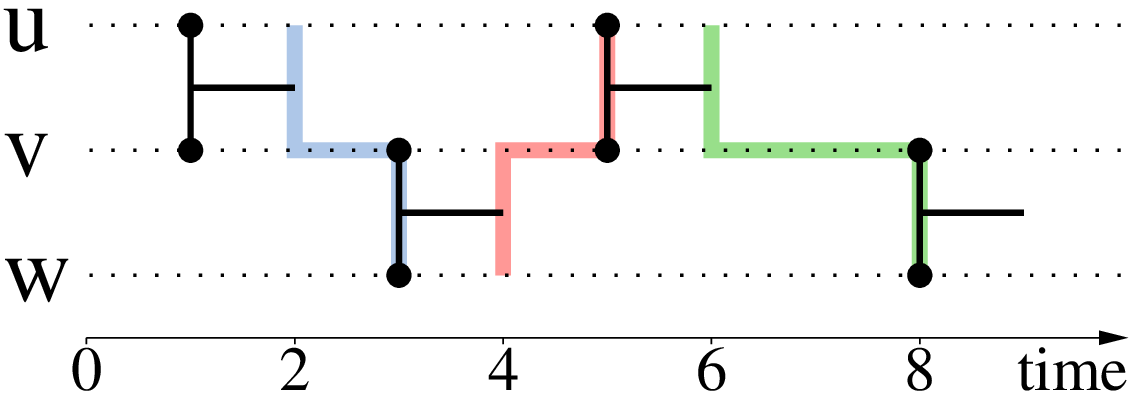}
\hfill
\includegraphics[scale=\myscale]{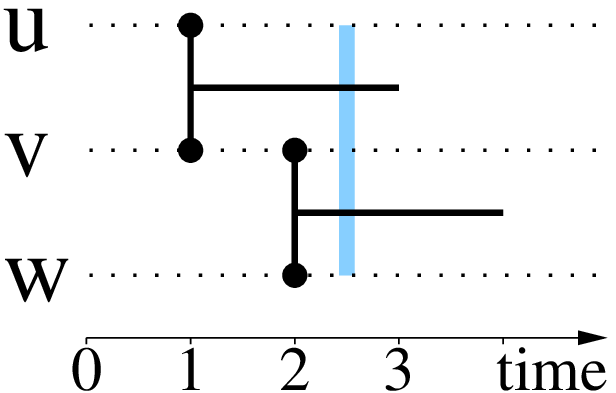}
\hfill\ 
\caption{
{\bf Basic examples for betweenness centrality computations in link streams.}
{\bf Left:} $L_1 = (T,V,E)$ with $T = [0,4]$, $V=\{u,v,w\}$, and $E = [1,2]\times\{uv\} \cup [2,3]\times\{vw\}$. We display in blue the unique \sfp\ from $u$ to $w$.
{\bf Center:}
$L_2 = (T,V,E)$ with $T = [0,10]$, $V=\{u,v,w\}$, and $E = ([1,2]\cup[5,6])\times\{uv\} \cup ([3,4]\cup[8,9])\times\{vw\}$. We display in blue and in green the two \sfp s from $u$ to $w$ (for different starting and arrival times), and in red the unique \sfp\ from $w$ to $u$.
{\bf Right:}
$L_3 = (T,V,E)$ with $T = [0,6]$, $V=\{u,v,w\}$, and $E = [1,3]\times\{uv\} \cup [2,4]\times\{vw\}$. We display in blue an instance of \sfp\ between $u$ and $w$.
}
\label{fig:betweenness-1}
\end{figure}

Let us consider for instance the case of $L_1$ defined in Figure~\ref{fig:betweenness-1} (left), and let us compute the betweenness $\betweenness(t,v)$ of $(t,v)$ for all $t$. To do so, we consider successively all possible pairs of nodes.

Let us begin with $u$ and $w$. There is a path from $(i,u)$ to $(j,w)$ only for $i$ in $[0,2]$ and $j$ in $[2,4]$. Then, there is a unique \sfp, and it is $(2,u,v),(2,v,w)$. It involves $v$ at time $2$ and only at this time. Therefore for all $i \in [0,2]$ and $j \in  [2,4]$, the value of $\bfrac{u}{w}$ is $1$ if $t=2$, and $0$ otherwise. These values are the same for paths from $w$ to $u$.

For all times $i$ and $j$, all \sfp s from $(i,u)$ to $(j,v)$, if any, are of the form $(k,u,v)$ for $k\in [\max(1,i),\min(2,j)]$. For $i<j$, there is an infinity of such paths and at most one involves $(t,v)$, leading to $\bfrac{u}{v} = 0$. If $i=j$ then there is a unique \sfp, and it involves $(t,v)$ only when $i=j=t$. Therefore, $\bfrac{u}{v}$ is different from $0$ only for $i=j=t$, while $i \in [0,2]$ and $j \in [\max(1,i),4]$, and so the contribution to $\betweenness(t,v)$ of paths from $u$ to $v$ is $0$. The same reasoning holds for paths from $v$ to $u$, from $w$ to $v$, and from $v$ to $w$.

Finally, \sfp s from $v$ to $v$ are empty sequences and so they do not involve $(t,v)$ for any $t$.

This leads to $\betweenness(2,v) = 2\cdot\int_0^2 \int_2^4 1 \diff j \diff i = 8$ and for all $t \ne 2$, $\betweenness(t,v) =0$.

\exampleskip

\exampleskip

Let us now consider the case of $L_2$, defined in Figure~\ref{fig:betweenness-1} (center), and let us first focus on the paths from $u$ to $w$. For any $i$, if $j<3$ then $(i,u) \notreaches (j,w)$. For $i\in[0,2]$ and $j\in[3,10]$, the unique \sfp\ from $(i,u)$ to $(j,w)$ is $(2,u,v),(3,v,w)$, in blue in the figure. For $i\in]2,10]$ and $j\in[0,8[$, $(i,u) \notreaches (j,w)$. For $i\in]2,6]$ and $j\in[8,10]$, the unique \sfp\ from $(i,u)$ to $(j,w)$ is $(6,u,v),(8,v,w)$, in green in the figure. Finally, for $i\in]6,10]$ and any $j$, $(i,u) \notreaches (j,w)$. Therefore, $\bfrac{u}{w}$ is different from $0$ only when $t\in[2,3]$, $i\in[0,2]$, $j\in[3,10]$ and when $t\in[6,8]$, $i\in]2,6]$, $j\in[8,10]$. It is then equal to 1.

Regarding paths from $(i,w)$ to $(j,u)$, the unique \sfp\ is $(4,w,v),(5,v,u)$ and it exists for $i\in[0,4]$ and $j\in[5,10]$. It involves $(t,v)$ when $t\in [4,5]$, leading to $\bfrac{w}{u} = 1$ for $t\in [4,5]$, $i\in[0,4]$, $j\in[5,10]$, and $0$ otherwise.

Like for $L_1$, the contribution of other pairs of nodes to $\betweenness(t,v)$ is $0$, and so we finally obtain
$\betweenness(t,v) = \int_0^2 \int_3^{10} 1 \diff j \diff i = 14$ for $t\in[2,3]$, 
$\betweenness(t,v) = \int_2^6 \int_8^{10} 1 \diff j \diff i = 8$ for $t\in[6,8]$, 
$\betweenness(t,v) = \int_0^4 \int_5^{10} 1 \diff j \diff i = 20$ for $t\in[4,5]$,
and $\betweenness(t,v) = 0$ otherwise.

\exampleskip

In the case of $L_3$ defined in Figure~\ref{fig:betweenness-1} (right), first notice that all \sfp s from $(i,u)$ to $(j,w)$ and from $(i,w)$ to $(j,u)$, if any, are of the form $(k,u,v),(k,v,w)$ or $(k,w,v),(k,v,u)$, respectively, with $k \in [2,3]$, $k\ge i$ and $k\le j$. Therefore, $\betweenness(t,v) = 0$ if $t \not\in [2,3]$. Moreover, $\bfrac{u}{w} = \bfrac{w}{u}$.


If $t \in [2,3]$, in the same way as for paths from $u$ to $v$ in $L_1$, we are in one of two cases: either there is an infinity of \sfp s from $(i,u)$ to $(j,w)$ and at most one of them involves $(t,v)$, or the fraction of values of $i$ and $j$ such that $\bfrac{u}{w}\ne 0$ is $0$.



Since, like in previous cases, the contribution of other pairs of nodes is $0$, and so we obtain $\betweenness(t,v) = 0$ in $L_3$ for all $(t,v)$.

However, let us consider the cluster $X = [2,3]\times\{v\}$. Then, $\Xbfrac{u}{w} = \Xbfrac{w}{u}$ is equal to 1
for $i\in[0,2]$ and $j\in[2,5]$,
for $i\in[2,3]$ and $j\in[i,5]$,
and it is equal to $0$ in all other cases.
Moreover, $\Xbfrac{u}{w} = \Xbfrac{w}{u}$ is equal
to $1$ for $i\in[2,3]$ and $j\in[i,5]$,
to $0.5$ for $i \in [0,1]$ and $j\in [3,5]$,
to $\frac{j-2}{j-1}$ for $i\in[0,1]$ and $j\in[2,3]$,
to $\frac{j-2}{j-i}$ for $i\in[1,2]$ and $j\in[2,3]$,
to $\frac{1}{3-i}$ for $i\in[1,2]$ and $j\in[3,5]$,
and it is equal to $0$ in all other cases. 
Likewise, $\Xbfrac{w}{v} = \Xbfrac{v}{w}$ is equal
to $1$ for $j\in[2,3]$ and $i\in[0,j]$,
to $0.5$ for $i \in [0,2]$ and $j\in [4,5]$,
to $\frac{1}{j-2}$ for $i\in[0,2]$ and $j\in[3,4]$,
to $\frac{3-i}{j-i}$ for $i\in[2,3]$ and $j\in[3,4]$,
to $\frac{3-i}{4-i}$ for $i\in[2,3]$ and $j\in[4,5]$,
and it is equal to $0$ in all other cases. 


We therefore obtain $\betweenness(X)
= 2 \cdot ( \int_0^2 \int_2^5 1 \diff j \diff i + \int_2^3 \int_i^5 1 \diff j \diff i )
+ 2 \cdot ( \int_2^3 \int_i^5 1 \diff j \diff i + \int_0^1 \int_3^5 0.5 \diff j \diff i + \int_0^1 \int_2^3 \frac{j-2}{j-1} \diff j \diff i + \int_1^2 \int_2^3 \frac{j-2}{j-i} \diff j \diff i + \int_1^2 \int_3^5 \frac{1}{3-i} \diff j \diff i )
+ 2 \cdot ( \int_2^3 \int_0^j 1 \diff i \diff j + \int_0^2 \int_4^5 0.5 \diff j \diff i + \int_0^2 \int_3^4 \frac{1}{j-2} \diff j \diff i + \int_2^3 \int_3^4 \frac{3-i}{j-i} \diff j \diff i + \int_2^3 \int_4^5 \frac{3-i}{4-i} \diff j \diff i )
= 17 + (2\ln(2)+10) + (2\ln(2)+10)
\sim 39.77$.


\exampleskip

\begin{figure}[!h]
\centering
\ \hfill
\includegraphics[scale=\myscale]{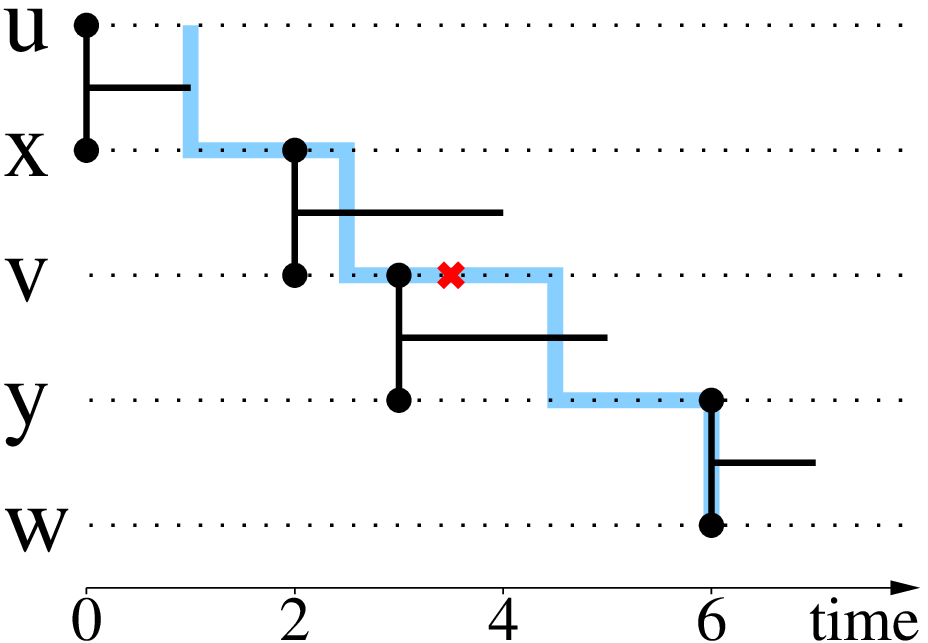}
\hfill
\includegraphics[scale=\myscale]{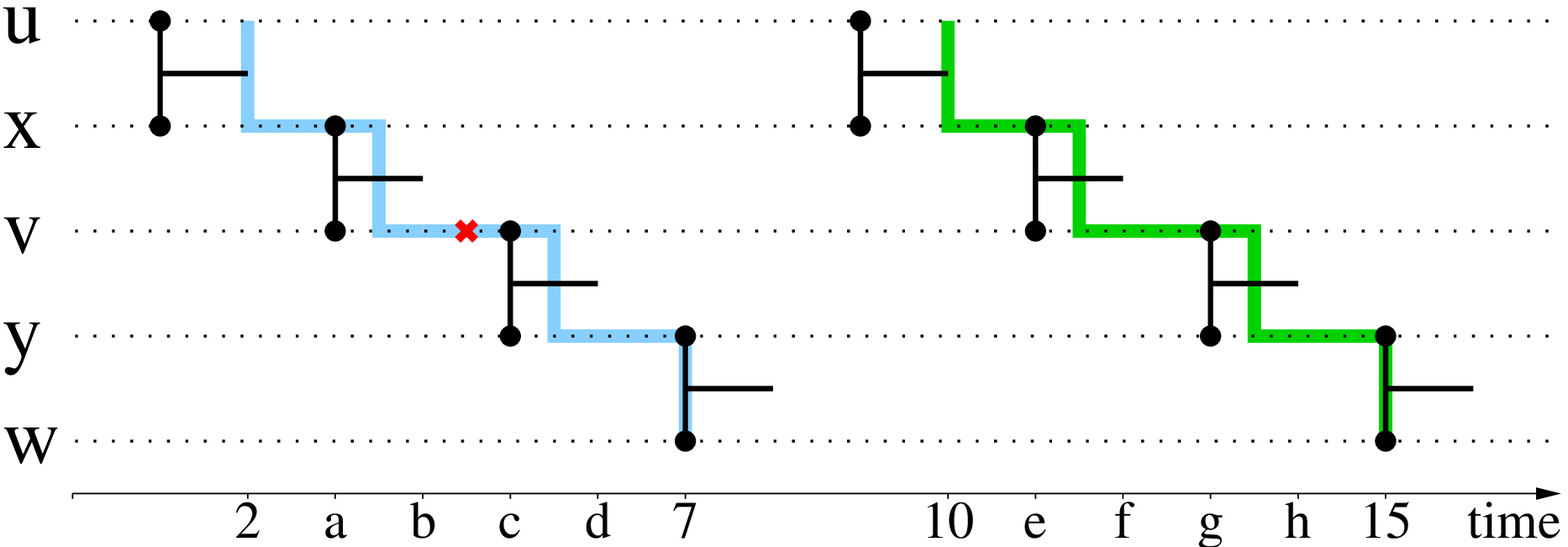}
\hfill\ 
\caption{
{\bf More examples for betweenness centrality computations in link streams.}
{\bf Left:}
$L_4 = (T,V,E)$ with $T = [0,8]$, $V=\{u,x,v,y,w\}$, and $E = [0,1]\times\{ux\} \cup [2,4]\times\{xv\} \cup [3,5]\times\{vy\} \cup [6,7]\times\{yw\}$. We display in blue an instance of \sfp\ from $u$ to $w$ and in red the element $(3.5,v)$.
{\bf Right:}
$L_5 = (T,V,E)$ with $T = [0,17]$, $V=\{u,x,v,y,w\}$, and $E = ([1,2]\cup[9,10])\times\{ux\} \cup ([a,b]\cup[e,f])\times\{xv\} \cup ([c,d]\cup[g,h])\times\{vy\} \cup ([7,8]\cup[15,16])\times\{yw\}$, for given values of $a$, $b$, $c$, $d$, $e$, $f$, $g$, and $h$ such that $2<a\le b<c\le d<7$ and $10<e\le f<g\le h<15$. The \sfp s from $u$ to $w$ all belong to two families: $(2,u,x),(k,x,v),(l,v,y),(7,y,w)$ with $k\in[a,b]$ and $l\in[c,d]$ (an instance is displayed in blue); and $(10,u,x),(m,x,v),(n,v,y),(15,y,w)$ with $m\in[e,f]$ and $n\in[g,h]$ (in green). We display in red an element $(t,v)$ with $t\in[b,c]$.
}
\label{fig:betweenness-2}
\end{figure}

Now let us consider $L_4$ defined in Figure~\ref{fig:betweenness-2} (left). We compute the contribution of $u$ and $w$ to the betweenness of $(3.5,v)$, displayed in red in the figure, \ie\ $\frac{\sigma((i,u),(j,w),(3.5,v))}{\sigma((i,u),(j,w))}$ for all $i$ and $j$. There is a \sfp\ from $(i,u)$ to $(j,w)$ only for $i\in[0,1]$ and $j\in[6,8]$, and it is always of the form $(1,u,x),(k,x,v),(l,v,y),(6,y,w)$ with $k\in[2,4]$, $l\in[3,5]$, and $l\ge k$. Among them, the ones involving $(3.5,v)$ are exactly those such that $k\in[2,3.5]$ and $l\in[3.5,5]$. This leads to the fraction $\frac{|[2,3.5]\times[3.5,5]|}{|[2,3]\times[3,5]| + \frac{1}{2}|[3,4]|^2 + |[3,4]\times[4,5]|} \sim 0.64$.

\exampleskip

Let us finally consider $L_5$ defined in Figure~\ref{fig:betweenness-2} (right). We compute the contribution of $u$ and $w$ to the betweenness of $(t,v)$, for a $t$ in $[b,c]$, like the one displayed in red in the figure. There are two families of \sfp s from $u$ to $w$ in this link stream: $(2,u,x),(k,x,v),(l,v,y),(7,y,w)$ with $k\in[a,b]$ and $l\in[c,d]$, that we call the {\em blue} family; and $(10,u,x),(m,x,v),(n,v,y),(15,y,w)$ with $m\in[e,f]$ and $n\in[g,h]$, that we call the {\em green} family. Notice that $(t,v)$, for a $t$ in $[b,c]$, is involved in all blue paths and in no green path. For $i\in[0,2]$ and $j\in[7,15[$ the \sfp s from $(i,u)$ to $(j,w)$ are the blue ones (they all involve $(t,v)$); for $i\in[0,2]$ and $j\in[15,17]$ they are both the blue and green ones (a fraction $\frac{(b-a)\cdot(d-c)}{(b-a)\cdot(d-c)+(f-e)\cdot(h-g)}$ of them involve $(t,v)$); for $i\in]2,10]$ and $j\in[15,17]$ they are the green ones (none of them involve $(t,v)$); and for all other values $i$ and $j$ there is no path from $(i,u)$ to $(j,w)$. This leads to 
$\int_0^2 \int_7^{15} 1 \diff j \diff i + \int_0^2 \int_{15}^{17} \frac{(b-a)\cdot(d-c)}{(b-a)\cdot(d-c)+(f-e)\cdot(h-g)} \diff j \diff i = 16 + 4 \cdot \frac{(b-a)\cdot(d-c)}{(b-a)\cdot(d-c)+(f-e)\cdot(h-g)}$.

\smallskip

In the computations above, we however assumed that $a \ne b$, $c \ne d$, $e \ne f$ and $g \ne h$ when we wrote that the fraction of blue paths in the set of all green and blue paths is $\frac{(b-a)\cdot(d-c)}{(b-a)\cdot(d-c)+(f-e)\cdot(h-g)}$. If $a=b$ or $c=d$, but $e \ne f$ and $g \ne h$ this still holds, as there are infinitely less blue paths than green ones ; the fraction of blue paths is $0$. If $a=b$ and $e=f$, but $c \ne d$ and $g \ne h$, however, the fraction above is undefined and the fraction of blue paths becomes $\frac{(d-c)}{(d-c)+(h-g)}$. Going further, if $a=b$, $c=d$, $e=f$ and $g=h$ then there is exactly one blue path and one green path, leading to a fraction of blue paths of $\frac{1}{2}$.

\section{Discrete versus continuous time}
\label{sec:continuous-vs-discrete}
\label{sec:discrete-vs-continuous}

All our examples and illustrations until now assumed that the set of time instants used in the definitions of stream graphs is an interval $[\alpha,\omega]$ of $\mathbb{R}$, thus bounded, infinite and continuous.
When we defined stream graphs in Section~\ref{sec:stream-graphs-and-link-streams}, we however claimed that our formalism is much more general, and may be used with different kinds of time modeling: bounded or unbounded, finite or infinite, continuous or discrete, and all combinations.

\begin{figure}[!h]
\centering
\includegraphics[scale=\myscale]{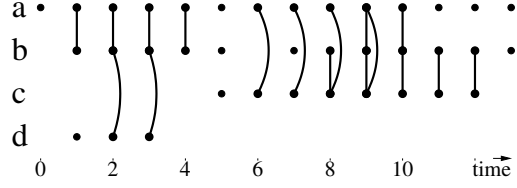}
\caption{
{\bf An example \myS\ of stream graph with discrete time.}
It is defined by $T = [0,13] \subseteq \mathbb{N}$, \ie\ $T = \{0,1,2,3,4,5,6,7,8,9,10,11,12,13\}$, $V = \{a,b,c,d\}$,
$\presence{a} = T$, $\presence{b} = \{1,2,3,4,5,7,8,9,10,11,12,13\}$, $\presence{c} = [5,12] = \{5,6,7,8,9,10,11,12\}$, $\presence{d} = [1,3] = \{1,2,3\}$,
$\presence{ab} = [1,4] \cup [9,10] = \{1,2,3,4,9,10\}$, $\presence{ac} = [6,9] = \{6,7,8,9\}$, $\presence{bc} = [8,12] = \{8,9,10,11,12\}$, $\presence{bd} = [2,3] = \{2,3\}$, and $\presence{ad} = \presence{cd} = \emptyset$.
}
\label{fig:discrete}
\end{figure}

Even with these various types of time sets, the formalism we developed in this paper applies directly (one just has to switch from integrals to sums in the case of discrete time). We illustrate this in this section by considering the situation where the set of time instants is an interval of $\mathbb{N}$ instead of $\mathbb{R}$, thus bounded, finite and discrete. See Figure~\ref{fig:discrete} for an illustration. We discuss more complex cases by the end of this section.

\notionskip

Let us consider \myS\ with $T = [\alpha,\omega] \subseteq \mathbb{N}$. The definitions of
$\cov{S} = \frac{|W|}{|T\times V|}$,
$n_v=\frac{|\presence{v}|}{|T|}$, $n = \frac{|W|}{|T|}$,
$m_{uv} = \frac{|\presence{uv}|}{|T|}$, $m = \frac{|E|}{|T|}$,
$k_t = \frac{|\V{t}|}{|V|}$, $k = \frac{|W|}{|V|}$,
$l_t = \frac{|\E{t}|}{|V \otimes V|}$, $l = \frac{|E|}{|V\otimes V|}$,
and $\uniformity(S) = \frac{\sum_{uv \in V\otimes V} |\presence{u} \cap \presence{v}|}{\sum_{uv \in V\otimes V} |\presence{u} \cup \presence{v}|}$ are directly applicable. For the example in Figure~\ref{fig:discrete}, we obtain for instance $\cov{S}=\frac{37}{14\cdot 4} \sim 0.66$, $n_a=1$, $n_d = \frac{3}{14} \sim 0.21$, $n=\frac{37}{14} \sim 2.6$, $k_0 = 0.25$, $k_1 = 0.75$ and $l_{10} = \frac{2}{6} \sim 0.33$.

The definition of density also holds (the node-based definition is identical and in the time-based definition the integral just needs to be replaced by a sum): $\delta(S) = \frac{\sum\limits_{uv\in V\otimes V}|\presence{uv}|}{\sum\limits_{uv\in V\otimes V}|\presence{u} \cap \presence{v}|} = \frac{\sum\limits_{t\in T} |\E{t}|}{\sum\limits_{t \in T} |\V{t}\otimes \V{t}|}$. In our example in Figure~\ref{fig:discrete}, $\delta(S) = \frac{\substack{0+1+2+2+1+0+1\\+1+2+3+2+1+1+0}}{\substack{0+3+3+3+1+3+1\\+3+3+3+3+3+3+1}} = 0.5$.

Going further, the definitions of substreams and clusters also apply. For instance, $C = \{(1,a), (1,b), (2,a), (2,b), (2,d)\}$ is a cluster of $S$ defined in Figure~\ref{fig:discrete}, and it induces the substream $S' = (T,V,C,E')$ with $E'=\{(1,ab),(2,ab),(2,bd)\}$.

As a consequence, the concepts of cliques, neighborhoods, degrees, and clustering coefficients, which depend only on the concepts above (cluster, density, number of nodes), are also directly applicable to this case. For instance, $\{9\}\times\{a,b,c\}$ is a maximal compact clique of $S$ defined in Figure~\ref{fig:discrete}, as well as $\{1,2,3,4\}\times\{a,b\}$, the neighborhood of $d$ is $\{(2,b),(3,b)\}$ and so $d$ has degree $\frac{2}{14} \sim 0.14$.

The quotient stream and the line stream of a discrete stream graph are also discrete stream graphs, with unchanged definitions. Likewise, the definition of $k$-cores is unchanged.

Paths in $S$ are now discrete objects, but this does not call for new definitions. For instance, in Figure~\ref{fig:discrete}, $(7,a,c),(8,c,b)$ is a path from $(0,a)$ to $(9,b)$. It involves exactly $(7,a)$, $(7,c)$, $(8,c)$, and $(8,b)$. It has length $2$ and duration $1$. It is not a shortest path from $(0,a)$ to $(9,b)$ since path $(9,a,b)$ is shorter. It is not a fastest path either, because path $(9,a,b)$ is faster. However, the path $(8,a,c),(8,c,b)$ has length $2$ and duration $0$, therefore it is a fastest path from $(0,a)$ to $(9,b)$ but not a shortest one.

Likewise, the definitions of connected clusters and components, as well as those of trees and cascades, that rely only on the concept of paths, directly translate to the discrete case.
The concepts of closeness and betweenness also do, but the betweenness relies now on a counting of discrete sets of (discrete) paths. Replacing the integral in its definition by a sum, it becomes:
$$
\betweenness(t,v)
= \sum_{(i,u)\in W, (j,w)\in W} \bfrac{u}{w}
$$
where
$\sigma((i,u),(j,w))$
is now the (finite) number of \sfp s from $(i,u)$ to $(j,w)$, and 
$\sigma((i,u),(j,w),(t,v))$ is the number of these path that involve $v$ at time $t$. Therefore, as before, $\bfrac{u}{w}$ is the fraction of all \sfp s from $(i,u)$ to $(j,w)$ that involve $(t,v)$.

In the case of Figure~\ref{fig:discrete}, for instance, $\sigma((0,a),(3,d)) = |\{((2,a,b),(2,b,d)), ((3,a,b),(3,b,d))\}| = 2$ paths. One of them involves $(2,b)$, and so $\frac{\sigma((0,a),(3,d),(2,b))}{\sigma((0,a),(3,d))} = 0.5$.


\notionskip

Finally, we have shown that considering a bounded discrete time interval does not call for new definitions and any change in our formalism: it directly applies to this case in a way very similar to the bounded continuous time case. This also holds for more complex time sets, including unbounded ones (then, a node $v$ has to be present a finite fraction of this infinite time set to satisfy $n_v \ne 0$) and/or discontinuous ones ($T$ may for instance be a collection of intervals of $\mathbb{R}$ or $\mathbb{N}$, or even parts of $\mathbb{Q}$). As such cases have limited practical and theoretical interest, we do not give more details here.

\section{$\Delta$-analysis and instantaneous links}
\label{sec:delta}

In some situations, directly studying the stream graph induced by a dataset makes little sense. If one considers phone calls or sexual contacts, for instance, nodes generally have only zero or one link at a time, leading to an instantaneous degree of $0$ or $1$. In the case of instant messaging or sensor-based measurements of proximity between individuals, links are instantaneous, leading to a density equal to $0$.

In such cases and in many others, one is generally interested in the fact that nodes interact regularly, typically at least once every $\Delta$ units of time, for a given $\Delta$. For instance, two individuals call each-other at least once a day, two sensors detect each-other at least once every ten seconds, etc.

Then, the usual approach consists in using this value of $\Delta$ as a parameter to define notions to describe the data, an approach that we call $\Delta$-analysis. For instance, one defines $n_\Delta$ and $m_\Delta$ as the expected number of nodes and links, respectively, present in a randomly chosen time interval of duration $\Delta$ in $T$. One defines the $\Delta$-degree $d_\Delta(v)$ of a node $v$ as its expected number of neighbors during a randomly chosen time interval of duration $\Delta$ in $T$. The $\Delta$-density is defined as follows\,\footnote{This is a generalization to stream graphs of the $\Delta$-density introduced in \cite{DBLP:conf/infocom/ViardL14} for link streams.}. Assume one takes a random time interval of duration $\Delta$ and two nodes involved in $S$ at some time during this interval, \ie\ a random triplet $(I,u,v)$ with $I=[t-\demidelta,t+\demidelta] \subseteq T$, $u$ and $v$ in $V$ such that $T_u \cap I$ and $T_v \cap I$ are nonempty. The $\Delta$-density of $S$ is the probability that these two nodes are linked together during this interval, \ie\ that $T_{uv} \cap I$ is nonempty.

\notionskip

The formalism we developed in this paper provides a more general way to deal with such cases, that we now present.

\notionskip

\begin{figure}[!h]
\centering
\ \hfill
\includegraphics[scale=\myscale]{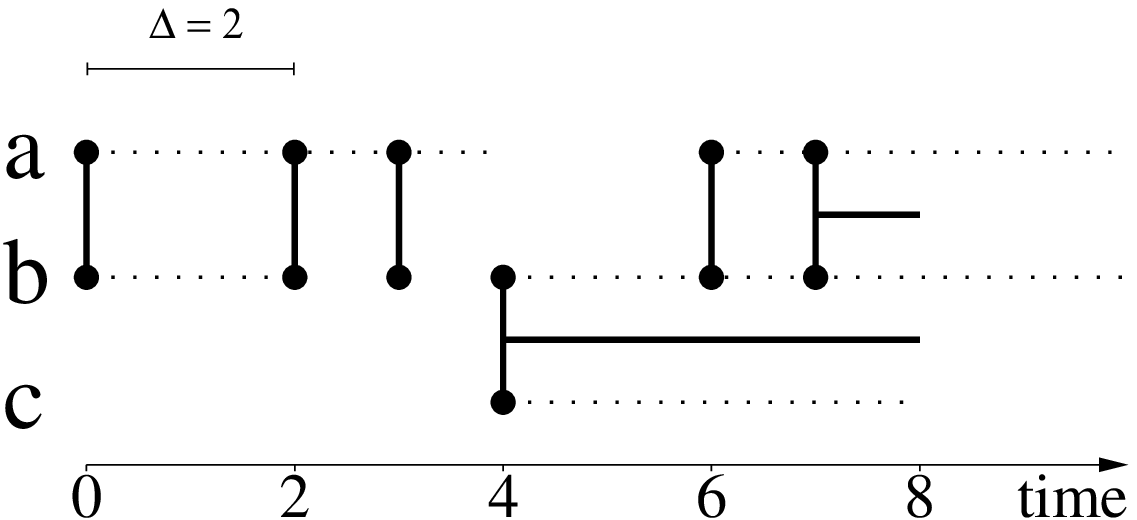}
\hfill
\includegraphics[scale=\myscale]{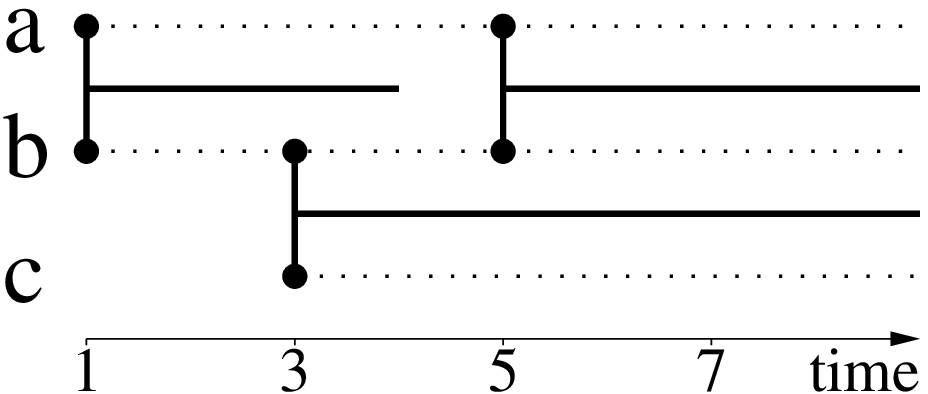}
\hfill \
\caption{
{\bf $\Delta$-analysis of a stream graph.} We display a stream graph $\myS$ (left) and the stream graph $S_\Delta=(T_\Delta,V,W_\Delta, E_\Delta)$ (right) derived from it with $\Delta = 2$. Here, $T=[0,10]$, $V=\{a,b,c\}$, $W=([0,4]\cup[6,10])\times\{a\} \cup ([0,2]\cup\{3\}\cup[4,10])\times\{b\} \cup [4,8]\times\{c\}$, and $E=(\{0,2,3,6\}\cup[7,8])\times\{ab\} \cup [4,8]\times\{bc\}$, leading to $T_\Delta=[1,9]$, $W_\Delta=[1,9]\times\{a,b\} \cup [3,9]\times\{c\}$ and $E_\Delta=([1,4]\cup[5,9])\times\{ab\} \cup [3,9]\times\{bc\}$.
Notice that $S$ contains instantaneous links, like for instance $(0,ab)$, and instantaneous nodes, like $(3,b)$.
}
\label{fig:delta}
\end{figure}

Given a stream graph $\myS$ with $T=[\alpha,\omega]$ and a value $\Delta\le\omega-\alpha$, we define $S_\Delta = (T_\Delta,V,W_\Delta, E_\Delta)$ as the stream graph such that $T_\Delta = [\alpha+\demidelta, \omega-\demidelta]$,
$W_\Delta = (T_\Delta \times V) \cap \bigcup_{(t,v)\in W} [t-\demidelta,t+\demidelta] \times \{v\} = \{ (t',v), t'\in T_\Delta, \exists (t,v)\in W \mbox{ s.t. } |t'-t| \le \demidelta \}$ and
$E_\Delta = (T_\Delta \times V\otimes V) \cap \bigcup_{(t,uv)\in E} [t-\demidelta,t+\demidelta] \times \{uv\} = \{ (t',uv), t'\in T_\Delta, \exists (t,uv)\in E \mbox{ s.t. } |t'-t| \le \demidelta \}$.
See Figure~\ref{fig:delta} for an illustration.

In other words, a node is present at time $t'$ in $S_\Delta$ whenever it is present in $S$ at a time $t$ in $[t'-\demidelta,t'+\demidelta]$, \ie\ $T_{\Delta v} = T_\Delta \cap \{ t', \exists t\in T_v, |t'-t| \le \demidelta \}$. Likewise, any two nodes are linked together at time $t'$ in $S_\Delta$ whenever they are linked together in $S$ at a time $t$ in $[t'-\demidelta,t'+\demidelta]$, \ie\ $T_{\Delta uv} = T_\Delta \cap \{ t', \exists t\in T_{uv}, |t'-t| \le \demidelta \}$.

\relationskip

We now show that the properties of $S_\Delta$ actually are equivalent to the $\Delta$-properties of $S$, and so one may conduct $\Delta$-analysis of $S$ by transforming it into $S_\Delta$ first, and then by using the formalism of this paper.

Let us define instantaneous versions of the $\Delta$-properties of $S$ cited above: for all $t$ in $[\alpha+\demidelta,\omega-\demidelta]$, $n_{\Delta t}$ is the number of distinct nodes present at some time in $[t-\demidelta,t+\demidelta]$: $n_{\Delta t} = |\{v\in V, \exists t', |t'-t| \le \demidelta \mbox{ and } (t',v)\in W\}|$. We define $m_{\Delta t}$ similarly, and $d_{\Delta t}(v)$ as the number of distinct nodes linked to $v$ at some time in $[t-\demidelta,t+\demidelta]$.

First notice that $n_{\Delta t}$ in $S$ is equal to $|V_t|$ in $S_\Delta$. Indeed, $|V_t|$ in $S_\Delta$ is equal to $|\{v\in V, \exists t'\in T, (t',v)\in W \mbox{ and } |t'-t|\le\demidelta\}| = n_{\Delta t}$. Likewise, $m_{\Delta t}$ in $S$ equals $|E_t|$ in $S_\Delta$, and $d_{\Delta t}(v)$ in $S$ equals $d_t(v)$ in $S_\Delta$.

Notice now that $n_\Delta = \frac{1}{|T_\Delta|}\cdot \int_{t\in T_\Delta} n_{\Delta t} \diff t$ in $S$. Therefore, it is equal to $\frac{1}{|T_\Delta|}\cdot \int_{t\in T_\Delta} |V_t| \diff t$  in $S_\Delta$, which is exactly $n$ in $S_\Delta$. Similar reasoning lead to the facts that $m_\Delta$ in $S$ is equal to $m$ in $S_\Delta$, and that $d_\Delta(v)$ in $S$ is equal to $d(v)$ in $S_\Delta$ for all $v$.

Going further, we have $\delta_\Delta(S) = \delta(S_\Delta)$. Indeed, $\delta(S_\Delta)$ is the probability that a random $(t,u,v)$ with $t\in T_\Delta$, $(t,u)\in W_\Delta$ and $(t,v)\in W_\Delta$ satisfies $(t,uv) \in E_\Delta$; and $\delta_\Delta(S)$ is the probability that a random $(I,u,v)$ with $I$ an interval of $T$ of duration $\Delta$, $u \in V$ and $v\in V$ such that $T_u \cap I \ne \emptyset$ and $T_v \cap I \ne \emptyset$ satisfies $T_{uv} \cap I \ne \emptyset$. A triplet $(t,u,v)$ satisfies the first set of constraints if and only if the triplet $(I,u,v)$ with $I=[t-\demidelta,t+\demidelta]$ fits the second set of constraints. In addition, it satisfies $(t,uv) \in E_\Delta$ if and only if $(I,u,v)$ satisfies $T_{uv} \cap I \ne \emptyset$. Therefore, the two probabilities are equal.


\notionskip

Finally, our approach makes it easy to conduct the $\Delta$-analysis of a stream graph $S$: it is equivalent to analyzing the stream-graph $S_\Delta$ with the general methods developed here, which go much further than the previously considered $\Delta$-properties.

This approach has another strength: one may use variable values of $\Delta$, which may be a function of time, depend on the involved nodes or links, or any other property. One may for instance consider that two colleagues are in contact whenever they meet each other at least once a week, but for holidays they remain in contact if they meet in the week before holidays and in the week after. It is easy to capture such modeling choice within our framework: they only change the way one builds $S_\Delta$ from $S$, and the analysis of $S_\Delta$ remains unchanged. Defining properties that would directly take into account such variations of $\Delta$ would be much more complex.

\section{Bipartite streams and other generalizations}
\label{sec:extensions}
\label{sec:generalizations}
\label{sec:bipartite}

An important strength of the graph formalism is that it may easily be extended to encompass richer, more complex cases. For instance, one may consider directed links by defining directed graphs $G=(V,E)$ with $E\subseteq V\times V$ instead of $E \subset V\otimes V$. One may allow loops ($vv\in E$ is possible), and/or consider multigraphs ($E$ is a multiset, thus several links between the two same nodes are possible). Going further, one may capture link strength or cost using weighted graphs, in which a weight is associated to each element of $E$ and/or $V$. One may combine these extensions by considering for instance directed weighted multigraphs.

Dealing with such graph generalizations calls for an update of classical graph concepts. For instance, the density of directed graphs must take into account the fact that the number of possible links changed; it also leads to notions of in- and out-degrees (the number of links towards and from a given node); etc. Some properties are non-trivial to extend (like for instance the density for weighted graphs) but most just need to be patched, thus giving a great expressivity and wide areas of applications to graphs.

{\bf Bipartite graphs are a particularly pervasive graph extension, and this section details this case as an illustration}: we show how a few key extensions of graph concepts to the bipartite case \cite{DBLP:journals/socnet/LatapyMV08} lead to similar extensions for bipartite stream graphs.

\notionskip

{\em
A bipartite graph $G=(\top,\bot,E)$ is defined by a set of top nodes $\top$, a set of bottom nodes $\bot$, and a set of links $E \subseteq \top \times \bot$: links may exist only between top and bottom nodes. This models data like client-product relations or affiliation networks: the considered nodes belong to two different sets and links may exist only between nodes in one set and nodes in the other set.

In $G$, $n_\top = |\top|$ and $n_\bot = |\bot|$ denote the number of top and bottom nodes. The definition of the number of links $m$ is the same as in classical graphs. The (bipartite) density of $G$ is then $\delta(G) = \frac{m}{n_\top \cdot n_\bot}$: it is the probability when one takes two nodes that may be linked together that they indeed are. Node neighborhoods and degrees are defined like in a classical (non-bipartite) graph. The average top and bottom degrees $d_\top$ and $d_\bot$ of $G$ are the average degrees of top and bottom nodes, respectively.

The top and bottom projections $G_\top = (\top,E_\top)$ and $G_\bot = (\bot,E_\bot)$ of $G$ are defined by $E_\top = \cup_{v\in \bot} N(v) \otimes N(v)$ and $E_\bot = \cup_{v\in \top} N(v) \otimes N(v)$, respectively. In other words, in $G_\top$ two (top) nodes are linked together if they have (at least) a (bottom) neighbor in common in $G$, and $G_\bot$ is defined symmetrically. If $v\in\top$ (resp. $v\in\bot$) then $N(v)$ always is a (not necessarily maximal) clique in $G_\bot$ (resp. $G_\top$).

Given a top node $v \in \top$ (the case of bottom nodes is symmetrical), let us denote by $G\setminus v$ the (bipartite) graph obtained by removing node $v$ and all its links from $G$: $G\setminus v = (\top\setminus\{v\},\bot,E\setminus(\{v\}\times\bot))$. The redundancy $rc(v)$ of $v \in \top$ is the density of the subgraph of $(G\setminus v)_\bot$ induced by its neighborhood $N(v)$ in $G$. In other words, it is the fraction of its pairs of neighbors that have (at least) another neighbor in common.
}

\graphstreamskip

\begin{figure}[!h]
\centering
\ \hfill
\includegraphics[scale=\myscale]{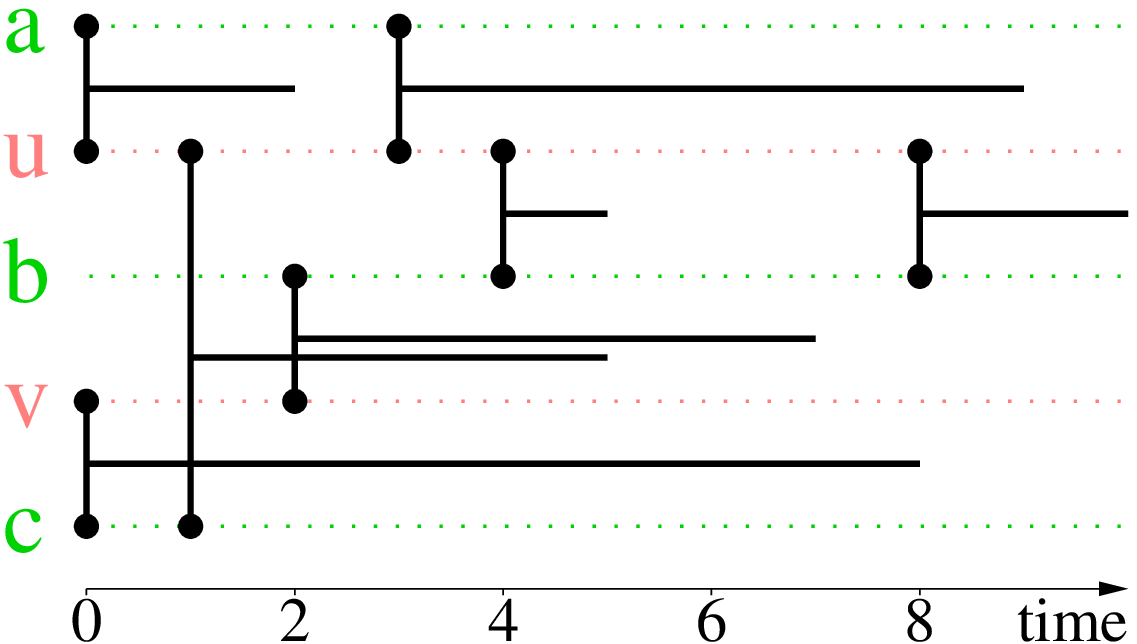}
\hfill
\includegraphics[scale=\myscale]{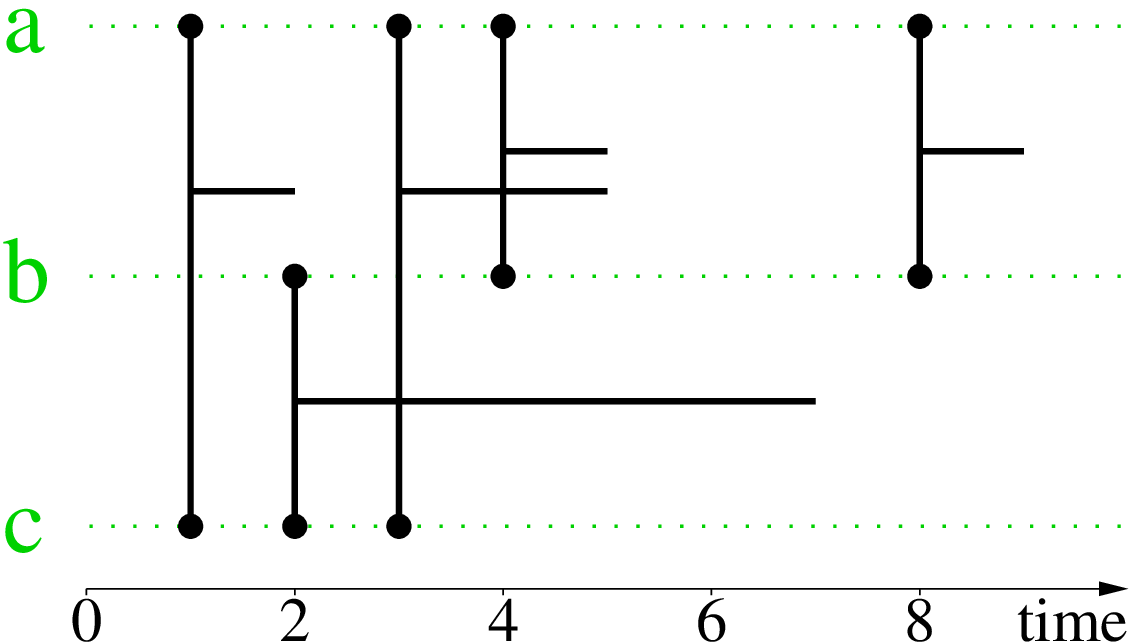}
\hfill\
\caption{
{\bf Left:} a bipartite link stream $L = (T,\top,\bot,E)$ with $T=[0,10]$, $\top=\{u,v\}$, $\bot=\{a,b,c\}$, and $E = ([0,2]\cup[3,9])\times\{(u,a)\} \cup ([4,5]\cup[8,10])\times\{(u,b)\} \cup [1,5]\times\{(u,c)\} \cup [2,7]\times\{(v,b)\} \cup [0,8]\times\{(v,c)\}$.
{\bf Right:} its $\bot$-projection $L_\bot$. For instance, $a$ and $c$ are linked together from time $3$ to $5$ because they both have $u$ in their neighborhood for this time period in $S$.
}
\label{fig:bipartite}
\end{figure}

We define a bipartite stream graph $S = (T,\top,\bot,W,E)$ from a set of top nodes $\top$, a set of bottom nodes $\bot$, a time span $T$ and two sets $W \subseteq T \times (\top\cup\bot)$ and $E \subseteq T \times \top \times \bot$ such that $(t,u,v) \in E$ implies $(t,u) \in W$ and $(t,v) \in W$. See Figure~\ref{fig:bipartite} (left) for an illustration.

In $S$, we extend $n = \frac{|W|}{|T|}$ into $n_\top = \frac{|W \cap (T\times\top)|}{|T|}$ and $n_\bot = \frac{|W \cap (T\times\bot)|}{|T|}$, the numbers of top and bottom nodes, respectively. We extend $k = \frac{|W|}{|V|}$ into $k_\top = \frac{|W \cap (T\times\top)|}{|\top|}$ and  $k_\top = \frac{|W \cap (T\times\top)|}{|\bot|}$ similarly, and we define $m$ and $l$ like for classical (non-bipartite) stream graphs.

We define the (bipartite) density of $S$ as $\delta(S) = \frac{\sum_{u\in\top,v\in\bot}|T_{uv}|}{\sum_{u\in\top,v\in\bot}|T_u\cap T_v|}$: it is the probability when one takes two nodes when they may be linked together that they indeed are. We define node neighborhoods and degrees like in a classical (non-bipartite) graph. We define the average top and bottom degrees $d_\top$ and $d_\bot$ of $S$ as the average degrees of top and bottom nodes, respectively, weighted by their presence time in the stream.

\notionskip

We define the top and bottom projections $S_\top = (T,\top,W_\top,E_\top)$ and $S_\bot = (T,\bot,W_\bot,E_\bot)$ of $S$ by $W_\top = W \cap (T\times\top)$, $W_\bot = W \cap (T\times\bot)$, $E_\top = \cup_{(t,v)\in W_\bot} \{(t,uw) \mbox{ s.t. } (t,u,v)\in E \mbox{ and } (t,w,v)\in E\}$ and $E_\bot = \cup_{(t,v)\in W_\top} \{(t,uw) \mbox{ s.t. } (t,v,u)\in E \mbox{ and } (t,v,w)\in E\}$, respectively. In other words, in $S_\top$ two (top) nodes are linked together at a given time instant if they have (at least) a (bottom) neighbor in common in $S$ at this time, and $S_\bot$ is defined symmetrically. See Figure~\ref{fig:bipartite} for an illustration. Notice that, if $v\in\top$ (resp. $v\in\bot$) then $N(v)$ always is a (not necessarily maximal) clique in $S_\bot$ (resp. $S_\bot$).

\notionskip

Given a top node $v \in \top$ (the case of bottom nodes is symmetrical), let us denote by $S\setminus v$ the (bipartite) stream graph obtained by removing node $v$ and all its links from $S$: $S\setminus v = (T,\top\setminus\{v\},\bot,W\setminus (T\times\{v\}),E\setminus(T\times \{v\}\times\bot))$. The redundancy $rc(v)$ of $v \in \top$ is the density of the substream of $(S\setminus v)_\bot$ induced by its neighborhood $N(v)$ in $S$. In other words, it is the fraction of its pairs of neighbors and time instants that have (at least) another neighbor in common at this time.

\relationskip

If $S$ is a graph-equivalent bipartite stream, then its corresponding graph also is bipartite. Moreover, the projections of $S$ are also graph-equivalent streams, and their corresponding graphs are the projections of the graph corresponding to $S$. In addition, the bipartite properties of $S$ are equivalent to the bipartite properties of its corresponding bipartite graph.

\section{Related work}
\label{sec:related}
\label{sec:related-work}


Studying interactions over time is crucial in a wide variety of contexts, leading to a huge number of papers dealing with various cases of interest. We cite for instance studies of phone calls \cite{Kovanen2013,Blondel2015}, contacts between individuals \cite{Barrat2013,Martinet2014,DBLP:conf/asunam/ViardLM15}, cattle exchanges \cite{Dutta2014,DBLP:conf/iccsa/PayenTL17}, messaging \cite{doi:10.1142/S0219525909002088,Gaumont2016}, or internet traffic \cite{Harshaw:2016:GTG:2897795.2897806,DBLP:conf/infocom/ViardL14}, but we could cite hundreds more. In each practical context, researchers and engineers face the challenge of analyzing the both temporal and structural nature of interactions, and they develop ad-hoc methods and tools to do so. Several surveys of these works are available from various perspectives \cite{livrelambiotte,SIZEMORE2017,doi:10.1162/NETN,SNIJDERS201044,Holme2015,livregeorge,HOLME201297,livredoreian}.

\medskip

The most classical approach consists in splitting time into slices and then building a graph, often called snapshot, for each time slice: its nodes and links represent the interactions that occurred during this time slice. One obtains a sequence of snapshots (one for each slice), and may study the time-evolution of their properties, see for instance \cite{DBLP:journals/corr/SikdarGM15,DBLP:journals/tkdd/LeskovecKF07,DBLP:journals/corr/abs-1102-0629,Gulyas2013,MEE3:MEE3236,Uddin2013TopologicalAO}, among many others. In \cite{DBLP:journals/snam/BatageljP16}, the authors even design a general framework to combine and aggregate wide classes of temporal properties, thus providing a unified approach for snapshot sequence studies. However, these approaches need time slices large enough to ensure that each snapshot captures significant information. But large slices lead to losses of temporal information, since all interactions within a same slice are merged. In addition, several or even varying slice durations may be relevant. As a consequence, choosing appropriate time slices is a research topic in itself \cite{DBLP:conf/conext/LeoCF15,quantifying2013,Krings2012,Scholtes2016,Caceres2013}. More importantly, key concepts like paths make little sense in this framework: paths within a slice do not respect the dynamics of interactions, and paths over several time slices are difficult to handle \cite{DBLP:conf/conext/LeoCF15}.

To avoid these issues, several authors propose to encode the full information into various kinds of augmented graphs. In \cite{DBLP:journals/paapp/CasteigtsFQS12,DBLP:journals/snam/BatageljP16,DBLP:journals/corr/abs-1102-0629} for instance, authors consider the graph of all nodes and links occurring within the data, and label each node and link with its presence times. In \cite{DBLP:conf/dsaa/WehmuthZF15,KOSTAKOS20091007,Michail2015,refId0}, the authors duplicate each node into as many copies as its number of occurrences (they assume discrete time steps); then, an interaction between two nodes at a given time is encoded by a link between the copies of these nodes at this time, and each copy of a node is connected to its copy at the next time step. In \cite{DBLP:conf/mobicom/WhitbeckACG12,doi:10.1063/1.3697996} and others, the authors build reachability graphs: two nodes are linked together if they can reach each other in the stream. With such encodings, some key properties of the stream are equivalent to properties of the obtained graph, and so studying this graph sheds light on the original data. However, concepts like density or clusters make little sense on such objects, and authors then resort to the time slicing approach \cite{DBLP:journals/corr/abs-1102-0629}.

All these approaches have a clear advantage: once the data is transformed into one or several graphs, it is possible to use graph tools and concepts to study the interactions under concern.
In the same spirit, various powerful methods for graph studies are extended to cope with the dynamics. This leads for instance to algebraic approaches for temporal network analysis \cite{DBLP:journals/snam/BatageljP16,DBLP:journals/corr/PraprotnikB16}, dynamic stochastic block models \cite{Xu2013,RSSB:RSSB12200,DBLP:conf/asunam/CorneliLR15,Corneli2016}, dynamic Markovian models \cite{Stadtfeld2017,doi:10.1177/0081175017709295,SOME:SOME099,SNIJDERS201044}, signals on temporal networks \cite{DBLP:journals/corr/HamonBFR15a}, adjacency tensors \cite{DBLP:conf/kdd/SunTF06,10.1371/journal.pone.0086028}, temporal networks studies with walks \cite{PhysRevE.85.056115,1367-2630-16-6-063023,Saramaki2015}, dynamic graphlets \cite{DBLP:journals/bioinformatics/HulovatyyCM16,Harshaw:2016:GTG:2897795.2897806} and temporal motif counting approaches \cite{1742-5468-2011-11-P11005,DBLP:conf/wsdm/ParanjapeBL17}. Clearly, these works extend higher-level methods to the temporal setting, whereas we focus here on the most basic graph concepts, in the hope that they will form a unifying ground to such works.

\medskip

Complementary to these approaches that extend methods, some works extend various graph concepts to deal with time, in a way similar to what we do here \cite{DBLP:journals/snam/BatageljP16,HOLME201297,Nicosia2013}.

In particular, path-related concepts received much attention because of their importance for spreading phenomena and communication networks, see for instance \cite{Holme2015,DBLP:conf/mobicom/WhitbeckACG12,Tang:2010:CTD:1672308.1672329,DBLP:conf/iccsa/PayenTL17}. Interestingly, although paths defined in these papers are similar to those we consider here, most derived concepts remain node-oriented. For instance most authors define the centrality of a given node and connected components as sets of nodes (without time information) \cite{DBLP:journals/snam/BatageljP16,Nicosia2013,DBLP:journals/corr/abs-1102-0629,DBLP:conf/mobicom/WhitbeckACG12,doi:10.1063/1.3697996,Tang:2010:CTD:1672308.1672329}. In \cite{DBLP:journals/advcs/CostaVWZS15}, the authors introduce a centrality for time instants. Since the centrality of nodes may greatly change over time \cite{DBLP:conf/asunam/MagnienT15}, it is important to define centralities of each node at each time instant. Some authors did so for various kinds of centralities \cite{e48b86952f3b4c04b534416f4a556b30,FLORES2017,refId0,Tang:2010:AIF:1852658.1852661,SIZEMORE2017} but, up to our knowledge, we are the first ones to consider paths from all nodes at all time instants to all other nodes at all other time instants. This has the advantage of fully capturing the dynamics of the data, in particular the fact that nodes are not always present.

Some works go beyond path-related notions and study dynamics of node and link presence, link repetitions, instantaneous degree, and triadic closure \cite{DBLP:conf/icwsm/ZignaniGRZZZ14,HERNANDEZORALLO2016160,Stadtfeld2017,Conan:2007:CPI:1365562.1365588,PhysRevE.81.055101,RSSB:RSSB12013,DBLP:conf/kdd/LeskovecBKT08,PhysRevE.64.025102,DBLP:journals/complexity/UddinKP16,DBLP:journals/snam/BatageljP16}.
However, up to our knowledge, there exists no previous generalization of density, neighborhood, or clustering coefficient that avoids time slicing.
Interestingly, a notion of degree very close to the one we propose here was introduced in the context of medical studies \cite{DBLP:journals/ijitdm/UddinHW14}. A notion close to average degree is introduced in \cite{Rozenshtein:2017:FDD:3058790.3046791} for dense dynamic sub-graphs searching. We also studied preliminary notions of density, cliques, quotient streams, and dense substreams in our own previous work \cite{DBLP:journals/tcs/ViardLM16,DBLP:journals/snam/GaumontML16,Gaumont2016,DBLP:conf/asunam/ViardLM15,DBLP:conf/infocom/ViardL14}.

\medskip



Finally, although there is a very rich body of works on temporal networks, dynamic graphs, longitudinal networks, time-varying graphs, relational event models, etc, none of these works aims at extending the basic graph theoretic language to the situation where time and structure are equally important, like we try to do here.


\section{Conclusion}
\label{sec:conclusion}


{\bf In this paper, we introduce a formalism to deal directly with the both temporal and structural nature of interactions over time.} We first define elementary concepts like numbers of nodes and links, density, clusters, and paths (Sections~\ref{sec:stream-graphs-and-link-streams} to~\ref{sec:clusters} and Section~\ref{sec:paths}). From them, we derive more advanced concepts like cliques, neighborhoods, degrees, clustering coefficients, and connected components (Sections~\ref{sec:cliques} to~\ref{sec:cdegree} and Section~\ref{sec:connectedness}), and we show how to go further by introducing quotient streams, line streams, $k$-cores and centralities (Sections~\ref{sec:quotient} to~\ref{sec:k-cores} and Section~\ref{sec:centralities}). Our formalism is able to cope with both discrete and continuous time (Section~\ref{sec:discrete-vs-continuous}), with both instantaneous links and links with durations (Section~\ref{sec:delta}), and we also consider the case where nodes have no dynamic, that we call link streams. Last but not least, our formalism may be extended to incorporate various features of the data, and we illustrate this with bipartite streams in Section~\ref{sec:bipartite}.

\notionskip

The strength of our approach is to rely on very basic (but non-trivial) innovations like non-integer numbers of nodes and links, symmetric roles for time instants and nodes, a simple and intuitive concept of density, an elementary definition of clusters, and paths that connect a node at a given time to a node at a given time. {\bf These basic concepts make it easy to define more advanced objects}: neighborhoods are clusters, degrees are fractional numbers of nodes in the neighborhoods, clustering coefficients are densities of neighborhoods, betweenness centralities are fractions of paths from any node at any time to any node at any time, etc. We demonstrate the strength of this approach by extending more advanced graph concepts such as quotient graph, trees, line graph, and $k$-cores, among others. Their definitions are mere retranscriptions of classical graph definitions into our formalism for stream graphs and link streams, and one may easily extend many other notions in this way.

In addition to this self-consistency, {\bf our formalism is consistent with graph theory} in a very strong and precise way: if one considers a stream graph with no dynamics (nodes are present all the time, and two nodes are either linked all the time or not at all), then the stream graph is equivalent to a graph and its stream properties are equivalent to the properties of the corresponding graph. As a consequence, our formalism is a generalization of graph theory, which provides a solid ground for generalizing other graph notions.

With our formalism, one is equipped with a wide set of concepts for describing data modeled as a stream graph or a link stream. It is natural to start with the description of how elementary metrics like $k_t$ (the fraction of nodes present at time $t$) evolve over time, and of distributions of values of $n_v$ (the fraction of time at which $v$ is present) for all nodes. One may then study the instantaneous degree distribution, the degree distribution of nodes, and the time-evolution of the time degree.  More advanced metrics and properties, such as connectedness, clustering coefficient or centralities, give finer insight on the data. Finally, just like graph concepts do for relations, {\bf our formalism provides a language for describing interactions over time} in an intuitive way, both at global and more local levels. Importantly, it does not require to choose a specific time scale for conducting such studies.

{\bf Data that would benefit from such an approach are countless,} but we believe that analysis of network traffic, mobility traces, and financial transactions are among the most promising ones, and we are working on such applications. Indeed, modeling such data with (directed, weighted) stream graphs and link streams captures most of their features, and progress in these fields is currently limited by the lack of appropriate modeling.

\notionskip

In order to conduct such real-world applications, it is crucial to design and implement convenient software able to efficiently compute the properties of large stream graphs. Work in this direction is in progress for the properties presented in this paper. However, it must be clear that some concepts raise serious algorithmic challenges. We worked for instance on clique and dense substream computations~\cite{DBLP:journals/tcs/ViardLM16,Gaumont2016}, and previous work exists on various problems, see for instance \cite{DBLP:journals/ijfcs/XuanFJ03,DBLP:journals/ijfcs/CasteigtsFMS15,DBLP:journals/paapp/CasteigtsFQS12,DBLP:journals/jisa/BhadraF12,Wu:2014:PPT:2732939.2732945}. In particular, the authors of \cite{DBLP:journals/paapp/CasteigtsFQS12} define a first complexity hierarchy for stream graphs. Still, most remains to be done in the design of efficient algorithms for stream graphs and the understanding of their complexity.

Another important direction is the design of models of stream graphs and link streams, which play a crucial role for simulations and proofs. In particular, an important approach in graph studies consists in generating uniformly at random graphs that have a prescribed set of properties. For instance, the Erd\"os-Renyi model generates graphs with prescribed size and density, while the configuration model generates graphs with prescribed size and degree distribution. The definitions we introduce in this paper (in particular for density and degree) open the way to the definition of models for generating stream graphs with prescribed properties, and to a more unified understanding of already existing models, like the ones defined in \cite{Zhao2013,Laurent2015,SOME:SOME099,DBLP:conf/kdd/LeskovecBKT08,SNIJDERS201044,Gulyas2013,PhysRevE.83.025102} for instance.

Last but not least, one may notice that stream graphs are not only generalizations of graphs. They actually lie at the crossroad of two very rich and powerful scientific areas: graph theory, as we have seen, and time series analysis. Indeed, if a stream graph has no dynamics then it is equivalent to a graph; if it has no structure then it is equivalent to a time series. As a consequence, we consider that a very promising direction for future work is to generalize time series concepts to stream graphs, in a way similar to what we did with graph concepts in this paper.

\bigskip
\noindent
{\small \bf Acknowledgments.} {\small This work is funded in part by the European Commission H2020 FETPROACT 2016-2017 program under grant 732942 (ODYCCEUS), by the ANR (French National Agency of Research) under grants ANR-15-CE38-0001 (AlgoDiv) and ANR-13-CORD-0017-01 (CODDDE), by the French program "PIA - Usages, services et contenus innovants" under grant O18062-44430 (REQUEST), and by the Ile-de-France program FUI21 under grant 16010629 (iTRAC). We warmly thank the many colleagues and friends who read preliminary versions of this work and provided invaluable feedback.}

\bibliographystyle{plain}
\bibliography{biblio}

\end{document}